\documentclass[cmp]{svjour}  
\usepackage{amssymb,latexsym,newlfont}
\usepackage{hike}

\renewcommand\makeheadbox{}
\begin{document}
\title{A Hiker's Guide to K3.\\
Aspects of $N=(4,4)$ Superconformal Field Theory
with Central Charge $c=6$}
\titlerunning{A Hiker's Guide to K3}
\author{
Werner Nahm\inst{1} \and
Katrin Wendland\inst{2}
}
\institute{Physikalisches Institut, Universit\a t Bonn,
Nu\ss allee 12, D-53115 Bonn, Germany\\
\email{werner@th.physik.uni-bonn.de}
\and
Physikalisches Institut, Universit\a t Bonn,
Nu\ss allee 12, D-53115 Bonn, Germany\\
\email{wendland@th.physik.uni-bonn.de}}
\authorrunning{W.~Nahm, K.~Wendland}
\date{}

\maketitle
\begin{abstract}
\noindent
We study the moduli space ${\cal M}$ of $N=(4,4)$ superconformal field
theories with  central charge $c=6$. After a slight emendation of its global
description we find the locations of various known models in the component
of ${\cal M}$ associated to $K3$ surfaces. Among them are the $\Z_2$ and
$\Z_4$ orbifold theories obtained from the torus component of ${\cal M}$.
Here, $SO(4,4)$ triality is found to play a dominant role. We obtain
the B-field values in direction of the exceptional divisors which arise
from orbifolding. 
We prove T-duality for the $\Z_2$ orbifolds and use it to derive the form of
${\cal M}$ purely within conformal field theory. For the Gepner model $(2)^4$
and some of its orbifolds we find the locations in ${\cal M}$ and prove
isomorphisms to nonlinear $\sigma$ models. In particular we prove that the
Gepner model $(2)^4$ has a geometric interpretation with Fermat quartic target
space.
\end{abstract}
This paper aims to make a contribution to a better understanding of the
$N=(4,4)$ superconformal field theories with left and right central charge
$c=6$. Ultimately, one would like to know their moduli space
${\cal M}$ as an algebraic space, their partition functions as functions on
${\cal M}$ and modular functions on the upper half plane, and an algorithm for
the calculation of all operator product coefficients, depending again on
${\cal M}$. This would constitute a good basis for the understanding of
quantum supergravity in six dimensions, and presumably for an investigation
of the more complicated physics in four dimensions.

The moduli space ${\cal M}$ has been identified with a high degree of
plausibility, though a number of details remain to be clarified.
It has two components, ${\cal M}^{tori}$ and ${\cal M}^{K3}$, one
16--dimensional associated to the four--torus and
one 80--dimensional associated to $K3$. The superconformal field theories in
${\cal M}^{tori}$ are well understood. One also understands some varieties of
theories which belong to ${\cal M}^{K3}$, including about 30 isolated
Gepner type models and varieties which contain orbifolds of theories in
${\cal M}^{tori}$. In the literature one can find statements concerning
intersections of these subvarieties, but not all of them are correct.
Indeed, their precise positions in ${\cal M}$ had not been studied up to now.
One difficulty is due to the fact that the standard description of
${\cal M}^{tori}$ is based on the odd cohomology of the torus, which does not
survive the orbifolding.

As varieties of superconformal theories ${\cal M}^{tori}$ and ${\cal M}^{K3}$
cannot intersect for trivial reasons. As ordinary conformal theories without
$\Z_2$ grading intersections are possible and will be shown to occur.

The plan of our paper is as follows.
In section \ref{modulispace} we will review known results 
following \cite{asmo94,as96}. We correct some of the details and add proofs
for well--known conjectural features. In section \ref{modtor} we explain the
connection between our description of ${\cal M}^{tori}$ in terms of the even
cohomology and the one given by Narain much earlier by odd cohomology
\cite{cent85,na86}.
Both are eight--dimensional, and they are related by $SO(4,4)$ triality.
Section \ref{orbifolds} deals with $\Z_2$ and $\Z_4$ orbifold
conformal field theories. We arrive at a description for the subvarieties of
these theories within ${\cal M}^{K3}$. In particular, we present a proof for
the well--known conjecture that orbifold conformal field theories tend to
give the value $B={1\over2}$ \cite[\para4]{as96} to the B-field in
direction of the exceptional divisors gained from the orbifold procedure and
determine the correct B-field values for $\Z_4$ orbifolds. 
Our results are in agreement with those of \cite{do97,blin97}, 
that were obtained in a different context. We calculate
the conjugate of torus T-duality under the $\Z_2$ orbifolding map to
${\cal M}^{K3}$ and find that it is a kind of squareroot of the Fourier--Mukai
T-duality on $K3$. This yields a proof of the latter and allows us to 
determine the form of ${\cal M}^{K3}$ purely within conformal field theory,
without having recourse to Landau-Ginzburg arguments.
We disprove the
conjecture that $\Z_2$ and $\Z_4$ orbifold moduli spaces meet
in the Gepner model $(2)^4$ \cite{eoty89}.
We show that the $\Z_4$ orbifold of the nonlinear $\sigma$ model on the
torus with lattice $\Lambda=\Z^4$ has a geometric interpretation on
the Fermat quartic hypersurface.

Section \ref{gepinmodsp} is devoted to the study of special points with
higher discrete symmetry groups in the moduli space, namely Gepner models
(actually $(2)^4$ and some of its orbifolds by phase symmetries). We stress
that our approach is different from the one
advocated in \cite{fkss90,fks92} where
massless spectra and symmetries of all Gepner models and their orbifolds
were matched to those of algebraic manifolds corresponding to these models.
The correspondence there was understood in terms of
Landau-Ginzburg models, a limit which we do not make use of at all. We
instead explicitly prove equivalence of the Gepner models under investigation
to nonlinear $\sigma$ models. This also enables us to give the precise location
of the respective models within the moduli space ${\cal M}^{K3}$.
We prove that the Gepner model $(2)^4$ is isomorphic to the
$\Z_4$ orbifold and therefore to the Fermat quartic model studied in the
previous section. We moreover find two meeting points of ${\cal M}^{K3}$
and ${\cal M}^{tori}$ generalizing earlier results for
bosonic theories \cite{kosa88} to the
corresponding $N=(4,4)$ supersymmetric models.
We find a meeting point of the moduli spaces of $\Z_2$ and $\Z_4$ orbifold
conformal field theories different from the one conjectured in \cite{eoty89}.
In section \ref{conc} we conclude by gathering the results and joining
them to a panoramic view of part of the moduli space (figure
\ref{modulispacepic}).

In the context of $\sigma$ models we must fix our
$\alpha^\prime$ conventions. For ease of notation we use the rather
unusual $\alpha^\prime = 1$, so T-duality for a bosonic
string compactified on a circle of radius $R$ reads $R\mapsto {1\over R}$.
We hoped to save  us a lot of factors of $\sqrt2$ this way.

Often, the left--right transformed analogue of some statement will not
be mentioned explicitly, in order to avoid tedious repetitions.
Fourier components of holomorphic fields are labelled by the energy,
not by its negative.

\section{The moduli space of $N=(4,4)$ superconformal field theories with
central charge $c=6$}\label{modulispace}
We consider unitary two dimensional superconformal quantum field theories.
They can be described as Minkowskian theories on the circle or equivalently
as euclidean theories on tori with parameter $\tau$ in the upper complex
halfplane. The worldsheet coordinates are called $\sigma_0,\sigma_1$.

The space of states ${\cal H}$ of a quantum field theory has a real structure
given by CPT. For any $N=(4,4)$ superconformal theory ${\cal H}$ contains
four--dimensional vector spaces $Q_l$ and $Q_r$ of real
left and right supercharges. Since we consider left and right
central charge $c=6$, we use the extension of the $N=(2,2)$
superconformal algebra by  an
$su(2)\oplus su(2)$ current algebra of level 1 \cite{aetal76}. 
The (3+3)-dimensional
Lie group generated by the corresponding charges will be denoted by
$SU(2)_l^{susy}\times SU(2)_r^{susy}$ and its $\{(\id,\id),(-\id,-\id)\}$
quotient
by $SO(4)^{susy}$. The commutant of $SU(2)_l^{susy}$
in $SO(Q_l)$ will be called $SU(2)_l$. Here and in the following we use the
notation $SO(W)$ for the special orthogonal group of a
real vector space $W$ with given scalar product.

One can identify $SU(2)_l^{susy}$ with $SU(2)_l$ by selecting
one vector in $Q_l$. The subgroup of $SO(Q_l)$ which fixes this vector is
an $SO(3)$ group with surjective projections to the two $SU(2)$ groups
modulo their centers and allows an identification of the images. Such an
identification seems to be implicit in many discussions in the literature,
but will not be used in this section.

We will consider canonical subspaces of ${\cal H}$ spanned by the states
with specified conformal dimensions $(h;\qu{h})$ which belong to some
irreducible representation of $SU(2)_l^{susy}\times SU(2)_r^{susy}$.
The latter are labelled by the charges $(Q;\qu{Q})$ with
respect to a Cartan torus of  $SU(2)_l^{susy}\times SU(2)_r^{susy}$. Since
any two Cartan tori are related by a conjugation, the spectrum does not depend
on the choice of this torus. Charges are normalized to integral values, as has
become conventional in the context of extended supersymmetry.

We assume the existence of a quartet of spectral flow
fields with $(h,Q;\qu{h},\qu{Q})=({1\over4},\eps_1;{1\over4},\eps_2),
\eps_i\in\{\pm1\}$. Operator products with each of them yield a combined
left+right spectral flow. Instead of using $N=(4,4)$ supersymmetry it
suffices to start with $N=(2,2)$ and this quartet. Indeed, the operator
product of a pair of quartet fields yields lefthanded flow operators with
$(h,Q;\qu{h},\qu{Q})=(1,\pm2;0,0)$, and analogously on the righthanded side
for another pair.
These enhance the $u(1)_l^{susy}\oplus u(1)_r^{susy}$ subalgebra of the
$N=(2,2)$ superconformal algebra to an $A_1^{(1)}\times A_1^{(1)}$ Kac-Moody
algebra. Thus the $N=(2,2)$ superconformal algebra is enhanced to $N=(4,4)$
\cite{eoty89}.

Our assumptions are natural in the context of superstring compactification.
There, unbroken extended spacetime supersymmetry is obtained from
$N=(2,2)$ worldsheet supersymmetry with spectral flow operators
\cite{se86,se87}. Thus our superconformal theories may be used as a background
for $N=4$ supergravity in six dimensions. Here, however, we concentrate on the
internal conformal field theory. External degrees of freedom are not taken into
account.

Let us give a brief summary on what is known about the moduli space ${\cal M}$
so far. The spaces of states 
of the conformal theories form a  bundle with local grading by finite
dimensional subbundles over
${\cal M}$. They can be decomposed into irreducible representations of the
left and right $N=4$ supersymmetries. The irreducible representations
are determined by their lowest weight values of $(h,Q)$.
These representations can be deformed continuously with respect to the
value of $h$,
except for the representations of non-zero Witten index, also called
\textsl{massless
representations} \cite{egta88,egta88b,ta89}. Apart from the vacuum
representation with $(h,Q)=(0,0)$, the lowest weight states of massless
representations are labelled by $(h,Q)=({1\over2},\pm1)$ in the Neveu-Schwarz
sector and by $(h,Q)=({1\over4},\pm1)$ or $(h,Q)=({1\over4},0)$ in the Ramond
sector. Let us enumerate the representations which are massless with respect
to both the left and the right handed side. Apart from the vacuum
we already mentioned the spectral flow operators with $(h,Q;\qu{h},\qu{Q})
=({1\over4},\eps_1;{1\over4},\eps_2), \eps_i\in\{\pm1\}$. They form a
vector multiplet under $SO(4)^{susy}$.
Since the vacuum is unique, there is exactly one multiplet of such fields. On
the other hand, the dimension of the vector space of real
$({1\over4},0;{1\over4},0)$ fields is not fixed a priori. We shall denote
it by $4+\delta$. With a slight abuse of notation, the orthogonal group of
this vector space will be called $O(4+\delta)$. These are all the
possibilities of massless representations in the Ramond sector. The
corresponding ground state fields describe the entire cohomology of
Landau-Ginzburg or $\sigma$ model descriptions of our theories \cite{lvw89}.

If in a given model
there is a field with $(h,Q;\qu{h},\qu{Q})=({1\over2},\pm1;0,0)$,
application of $su(2)_l$ and supersymmetry operators yields
four lefthanded Majorana fermions and the corresponding abelian
currents. As we shall see below, this suffices to show that
the model has an interpretation as nonlinear $\sigma$ model
on a torus, with the currents as generators of translation and the fermions as
parallel sections of a flat spin bundle.
Such models have $\delta=0$ and constitute the
component ${\cal M}^{tori}$ of ${\cal M}$.

The vector space ${\cal F}_{1/2}$ spanned by the fields with
$(h,Q;\qu{h},\qu{Q}) =({1\over2},\eps_1;{1\over2},\eps_2), \eps_i\in\{\pm1\}$
is obtained from the $({1\over4},0;{1\over4},0)$ Ramond fields by spectral
flow. Thus it gives an irreducible $4(4+\delta)$--dimensional representation of
$su(2)_l^{susy}\oplus su(2)_r^{susy}\oplus o(4+\delta)$.
It determines the supersymmetric deformations of the theory, as will be
considered below.

The massless representations cannot be deformed, so $\delta$
is constant over the generic points of a connected component
of ${\cal M}$ and can only increase over nongeneric ones. 
Tensor products of a massive lefthanded representation with a
righthanded massless representation cannot be deformed either, since
$h-\qu{h}$ must remain intergral. The span of such tensor products in
the space of states  yields a string theoretic generalization 
$\mathcal E$ of the
elliptic genus \cite{scwa86,scwa86b}, which is constant for all theories
within a connected component of ${\cal M}$. 
Since for $c=6$ and theories with merely
integer charges $\mathcal E$ is a theta
function of level $2$ and characteristic $q^{-1}$, by its properties
under modular transformations one can show that $\mathcal E$ is a 
multiple of the elliptic genus $\mathcal E_{K3}$ of a 
$K3$ surface. According to their charges, the numbers of 
$({1\over4},{1\over4})$ fields can be arranged into a Hodge diamond
$$
\begin{array}{ccccc} &&1\\ &n_l&&n_r\\ 1&&4+\delta&&1\\
&n_r&&n_l \\ &&1 \end{array}
$$
where by the above $n_l\in\{0,2\}$ also yields the number of 
left handed Dirac fermions. The uniqueness of the left and 
right elliptic genera shows $n_l=n_r$ and $\delta=16-8n_l$.
Moreover,
left and righthanded elliptic genera 
have the same power series expression.
They vanish over ${\cal M}^{tori}$. 
In particular, as was anticipated
above, the existence of one field with $(h,Q;\qu h,\qu Q)=({1\over2},\pm1;0,0)$
suffices to show that the theory is toroidal.
The elliptic genus on $\mathcal M$ is interpreted
as index of a supercharge acting on the loop space of $K3$ \cite{wi87,wi88}.
We call one
of our conformal field theories \textsl{associated to torus or $K3$},
depending on the elliptic genus. For the theories associated to $K3$ one
has $\delta=16$.

To understand the local structure of the  moduli space ${\cal M}$ we
must determine the tangent space ${\cal H}_1$ in a given point of ${\cal M}$,
i.e. describe the deformation moduli of a given theory. This space
consists of real fields of dimensions $h=\qu{h}=1$ in the space of states
${\cal H}$ over the chosen point. The Zamolodchikov metric \cite{za86} on the
space of such fields establishes on
${\cal M}$ the structure of a Riemannian manifold,
with holonomy group contained in
$O({\cal H}_1)$. To preserve the supersymmetry algebra, ${\cal H}_1$ must
consist of $SO(4)^{susy}$ invariant fields in
the image of ${\cal F}_{1/2}$ under $(Q_l)_{1/2}\otimes (Q_r)_{1/2}$, where
the latter subscripts denote Fourier components.
Accordingly, ${\cal F}_{1/2}\oplus {\cal H}_1$ yields a well--known
representation of the $osp(2,2)$ superalgebra spanned by $(Q_l)_{\pm 1/2}$,
$su(2)_l^{susy}$ and the Virasoro operator $L_0$.
In particular, ${\cal H}_1$ should be
$4(4+\delta)$--dimensional and form an irreducible representation of
$su(2)_l\oplus su(2)_r\oplus o(4+\delta)$.
We shall assume that all elements of ${\cal H}_1$ really give
integrable deformations, as has been shown to all orders in perturbation
theory \cite{di87}. Note, however, that there is no complete proof yet.

The holonomy group of ${\cal M}$ projects
to an $O(4+\delta)$ action on the uncharged massless Ramond
representations and to an $SO(4)$ action on $Q_l\otimes Q_r$.
Thus its Lie algebra is contained in
$su(2)_l\oplus su(2)_r\oplus o(4+\delta)$.
The two Lie algebras are equal for ${\cal M}^{tori}$ and one expects
the same for $\delta=16$. Below we shall find
an isometry from ${\cal M}^{tori}$ to a subvariety of ${\cal M}^{K3}$,
such that the holonomy Lie algebra of the latter space is at least
$su(2)_l\oplus su(2)_r\oplus so(4)$. Moreover, this isometry shows that
${\cal M}^{K3}$ is not compact.
Since one has the inclusion
$$
su(2)\oplus su(2)\oplus o(4+\delta)
\cong sp(1)\oplus sp(1)\oplus o(4+\delta)
\hookrightarrow sp(1)\oplus sp(4+\delta),
$$
the moduli space of $N=(4,4)$ superconformal field theories
with $c=6$ associated to torus or $K3$ is a quaternionic K\a hler manifold
of real dimension $4(4+\delta)$.
To determine its local structure, recall that we are looking
for a noncompact space. By Berger's classification of
quaternionic K\a hler manifolds \cite{be55} it can only be reducible or
quaternionic symmetric \cite[Th.~9]{si62}. Because non--Ricci flat
quaternionic K\a hler manifolds are (even locally) de Rham irreducible
\cite{wo65}, this means that it can only be
Ricci flat or quaternionic symmetric. The former is excluded
because geodesic submanifolds on which all holomorphic sectional curvatures
are negative and bounded away from zero have been found
\cite{pest90,cfg89,ce90}. Hence the moduli space must locally be the 
Wolf space
\beqn{modallg}
{\cal T}^{4,4+\delta} \>=\>
O^+(4,4+\delta;\R) \big/ SO(4)\times O(4+\delta)\e 
\>\cong\>
SO^+(4,4+\delta;\R) \big/ SO(4)\times SO(4+\delta) ,
\eeqn
i.e. one component of 
the Grassmannian of oriented spacelike four--planes
$x\subset \R^{4,4+\delta}$ \cite{ce91}, reproducing Narain's and Seiberg's
previous results \cite{cent85,na86,se88}.
Here $SO^+(W)$ denotes the identity component of the
special orthogonal group $SO(W)$
of a vector space $W$ with given scalar product.
The space of maximal positive definite subspaces of $W$ has two components,
and $O^+(W)$ denotes the subgroup of elements of $O(W)$ which do not 
interchange these components. Note that for positive definite $W$ we have
$SO(W)=O^+(W)$. The Zamolodchikov metric on ${\cal T}^{4,4+\delta}$
is the group invariant one.

From the preceeding discussion,
$x$ can be interpreted as the $SO(4)^{susy}$
invariant part of the tensor product of $Q_l\otimes Q_r$ with the
four-dimensional space of charged Ramond ground states.
Note that the action of
$so(4)=su(2)_l\oplus su(2)_r$ discussed above
generates orthogonal transformations of the four--plane
$x\in{\cal T}^{4,4+\delta}$ corresponding to the theory under inspection,
whereas $o(4+\delta)$ acts on its orthogonal complement.

We repeatedly used the splitting $so(4)=su(2)_l\oplus su(2)_r$.
Consider the antisymmetric product $\Lambda^2x$
of the above four--plane $x$. We choose the
orientation of $x$ such that $su(2)_l$ fixes the anti--selfdual part
$(\Lambda^2x)^-$ of $\Lambda^2x$ with respect to the  group invariant 
metric on $O^+(4,4+\delta;\R)$.
When the theory has a parity
operation which interchanges $Q_l$ and $Q_r$, this induces a change of
orientation of $x$. The choice of an $N=(2,2)$ subalgebra within the $N=(4,4)$
superconformal algebra corresponds to the
selection of a Cartan torus $u(1)_l\oplus u(1)_r$ of
$su(2)_l\oplus su(2)_r$. This induces the choice of an oriented
two--plane in $x$. The rotations of $x$
in this two--plane are generated by $u(1)_{l+r}$,
those perpendicular to the plane by $u(1)_{l-r}$. Thus the moduli space
of $N=(2,2)$ superconformal field theories with central charge $c=6$
is given by a Grassmann bundle over ${\cal M}$, with fibre
$SO(4)/(SO(2)_{l+r}\times SO(2)_{l-r})\cong \S^2\times\S^2$.

Generic examples for our conformal theories are the nonlinear
$\sigma$ models with the oriented four--torus or the $K3$ surface as target
space $X$. In the $K3$ case, the existence of these quantum field theories has
not been proven yet, but their conformal dimensions and operator product
coefficients  have a well defined perturbation
theory in terms of inverse powers
of the volume. We tacitly make the assumption that a rigorous treatment is
possible and warn the reader that many of our statements depend on this
assumption.

A nonlinear $\sigma$ model on $X$ assigns an action to any twocycle on $X$.
This action is the sum of the area of the cycle for a given Ricci flat metric
plus the image of the cycle under a
cohomology element $B\in H^2(X,\R)$. Since integer shifts of the action
are irrelevant, the physically relevant B-field is the projection of B to
$H^2(X,\R)/H^2(X,\Z)$.
Thus the parameter space of nonlinear $\sigma$ models has the form
$\{\mathit{Ricci\, flat\, metrics}\}\times\{\mathit{B-fields}\}$. The corresponding
Teichm\"uller space is
\beq{teich}
{\cal T}^{3,3+\delta}\times\R^+\times H^2(X,\R).
\eeq
Its elements will be denoted by $(\Sigma,V,B)$. The first factor of the
product is the Teichm\"uller space of Ricci flat metrics of volume 1 on $X$,
the second parametrizes the volume, and the last one represents the
B-field. The Zamolodchikov metric gives a warped product structure to this
space. Worldsheet parity transformations $(\sigma_0,\sigma_1) \mapsto
(-\sigma_0,\sigma_1)$ change the sign of the cycles, or equivalently the
sign of $B$, which yields an automorphism of the parameter space.

Target space parity for $B=0$ yields a specific
worldsheet parity transformation and thus an identification of $su(2)_l$ with
$su(2)_r$. The corresponding diagonal Lie algebra $su(2)_{l+r}$ generates an
$SO(3)$ subgroup of $SO(4)$. Under the action of this subgroup $x$
decomposes into a line and its orthogonal three--plane $\Sigma\subset x$. The
$\S^2\times\S^2$ bundle over ${\cal M}$ now has a diagonal $\S^2$ subbundle.
Each point in the fibre corresponds to the choice of an $SO(2)$ subgroup of
$SO(3)$ or a subalgebra $u(1)_{l+r}$ of $su(2)_{l+r}$. Geometrically this
yields a complex structure in the target space. Thus the $\S^2$ bundle over
the $B=0$ subspace of ${\cal M}$ is the bundle of complex structures over
the moduli space of Ricci flat metrics on the target space.

Recall some basic facts about the Teichm\u ller space ${\cal T}^{3,3+\delta}$
of Einstein metrics on an oriented four--torus or $K3$ surface $X$.
We consider the vector space $H^2(X,\R)$ together with its
intersection product, such that $H^2(X,\R)\cong\R^{3,3+\delta}$. In other
words, positive definite subspaces have at most dimension three, negative
definite ones at most dimension $3+\delta$.
On $K3$ this choice of sign determines a canonical orientation.
When one wants to study ${\cal M}^{tori}$ by itself, the choice of a torus
orientation is superfluous. Our main interest, however, is the study of
torus orbifolds. For a canonical blow--up of the resulting singularities
one needs an orientation. The effect of a change of orientation on the torus
will be considered below.

Metric and orientation on $X$ define a Hodge star operator, which
on $H^2(X,\R)$ has
eigenvalues +1 and -1. The corresponding eigenspaces of dimensions three and
$3+\delta$ are positive and negative definite, respectively.
Let $\Sigma\subset H^2(X,\R)$ be the positive definite three--plane obtained in
this way. The orientation on $X$ induces an orientation on $\Sigma$. One can
show that Ricci flat metrics are locally uniquely specified by $\Sigma$,
apart from a scale factor given by the volume. Since the Hodge star operator in
the middle dimension does not change under a rescaling of the metric, the
volume $V$ must be specified separately.
It follows that ${\cal T}^{3,3+\delta}\times \R^+$ is the
Teichm\u ller space of Einstein metrics on $X$.
Explicitly, we have
\beq{teichmet}
{\cal T}^{3,3+\delta}=O^+(H^2(X,\R)) \big/ SO(3)\times O(3+\delta) .
\eeq
The $SO(3)$
group in the denominator is to be interpreted as $SO(\Sigma_0)$ for some
positive definite reference three--plane in $H^2(X,\R)$, while $O(3+\delta)$
is the corresponding group for the orthogonal complement of $\Sigma_0$.
Equivalently, ${\cal T}^{3,3+\delta}$ could have been written as
$SO^+(H^2(X,\R)) \big/ SO(3)\times SO(3+\delta)$. 
We choose the description \req{teichmet} for later
convenience in the construction of the entire moduli space.

For higher dimensional Calabi-Yau spaces the $\sigma$ model description
works only for large volume due to instanton corrections.  In our case,
however, the metric on the moduli space does not receive corrections
\cite{nasu95}. Therefore the Teich\-m\"ul\-ler
space \req{teich} of $\sigma$ models on $X$ should
be a covering of a component of ${\cal M}$, thus isomorphic to
the Teichm\"uller space ${\cal T}^{4,4+\delta}$
obtained in \req{modallg}.
Indeed, for $\delta=16$ a natural isomorphism
\beq{decomp}
{\cal T}^{4,4+\delta}
\cong {\cal T}^{3,3+\delta}\times\R^+\times H^2(X,\R)
\eeq
was given in \cite{asmo94,as96}, with a correction and clarification by
\cite{rawa98,di99}. The same construction actually works for $\delta=0$, too.
It uses the identification
$$
{\cal T}^{4,4+\delta} =
O^+(H^{even}(X,\R)) \big/ SO(4)\times O(4+\delta) ,
$$
where $SO(4)$ is to be interpreted as $SO(x_0)$ for some
positive definite reference four--plane in $H^{even}(X,\R)$, while
$O(4+\delta)$ is the corresponding group for the orthogonal complement of
$x_0$. In other words, the elements of ${\cal T}^{4,4+\delta}$ are interpreted
as positive definite oriented four--planes
$x\subset H^{even}(X,\R)$ by $H^{even}(X,\R)\cong\R^{4,4+\delta}$.
Note that all the cohomology of $K3$ is even, whereas
$H^{odd}(X,\R)\cong\R^{4,4}$ when $X$ is a four--torus.

To explicitly realize the isomorphism \req{decomp}
one also needs the positive generators $\upsilon$ of $H^4(X,\Z)$ and
$\upsilon^0$ of $H^0(X,\Z)$, which are Poincar\'e
dual to points and to the whole oriented
cycle $X$, respectively.
They are nullvectors in
$H^{even}(X,\R)$ and satisfy $\langle\upsilon,\upsilon^0\rangle=1$.
Thus over $\Z$ they  span
an even, unimodular lattice isomorphic to the standard hyperbolic lattice
$U$ with bilinear form
$$\left(\begin{array}{cc}0&1\\1&0\end{array}\right).$$

Now consider a triple $(\Sigma,V,B)$ in the right hand side of \req{decomp}.
Define
\beqn[l]{expdecomp}
\xi: \Sigma\rightarrow H^{even}(X,\R), \quad
\xi(\sigma):=\sigma - \langle B,\sigma\rangle\upsilon, \e
\quad x:=\span_\R\left( \xi\left(\Sigma\right), \;
\xi_4:=\upsilon^0+B+\left(V-\inv[\|B\|^2]{2}\right)\upsilon\right) .
\eeqn
Then $\wt{\Sigma}=\xi(\Sigma)$ is
a positive definite oriented three--plane  in
$H^{even}(X,\R)$, and the vector
$\xi_4$ is orthogonal to $\wt{\Sigma}$. Since
$\|\xi_4\|^2=2V$, it has positive square.
Together, $\wt{\Sigma}$ and $\xi_4$ span an
oriented four--plane $x\subset H^{even}(X,\R)$. Obviously, the map
$(\Sigma,V,B)\mapsto x$ is
invertible, once $\upsilon$ and $\upsilon^0$ are given.

To describe the projection from
Teichm\"uller space to ${\cal M}$ we need to consider the lattices
$H^2(X,\Z)$ and $H^{even}(X,\Z)$. They are
even, unimodular, and have signature $(p,p+\delta)$
with $p=3$ and $p=4$, respectively.
Such lattices are isometric to
$\Gamma^{p,p+\delta}= U^p\oplus (E_8(-1))^{\delta/8}$.
Here each summand is a free $\Z$ module, $E_8$ has as
bilinear form the Cartan matrix of $E_8$, and for any lattice $\Gamma$ we
denote by $\Gamma(n)$ the same $\Z$ module $\Gamma$
with quadratic form scaled by $n$.

We now consider the projection from
Teichm\"uller space to ${\cal M}$. First we have to identify all points
in ${\cal T}^{3,3+\delta}$ which yield the same Ricci flat metric.
This means that we have to quotient the Teichm\"uller  space  \req{teichmet}
by the so--called
\textsl{classical symmetries}. The projection is given by
\beq{einsteinmod}
O^+(H^2(X,\Z))\big\backslash {\cal T}^{3,3+\delta}
\eeq
\cite{koto87}.
Here we use the notation $O^+(\Gamma)$ for the intersection of $O^+(W)$
with the automorphism group of a lattice $\Gamma\subset W$.
The interpretation of the quotient space \req{einsteinmod} as moduli space of
Einstein metrics of volume 1 on $X$ is straightforward in the torus case,
but for $X=K3$ one has to include orbifold limits
(see section \ref{orbifolds}). The corresponding $\sigma$
models are not expected to exist for all values of $B$ \cite{wi95}.
To simplify the
discussion we include such \textsl{conifold points} in ${\cal M}$.
On ${\cal T}^{4,4+\delta}$ the group of classical symmetries lifts
by \req{expdecomp}
to the subgroup of $O^+(H^{even}(X,\Z))$ which fixes both lattice vectors
$\upsilon$ and $\upsilon^0$.

Next we consider the shifts of $B$ by elements $\lambda\in H^2(X,\Z)$,
which neither change the physical content. One easily calculates
that this also yields a left action on ${\cal T}^{4,4+\delta}$
by a lattice automorphism in $O^+(H^{even}(X,\Z))$, generated by
$w \mapsto w-\langle w,\lambda\rangle \upsilon$ for
$\langle w,\upsilon\rangle =0$ and
$\upsilon^0 \mapsto \upsilon^0+\lambda -{\|\lambda\|^2\over2}\upsilon$. These
transformations fix $\upsilon$  and shift $\upsilon^0$ to arbitrary
nullvectors dual to $\upsilon$. Thus the choice of $\upsilon^0$ is
physically irrelevant.

We shall argue that the projection from Teichm\"uller space to ${\cal M}$
is given by
\beq{modallgglobal}
{\cal T}^{4,4+\delta}\longrightarrow
O^+(H^{even}(X,\Z))\big\backslash {\cal T}^{4,4+\delta}.
\eeq
The group $O^+(H^{even}(X,\Z))$
acts transitively on pairs of primitive lattice vectors of equal
length \cite{lope81,ni80}. Thus \req{modallgglobal} would imply that
different choices of
$\upsilon,\upsilon^0$ are equivalent. 
Anticipating this result in general,
we call the choice of an arbitrary primitive
nullvector $\upsilon\in H^{even}(X,\Z)$ a \textsl{geometric interpretation}
of a positive oriented four--plane $x\subset H^{even}(X,\Z)$.
Such a choice yields
a family of $\sigma$ models with physically equivalent data
$(\Sigma,V,B)$. A conformal field theory
has various different geometric interpretations, and the choice of $\upsilon$
is comparable to a choice of a chart of ${\cal M}$.

Aspinwall and Morrison also identify theories which are related by the
worldsheet parity transformation \cite{asmo94}. 
We regard the latter as a symmetry of
${\cal M}$. It is given by  change of orientation of the four--plane $x$ or
equivalently by a conjugation of $O^+(H^{even}(X,\R))$ with an element of
$O(H^{even}(X,\R))-O^+(H^{even}(X,\R))$ which transforms the lattice
$H^{even}(X,\Z)$ and the reference four--plane $x_0$ into themselves.
To stay in the classical context, one may choose an element which fixes
$\upsilon$ and $\upsilon^0$. More canonically, parity corresponds to
$(\upsilon, \upsilon^0)\mapsto (-\upsilon,-\upsilon^0)$. The latter
induces $\xi_4\mapsto-\xi_4$ and $(\Sigma,V,B)\mapsto(\Sigma,V,-B)$.

Let us consider the general pattern of identifications.
When two points in Teich\-m\"ul\-ler space are identified the same is true for
their tangent spaces. Higher derivatives can be treated by perturbation
theory in terms of tensor products of the tangent spaces ${\cal H}_1$.
Assuming the convergence of the perturbation expansion in
conformal field theory, any such isomorphism
can be transported to all points of ${\cal T}^{4,4+\delta}$. Therefore
$\sigma$ model isomorphisms are given
by the action of a group ${\cal G}^{(\delta)}$ on this space. In the previous
considerations we have found a subgroup of ${\cal G}^{(\delta)}$.

Below we shall prove that the interchange of $\upsilon$
and $\upsilon^0$,  which is the 
Fourier-Mukai transform \cite{rawa98},
also belongs to ${\cal G}^{(\delta)}$. When $B=0$, this  
yields the map $(\Sigma,V,0)\mapsto (\Sigma,V^{-1},0)$. 
In the torus case, it is known as
T-duality and it seems natural to extend this name to $X=K3$. We will not    
use the name mirror symmetry for this transformation.

It is obvious that classical symmetries, integral B-field shifts, and
T-duality  generate all of $O^+(H^{even}(X,\Z))$. Thus
${\cal G}^{(\delta)}$ contains all of this group. As argued in
 \cite{asmo94,as96}, it cannot be larger, since otherwise the quotient of
${\cal T}^{4,4+\delta}$ by ${\cal G}^{(\delta)}$ plus the parity automorphism
would not be Hausdorff \cite{al66}. For a proof of the Hausdorff property
of ${\cal M}$ one will need some features of the superconformal field theories,
which should be easy to verify once they are somewhat better understood. First,
one has to check that all fields are generated by the iterated operator
products of a finite dimensional subspace of basic fields.
Next one has to show that the operator product coefficients
are determined in terms of a finite number of basic coefficients, and that
the latter are constrained by algebraic equations only. This would show that
${\cal M}$ is an algebraic space. In particular, every point has a
neighborhood which contains no isomorphic point. All
of these features are true in the known examples of conformal field theories
with finite effective central charge, in particular for the unitary theories.
They certainly should be true in our case.

In the context of $\sigma$ models it often is useful to choose a complex
structure on $X$.
When such a structure is given, the real and imaginary parts
of any generator of $H^{2,0}(X,\C)$ span an oriented two-plane
$\Omega\subset\Sigma$. Conversely, any such subspace $\Omega$
defines a complex structure. This means that the choice
of an Einstein metric is nothing but the choice of
an $\S^2$ of complex structures on $X$, in other words a
\textsl{hyperk\a hler structure}.
In terms of cohomology, $\Omega$ specifies
$H^{2,0}(X,\C)\oplus H^{0,2}(X,\C)$. The orthogonal complement of
$\Omega$ in $H^2(X,\R)$ yields $H^{1,1}(X,\R)$. Any vector
$\omega\in H^{1,1}(X,\R)$ of positive norm yields a
K\a hler class compatible with the complex structure and the
hyperk\a hler structure $\Sigma$ spanned by $\Omega$ and $\omega$.

Since $H^2(X,\Z)$ is torsionfree for tori and $K3$ surfaces,
the N\'eron-Severi group $NS(X)$  can be identified with 
$Pic(X):=H^2(X,\Z)\cap H^{1,1}(X,\R)$, the
\textsl{Picard lattice} of $X$.
By a result of Kodaira's, $X$ is algebraic, if $NS(X)$
contains an element $\rho$ of positive
length squared \cite{ko64}.
Given a hyperk\a hler
structure $\Sigma$ we can always find $\Omega\subset\Sigma$
such that $X$ becomes an algebraic surface. It suffices to choose
$\omega$ as the projection of $\rho$ on $\Sigma$ and $\Omega$ as the
corresponding orthogonal complement. The projection is non-vanishing,
since the orthogonal complement of $\Sigma$ in  $H^2(X,\R)$ is
negative definite. Varying $\rho$ one obtains a countable infinity of
algebraic structures on $X$. Thus the occasionally encountered interpretation
of moduli of conformal field theories as corresponding to nonalgebraic
deformations of $K3$ surfaces does not make sense (this was already pointed
out  in \cite{ce91} by different arguments).

The choice of $\Omega\subset \Sigma$ lifts to a corresponding
choice of a two-plane $\wt{\Omega}\subset x$. As discussed above this selects
a $(2,2)$ subalgebra of the $(4,4)$ superalgebra.
We will refer to the choice of such a two--plane as
fixing a complex structure. More precisely,
the two--plane $\wt{\Omega}$ specifies a complex structure
\textsl{in every geometric
interpretation} of the  conformal field theory.
\subsection{Moduli space of theories associated to tori}
\label{modtor}
Originally, Narain determined the moduli space ${\cal M}^{tori}$
of superconformal
field theories associated to tori
by explicit construction of nonlinear $\sigma$ models
\cite{cent85,na86}. With the above formalism we can
reproduce his description as follows.

Let us consider tori of arbitrary dimension $d$. We change the notation by
transposing the group elements, which exchanges left and
right group actions. This yields
$$
{\cal M}^{Narain}=O(d)\times O(d)\big\backslash O(d,d)/O(\Gamma^{d,d}).
$$
This moduli space has a symmetry given by worldsheet parity. We shall see
that its action on $O(d,d)$ exchanges the two $O(d)$ factors.
For later convenience we are going to use the cover 
$SO(d)\times SO(d)\big\backslash SO^+(d,d)/SO^+(\Gamma^{d,d})$
of ${\cal M}^{Narain}$. For even $d$ this is a four--fold cover, for odd
$d$ a two--fold one.
The $\R$--span of $\Gamma^{d,d}$ is naturally isomorphic to
$\R^d \oplus (\R^d)^\ast$, where $\R^d$ is considered as an isotropic subspace
and $W^\ast$ denotes the dual of a vector space $W$,
and analogously for lattices.
Thus $O(d,d)$ can be considered as the  orthogonal group of a vector
space with elements $(\alpha, \beta)$, $\alpha,\beta\in \R^d$
and scalar product
$$(\alpha, \beta)\cdot (\alpha', \beta')=
\alpha\cdot\beta'+\alpha'\cdot\beta.$$
There is a canonical maximal positive definite $d$-plane given by
$\alpha=\beta$ in $\R^d \oplus (\R^d)^\ast=\R^{d,d}$.
The group $SO(d)\times SO(d)$ is supposed to describe
rotations in this $d$-plane and in its orthogonal complement.
In this description, the parity transformation consists of interchanging
these two orthogonal $d$-planes, plus a sign change of the  bilinear form
on $\R^{d,d}$.

Now we use the isometry
\beqn{geotor}
V: SO(d)\big\backslash GL^+(d)\times Skew(d\times d,\R)
\>\longrightarrow\> SO(d)\times SO(d)\big\backslash SO^+(d,d)
\;\cong\;{\cal T}^{d,d}
\eeqn
given by
\beq{vector}
V(\Lambda,B) =
\left( \begin{array}{cc}(\Lambda^T)^{-1} &0\\0& \Lambda\end{array}\right)
\left( \begin{array}{cc} \id&-B\\0&\id \end{array}\right).
\eeq
We identify $\Lambda\in GL^+(d)$ with the image of $\Z^d$ under $\Lambda$.
Finally we change coordinates by
$p_l:=(\alpha+\beta)/\sqrt2,\; p_r:=(\alpha-\beta)/\sqrt2$, such that
the scalar product becomes
\beq{tormet}
(p_l;p_r)\cdot (p_l^\prime;p_r^\prime)
:=  p_lp_l^\prime  -  p_r p_r^\prime.
\eeq
This means that the positive definite $d$-plane is given by $p_r=0$
and its orthogonal complement by $p_l=0$. Altogether,
with $\wt{B}:=(\Lambda^T)^{-1}B\Lambda^{-1}$ a point in
${\cal M}^{tori}$ is now described by the lattice
\beqn{chargelattice}
\Gamma(\Lambda,B)
\>=\> \left\{ \vphantom{\inv{\sqrt2}}
\left(p_l(\lambda,\mu); p_r(\lambda,\mu)\right)\right.\e
\>\> \left.
\hphantom{p_l}
:=\inv{\sqrt2}
\left.\left(\mu-\wt{B}\lambda+\lambda; 
\mu-\wt{B}\lambda-\lambda\right)
\right| (\lambda,\mu)\in \Lambda\oplus\Lambda^\ast \right\}.
\eeqn
The corresponding $\sigma$ model has the real torus
$T=\R^d/\Lambda$
as target space
and $B\in  H^2(T,\R)\cong Skew(d\times d,\R)$ as B-field.
Introducing $d$ Majorana fermions $\psi_1,\dots,\psi_d$
as superpartners of the
abelian currents $j_1,\dots,j_d$
on the torus one constructs an $N=(2,2)$ superconformal
field theory with central charge $c=3d/2$ which will be
denoted by ${\cal T}(\Lambda,B)$.
From equation $\req{vector}$ it is clear that integral shifts of $B$
and lattice automorphisms yield isomorphic theories.

The theory is specified by its charge lattice $\Gamma(\Lambda,B)$.
Namely, to any pair $(\lambda,\mu)\in\Lambda\oplus\Lambda^\ast$ there
corresponds a vertex operator $V_{\lambda,\mu}$ with charge
$\left(p_l(\lambda,\mu); p_r(\lambda,\mu)\right)$ with respect to
$(j_1,\dots,j_d;\qu{\jmath}_1,\dots,\qu{\jmath}_d)$
and with dimensions $(h;\qu{h})=({1\over 2}p_l^2;{1\over 2}p_r^2)$.
Thus $h$ and $-\qu{h}$ are the squares of the projections of $(p_l;p_r)$
to the positive definite $d$-plane and its orthogonal complement, respectively.
In this description, the parity operation is represented by the interchange
of the latter two planes plus a sign change in the quadratic form on
$\R^{d,d}$.
The transformations which exchange the sheets of our covering of 
Narain's moduli space ${\cal M}^{Narain}$ are given by target space orientation
change and T--duality, as can be read off from equation \req{chargelattice}.

The partition function of this theory is
\beqn{torpartition}
Z(\tau,z)
\>=\> Z_{\Lambda,B}(\tau)
\cdot\inv{2} \sum_{i=1}^4 \left| {\theta_i(\tau,z)\over \eta(\tau)} \right|^d ,
\edd
Z_{\Lambda,B}(\tau)
\>=\> {1\over \left|\eta(\tau)\right|^{2d}}
\sum_{(\lambda,\mu)\in \Lambda\oplus\Lambda^\ast}
q^{{1\over 2}(p_l(\lambda,\mu))^2} \bar q^{{1\over 2}(p_r(\lambda,\mu))^2},
\eeqn
where $q=\exp(2\pi i\tau)$ and analogously for
$\qu{q}$.
The functions $\theta_j(\tau,z), j=1,\dots, 4$
are the classical theta functions
and $\eta(\tau)$ is the Dedekind eta function.
For ease of notation we will write
$\eta=\eta(\tau), \theta_j(z)=\theta_j(\tau,z)$, and
$\theta_j=\theta_j(\tau,0)$ in the following.

By considering ${\cal H}_1$ one easily checks that all theories in
${\cal M}^{tori}$ are described by some even unimodular lattice $\Gamma$.
We want to show that every such lattice has a $\sigma$ model interpretation
$\Gamma=\Gamma(\Lambda,B)$ (see also \cite{as96}). Choose a maximal nullplane
$Y\subset\R^{d,d}=\R^d\oplus(\R^d)^\ast$
such that $Y\cap\Gamma\subset\Gamma$
is a primitive sublattice. Apply an $SO(d)\times O(d)$ 
transformation such that the
equation of this plane becomes $\beta=0$. Put $Y\cap\Gamma=(\Lambda^\ast,0)$.
Next choose a dual nullplane $Y^0$ such that $Y\oplus Y^0=\R^{d,d}$ and
$Y^0\cap\Gamma\subset\Gamma$ is a primitive lattice, too.
Existence of $Y^0$ can be shown by  a Gram type algorithm. Then
$Y^0 = \{ (-B\beta,\beta)\mid \beta\in\R^d\}$
for some skew matrix $B$, and $\Gamma=\Gamma(\Lambda,B)$.
Note that different choices of $Y^0$  merely correspond to
translations of $B$ by integral matrices. So the geometric interpretation is
actually fixed by the choice of $Y$ alone as soon as $B$ is viewed as an
element of $Skew(d)/Skew(d\times d;\Z)$.

In this interpretation, $\R^d$ is identified with the cohomology
group $H^1(\R^d/\Z^d,\R)$ of the reference torus $T=\R^d/\Z^d$.
In addition to its defining representation, the double cover of the
group $SO^+(d,d)$
also has half-spinor representations, namely its images in
$SO^+(H^{odd}(T,\R))$ and in $SO^+(H^{even}(T,\R))$.
For $d=4$ one has the obvious isomorphism
$SO^+(4,4)\cong SO^+(H^{odd}(T,\R))$, which together with
$SO^+(4,4)\cong SO^+(H^{even}(T,\R))$ yields the celebrated $D_4$
triality \cite[I.8]{lami89}. It is the latter automorphism which we
will need in this paper, since the odd cohomology of $X$ does not survive
orbifold maps.

Note that for $Spin(4,4)$
representations on $\R^{4,4}$ there is the same triality relation
as for $Spin(8)$ representations on $\R^8$, i.e. an ${\cal S}_3$
permuting the vector representation, the chiral and the antichiral
Weyl spinor representation.
The role of triality is already visible upon comparison of the geometric
interpretations, where
the analogy between choices of nullplanes $Y,Y^0$ as described above
and nullvectors $\upsilon,\upsilon^0$ in \req{expdecomp} is apparent.
Indeed, part of the triality manifests itself in a one to one correspondence
between maximal isotropic subspaces $Y\subset\R^{4,4}$ and
null Weyl spinors $\upsilon$ such that
$Y=\{ y\in\R^{d,d} \mid c(y)(\upsilon) = 0 \}$
where $c$ denotes
Clifford multiplication on the spinor bundle \cite{butr88}.
One can regard this as further justification
for the interpretation of $\upsilon$ as volume form which generates $H^4(T,\Z)$
in our geometric interpretation.
Recall also that in both cases  different choices of $Y^0,\upsilon^0$
correspond to B-field shifts by integral forms.

We now  explicitly describe the isomorphism \req{geotor} to show that it
is a triality automorphism. First compare \req{geotor}
to \req{decomp} and notice that $Skew(4)\cong \R^{3,3}$ which will simply
be written $Skew(4)\ni B\mapsto b\in\R^{3,3}$ in the following. Moreover,
because $|\!\det\Lambda|$ is the volume of the torus $T=\R^d/\Lambda$,
we can decompose 
$SO(4)\backslash GL^+(4)\cong SO(4)\backslash SL(4)\times\R^+$.
Now let $T_{\Lambda_0}=\R^4/\Lambda_0$
where $\Lambda_0$ is a lattice of determinant 1 and is viewed as element
of $SL(4)$.
Consider the induced representation $\rho$ of $SL(4)$ on the exterior product
$\Lambda^2(\R^4)$ which defines an isomorphism
$\Lambda^2(\Lambda_0)\cong H_2(T_{\Lambda_0},\Z)$
for every $\Lambda_0\in SL(4)$.
Because $\rho$
commutes with the action of the Hodge star operator $\ast$ and
$\ast^2=\id$ on twoforms, $SL(4)$ is actually  represented by
$SO^+(3,3)$. In terms of coordinates as in \req{vector} and with
$\Lambda=V^{1/4}\Lambda_0=(\lambda_1,\dots,\lambda_4), V=|\!\det\Lambda|$,
we can write
\beqn{rho}
\rho\left(\Lambda_0\right) \>=\> V^{-1/2} \left(
\lambda_1\wedge\lambda_2\, ,\, \lambda_1\wedge\lambda_3\, ,\,
\lambda_1\wedge\lambda_4\, ,\,\lambda_3\wedge\lambda_4\, ,\,
\lambda_4\wedge\lambda_2\, ,\, \lambda_2\wedge\lambda_3
\right)\e
\>\>
\hphantom{\rho\left(\Lambda_0\right) = V^{-1/2} \left(
\lambda_2\lambda_2\wedge\lambda_3\right)}
\in SO^+\left(H_2(T,\R)\right) \cong SO^+(3,3).
\eeqn
Because $SO^{+}(3,3)\cong SL(4)/\Z_2$ and
$SO(3)\times SO(3)/\Z_2\cong SO(4)$ we find\linebreak
$SO(4)\big\backslash  SL(4) \cong {\cal T}^{3,3}$ and  all in all  have
\beq{decomptor}
{\cal T}^{4,4} \stackrel{\req{geotor}}{\cong}
SO(4)\big\backslash GL^+(4)\times Skew(4)
\stackrel{\cong}{\longrightarrow }
{\cal T}^{3,3}\times\R^+\times\R^{3,3}
\stackrel{\req{decomp}}{\cong} {\cal T}^{4,4}.
\eeq
By \req{decomptor} the geometric interpretation
of a superconformal field theory is translated from a description in
terms of the lattice of the underlying torus, i.e. in terms of
$\Lambda\cong H_1(T_\Lambda,\Z)$,
to a description in terms of $H_2(T_\Lambda,\Z)\cong \Lambda^2(\Lambda)$.
This translation is essential for understanding  the relation between
the moduli spaces ${\cal M}^{tori}$ and ${\cal M}^{K3}$.
To actually arrive at the description  \req{decomp} in
terms of hyperk\a hler structures, i.e. in terms of $H^2(T,\Z)$, we have
to apply Poincar\'e duality or use the dual lattice $\Lambda^\ast$
instead of $\Lambda$. This distinction will no longer be relevant after 
theories related by T--duality have been identified.

We insert the coordinate expressions in \req{vector} and \req{expdecomp}
into \req{decomptor},
write $\Lambda=V^{1/4}\Lambda_0, V=|\!\det\Lambda|$ as before and arrive at
\beq{triality}
V(\Lambda,B) \longmapsto S(\Lambda,B)
= \left(\begin{array}{c|c|c}
V^{1/2}&0&0\\[3pt]\hline&&\\[-6pt]
0&\rho(\Lambda_0)&0\\[3pt]\hline&&\\[-6pt] 0&0&V^{-1/2}
\end{array}\right)
\left(\begin{array}{c|c|c}
1&0&0\\[6pt]\hline&&\\[-6pt] b&\id&0\\[3pt]\hline&&\\[-6pt]
-{\|B\|^2\over2}&-b^T&1
\end{array}\right).
\eeq
Observe that \req{triality} is a homomorphism
${\cal T}^{4,4}\rightarrow{\cal T}^{4,4}$ and thus gives a natural explanation
for the quadratic dependence on $B$ in \req{expdecomp}. Moreover,
\req{triality} reveals the structure of the warped product \req{decomp}
alluded to before. But above all on Lie algebra level one can now
easily read off that \req{triality} is the triality automorphism
exchanging the two half spinor representations $V$ and $S$.
Namely, let $h_1,\dots,h_4$ denote generators of the Cartan subalgebra
of $so(4,4)$. Here $h_i$ generates dilations of the radius $R_i$ of
our torus in direction $\lambda_i$. Since $\exp(\theta h_i)$ scales
$V^{\pm1/2}$ by $e^{\pm\theta/2}$ and with
\req{rho} one then finds that \req{triality} indeed is induced by
the triality automorphism which acts on the Cartan subalgebra by
$$
\begin{array}{ll}\displaystyle
h_1\mapsto\inv{2}(h_1+h_2+h_3+h_4), \> h_2\mapsto\inv{2}(h_1+h_2-h_3-h_4),\e
h_3\mapsto\inv{2}(h_1-h_2+h_3-h_4), \> h_4\mapsto\inv{2}(h_1-h_2-h_3+h_4):
\end{array}\hspace*{-1em}
\begin{array}{c}
\setlength{\unitlength}{1em}
\begin{picture}(7,5)(-5,-2)
\thicklines
\multiput(0,0)(0,-2){2}{\circle*{0.5}}
\multiput(-2,2)(4,0){2}{\circle*{0.5}}
\put(0,0){\line(0,-1){2}}
\put(0,0){\line(1,1){2}}
\put(0,0){\line(-1,1){2}}
\put(-0.3,0){\makebox(0,0)[rc]{$\scriptstyle h_2-h_3$}}
\put(-0.3,-2.1){\makebox(0,0)[rc]{$\scriptstyle h_2-h_4$}}
\put(-2.3,2){\makebox(0,0)[rb]{$\scriptstyle h_2+h_4$}}
\put(2.3,2){\makebox(0,0)[lb]{$\scriptstyle h_1-h_3$}}
\qbezier(-1.4,2.2)(0,3)(1.4,2.2)
\put(-1.4,2.2){\vector(-2,-1){0.2}}
\put(1.4,2.2){\vector(2,-1){0.2}}
\end{picture}\hphantom{blabla}
\setlength{\unitlength}{1ex}
\end{array}
$$
Note that triality interchanges the outer automorphisms of
$SO^+(4,4)$ related to worldsheet parity and target space orientation.

Triality considerations have a long history in superstring and supergravity
theories, see for example \cite{sh80,cu82,gool86}. Concerning recent work, 
as communicated to us by N.~Obers,
$SO(4,4)$ is crucial in the conjectured duality
between heterotic strings on the fourtorus and type IIA on $K3$ 
\cite{obpi99a,kop00}.
In connection with the calculation of $G(\Z)$ invariant string theory
amplitudes one can use triality to write down new identities for
Eisenstein series \cite{obpi99a,obpi99b}.

We now come to a concept which is of major importance in the context of
Calabi-Yau compactification and nonlinear $\sigma$ models,
namely the idea of \textsl{large volume limit}.
A precise notion is necessary
of how to associate a unique geometric interpretation
to a theory
described by an even self dual lattice $\Gamma$
when parameters of volume go to infinity.
Intuitively, because of the uniqueness condition, this should describe the
limit where all the radii of the torus in this particular
geometric interpretation are large. Because in the charge lattice
\req{chargelattice} $\lambda\in\Lambda$ and $\mu\in\Lambda^\ast$ are
interpreted as winding and momentum modes,
the corresponding nullplane $Y$  should
have the property
\beqn{largevolume}
Y\cap\Gamma
\>=\>
\span_\Z \left\{ \left.
\inv{\sqrt2}(\mu;\mu)\in\Gamma \right| \|\mu\|^2\ll 1\right\}\e
\>\subset\>
\span_\Z \left\{ \left.\vphantom{\inv{\sqrt2}}
(p_l;p_r)\in\Gamma \right| \| p_l\|^2\ll1, \| p_r\|^2\ll1\right\}
=: \wt{\Gamma}.
\eeqn
Because $\| p_l\|^2-\| p_r\|^2\in\Z$, for $(p_l;p_r)\in \wt{\Gamma}$
we have  $\| p_l\|^2=\| p_r\|^2$. This shows $Y\cap\Gamma=\wt{\Gamma}$
because any $(p_l;p_r)\not\in Y^\perp=Y$ must have large
components. Moreover, if a maximal isotropic plane $Y$ as in
\req{largevolume}  exists, then it is uniquely defined, thus yielding
a sensible
notion of large volume limit.
Large volume and small volume limits are exchanged by T--duality.

For our embedding of torus orbifold theories into the $K3$ moduli space
${\cal M}^{K3}$ we have to keep target space orientation. We also want to
keep the left--right distinction in the conformal field theory.
Torus T--duality just yields a reparametrization of the theory and should be
divided out of the moduli space. Thus for us the relevant moduli space
of torus theories is given by
\beq{torms}
{\cal M}^{tori}=SO(d)\times O(d)\big\backslash O^+(d,d)/O^+(\Gamma^{d,d}).
\eeq
Notice that this is a double cover of ${\cal M}^{Narain}$.
\subsection{Moduli space of theories associated to $K3$ surfaces}\label{k3mod}
We now give some more details about
the moduli space of conformal field theories
associated to $K3$ which we will
concentrate on for the rest of the paper, namely
\beq{k3ms}
{\cal M}^{K3} =
O^+(H^{even}(X,\Z))\big\backslash {\cal T}^{4,20}
\eeq
by \req{modallgglobal}. For other presentations see
\cite{asmo94,rawa98,di99}.

In the decomposition \req{decomp} we determine the product metric such that
it becomes an isometry. In particular, it faithfully relates
moduli of the conformal field theory to deformations of geometric objects.
Recall that the structure of the tangent space ${\cal H}_1$
of ${\cal M}^{K3}$ in a given
superconformal field theory
is best understood by examining the $({1\over2},{1\over2})$-fields
in ${\cal F}_{1/2}$. In our case we have related it to the
$su(2)_l^{susy}\oplus su(2)_r^{susy}$ invariant subspace of the tensor
product $Q_l\otimes Q_r\otimes {\cal H}_{1/4}^{(4)}\otimes
{\cal H}_{1/4}^{(0)}$, where ${\cal H}_{1/4}^{(4)}$ denotes the charged and
${\cal H}_{1/4}^{(0)}$ the uncharged Ramond ground states. The invariant
subspace of $Q_l\otimes Q_r\otimes {\cal H}_{1/4}^{(4)}$ yields a
four--plane with an orthogonal group generated by $su(2)_l\oplus su(2)_r$.
When a frame in $Q_l\otimes Q_r$ is chosen, the latter tensor product factor
can be omitted. The description of ${\cal M}$ implies that ${\cal
H}_{1/4}^{(4)}\oplus {\cal H}_{1/4}^{(0)}$ has a natural non-degenerate
indefinite metric and remains invariant under deformations, but it has not
been understood how this comes about. In terms of the four--plane $x\in{\cal
T}^{4,20}$ giving the location of our theory in moduli space, specific vectors
in the tangent space  $T_x{\cal T}^{4,20}$  are described
by infinitesimal deformations of one generator $\xi\in x$ in direction
$x^\perp$ that leaves $\xi^\perp\cap x$ invariant.

To formulate this in terms of a geometric interpretation $(\Sigma,V,B)$
specified by
\req{expdecomp}, pick a basis $\eta_1,\dots,\eta_{19}$ of
$\Sigma^\perp\subset H^2(X,\R)\cong\R^{3,19}$. Then
$x^\perp$ is spanned by
$\{\eta_i-\langle \eta_i,B\rangle\,\upsilon; i=1,\dots,19\}$ and
$\eta_{20}:=\upsilon^0+B-({\|B\|^2\over2}+V)\upsilon$.
In each of the $SO(4)$ fibres of ${\cal H}_1$ over
$\eta_i-\langle \eta_i,B\rangle\,\upsilon, i=1,\dots,19$
we find a threedimensional
subspace deforming generators of $\Sigma$ by
$\eta_i$, as well as the deformation of $B$ in direction of $\eta_i$.
The fibre over $\eta_{20}$ contains B-field deformations in direction
of $\Sigma$ and the deformation of volume. All in all, a
$3\cdot19=57$ dimensional subspace of ${\cal H}_1=T_x{\cal M}^{K3}$
is mapped onto
deformations of $\Sigma$ by $(1,1)$-forms
$\eta\in\Sigma^\perp\cap H^2(X,\R)\subset H^{1,1}(X,\R)$, no matter
what complex structure we pick in $\Sigma$. The $23$ dimensional complement
of this subspace is given by $19+3$ deformations of the B-field
by forms $\eta\in H^2(X,\R)$ and the volume deformation.

One of the most valuable tools for
understanding the structure of the moduli space is the study of
symmetries. So the next question to be answered
is how to translate symmetries of our superconformal
field theory to its geometric interpretations.
Those symmetries  which commute with the
$su(2)_l\oplus su(2)_r$ action leave the four--plane $x$ invariant and
are called \textsl{algebraic symmetries}.
When the $N=(4,4)$ supersymmetric theories are constructed in terms of
$(2,2)$ supersymmetric theories one has a natural framing. In this
context, algebraic symmetries are those which leave the entire vector space
$Q_l\otimes Q_r$ of supercharges invariant. More generally, any abelian 
symmetry group  of our theory projects to a $u(1)_l\oplus u(1)_r$
subgroup of $su(2)_l\oplus su(2)_r$ and fixes the
corresponding $N=(2,2)$ subalgebra. When corresponding supercharges are
fixed, the abelian symmetry group acts diagonally on the charge generators
$J^\pm,\qu{J}^\pm$ of $su(2)_l^{susy}\oplus su(2)_r^{susy}$. The algebraic
subgroup of this symmetry group is the one which fixes these charges.

If the primitive nullvector $\upsilon$ specifying our geometric
interpretation $(\Sigma,V,B)$ is invariant upon the induced action of an
algebraic symmetry we call the latter a
\textsl{classical symmetry} of the geometric interpretation $(\Sigma,V,B)$.
Because a classical symmetry $\alpha^\ast$ fixes $x$ by definition
we get an induced automorphism
of $H^2(X,\R)$ which leaves $\Sigma\subset H^2(X,\R)$ and
$B\in H^2(X,\R)/H^2(X,\Z)$ invariant. Moreover, because $\xi_4$ in
\req{expdecomp} is invariant as well,
$\eta_{20}=\upsilon^0+B-({\|B\|^2\over2}+V)\upsilon$ is fixed.
Thus $\alpha^\ast$ acts trivially on
moduli of volume and B-field deformation in direction of $\Sigma$.
Because $\alpha^\ast$ acts as automorphism on
$H^{1,1}(X,\R)=\Omega^\perp\cap H^2(X,\R)$ for any choice of complex
structure $\Omega\subset\Sigma$ on $X$ leaving the onedimensional
$H^{1,1}(X,\R)\cap\Sigma$ invariant, all in all, $x\mapsto(\Sigma,V,B)$  maps
the action of $\alpha^\ast$  to an automorphism of $H^2(X,\R)$
which on $H^{1,1}(X,\R)$ has exactly the same spectrum
as $\alpha^\ast$ on $({1\over2},{1\over2})$-fields with
charge, say, $Q=\qu{Q}=1$.

If the integral action of $\alpha^\ast$ on $H^2(X,\C)$ is induced
by an automorphism $\alpha\in Aut(X)$ of finite order
of the $K3$ surface $X$, then by
definition, because $\alpha^\ast$ acts trivially on $H^{2,0}(X,\C)$,
$\alpha$ is an \textsl{algebraic automorphism}  \cite{ni80b}.
This notion
of course only makes sense after a choice of complex structure,
or in conformal field theory language an
$N=(2,2)$ subalgebra of the $N=(4,4)$ superconformal algebra fixing generators
$J,J^\pm,\qu{J},\qu{J}^\pm$ of $su(2)_l\oplus su(2)_r$. Still, because we
always assume the metric to be invariant under $\alpha^\ast$ as well,
i.e. $\Sigma\subset H^2(X,\R)^{\alpha^\ast}$, this is no further restriction.
On the other hand, given an algebraic automorphism
$\alpha$ of $X$
which induces an automorphism of $H^2(X,\R)$ that
leaves the B-field invariant,
$\alpha$  induces a  symmetry of our conformal field theory
which leaves $J,J^\pm,\qu{J},\qu{J}^\pm$ invariant.
This gives a precise notion of how to
continue such an algebraic automorphism to the conformal field theory level.

We are thus naturally led to a discussion of algebraic automorphisms
of $K3$ surfaces, which are  mathematically well understood thanks to
the work of Nikulin \cite{ni80b} for the abelian and Mukai \cite{mu88}
for the general case.
The first to explicitly take advantage of their special properties
in the context of conformal field theory was
P.S.~Aspinwall \cite{as95}.
From \cite[Th.~4.3,4.7,4.15]{ni80b}
one can deduce the following consequence of the global Torelli theorem:
\btheo{algautocriterion}
Let $g$ denote an automorphism of $H^2(X,\C)$ of finite order
which maps forms corresponding to effective divisors of self intersection
number $-2$ in $Pic(X)$ to forms corresponding to effective divisors. Then $g$
is induced by an algebraic automorphism
of $X$ iff
$\left( H^2(X,\Z)^g \right)^\perp \cap H^2(X,\Z) \subset Pic(X)$
is negative definite
with respect to the intersection form
and does not contain elements of length squared $-2$.
\etheo
If for a geometric interpretation $(\Sigma,V,B)$ of
$x\in O^+(H^{even}(X,\Z))\backslash{\cal T}^{4,20}$
we have classical symmetries which act effectively on
what we read off as $H^2(X,\C)$
but are not induced by an algebraic automorphism of
the $K3$ surface $X$
by theorem \ref{algautocriterion}, then our interpretation
of $x$ as giving a superconformal field theory breaks down.
Such points should be conifold points of the moduli space
${\cal M}^{K3}$, characterized by too high an amount of
symmetry. One can regard Nikulin's theorem \ref{algautocriterion}
as harbinger of Witten's result that in points of enhanced symmetry
on the moduli space of type IIA string theories
compactified on $K3$ the conformal  field theory description breaks down
\cite{wi95}.

By abuse of notation we will often renounce to distinguish between
an algebraic automorphism on $K3$ and its induced action on  cohomology.

From Mukai's work \cite[Th.~1.4]{mu88} one may learn that the induced
action of any
algebraic automorphism group $G$ on the total rational cohomology
$H^\ast(X,\Q)$ is a \textsl{Mathieu representation} of $G$ over $\Q$,
i.e. a representation with character
\beq{mathieu}
\chi(g) = \mu(\ord(g)) , \mbox{ where for }
n\in \N: \;
\mu(n) := {24\over n\prod\limits_{
\stackrel{p\, \mbox{\tiny prime,}}{\scriptscriptstyle p|n} } (1+{1\over p})}.
\eeq
It follows that
\beq{mukainum}
\dim_\Q H^\ast(X,\Q)^G =
\mu(G):= {1\over |G|} \sum_{g\in G} \mu(\ord(g))
\eeq
\cite[Prop.~3.4]{mu88}. We remark that because $G$ acts algebraically,
we have\linebreak
$\dim_\Q H^\ast(X,\Q)^G = \dim_\R H^\ast(X,\R)^G= \dim_\C H^\ast(X,\C)^G$.
By definition of algebraic automorphisms $H^\ast(X,\C)^G\supset$
$H^0(X,\C)\oplus H^{2,0}(X,\C)\oplus H^{0,2}(X,\C)
\oplus H^{2,2}(X,\C)$, so
\beq{invdim}
\mu(G)-4 = \dim_\R H^{1,1}(X,\R)^G .
\eeq
Moreover, from theorem \ref{algautocriterion} we know that
$\left(H^2(X,\R)^G\right)^\perp\subset H^{1,1}(X,\R)$ is negative definite,
and because
$H^{1,1}(X,\R)$ has signature $(1,19)$,
we may conclude that it contains an invariant element
with positive length squared. Thus
$\mu(G)\geq 5$ for
every algebraic automorphism group $G$ \cite[Th.~1.4]{mu88}.
Moreover \cite[Cor.~3.5, Prop.~3.6]{mu88},
\beq{mukabsch}
G\neq\{\id\} \Longrightarrow \mu(G)\leq 16 .
\eeq
Finally let us consider the special case of
an algebraic automorphism $\alpha$ of order 4, which will be useful
in due course. By
$n_k$ we denote the multiplicity of the eigenvalue $i^k$ of
the induced action $\alpha^\ast$ on $H^{1,1}(X,\C)$. Because by
\req{mathieu} and \req{mukainum} $\mu(\Z_4)=10$ and $\mu(\Z_2)=16$,
using \req{invdim} we find
$n_0=10-4=6, n_2=16-4-n_0=6$. The
automorphism $\alpha^\ast$ acts on the lattice $H^2(X,\Z)$, so
it must have integer trace.
On the other hand $20=\dim_\C H^{1,1}(X,\C) = n_0+n_1+n_2+n_3$, hence
\beq{fourspec}
n_0=n_2=6,\quad n_1=n_3=4 .
\eeq

\section{Special subspaces of the moduli space: Orbifold theories}
\label{orbifolds}
This section is devoted to the study of theories which have a geometrical
interpretation on an orbifold limit of $K3$.
We begin by giving a  short account on the
relevant features of the orbifold construction,
for details the reader is referred to the vast literature, e.g.
\cite{dhvw86,di87}.

On the geometric side, the $\Z_l$ orbifold construction
of $K3$ can be described as follows \cite{wa88}:
Consider a fourtorus $T$, where
$T=T^2\times \wt{T}^2$  with two
$\Z_l$ symmetric twotori $T^2=\C/L,\,\wt{T}^2=\C/\wt{L}$
which need not be orthogonal.
Let $\zeta\in\Z_l$  act algebraically on
$(z_1,z_2)\in T^2\times \wt{T}^2$ by $(z_1,z_2)
\mapsto(\zeta z_1,\zeta^{-1}z_2)$.
Mod out this symmetry and blow up the resulting singularities;
that is, replace each singular point  by a chain of exceptional divisors,
which in the case of $\Z_l$-fixed points have as intersection
matrix the Cartan matrix of  $A_{l-1}$. In particular,
the exceptional divisors themselves are
rational curves, i.e. holomorphically embedded
spheres with self intersection number $-2$. In terms of the homology
of the resulting surface $X$ these rational curves are elements
of $H_2(X,\Z)\cap H_{1,1}(X,\C)$.
To translate to cohomology we work with their Poincar\'e duals, which
now are elements of $Pic(X)$ with length squared $-2$.
One may check that for $l\in\{2,3,4,6\}$ this procedure changes the
Hodge diamond by
$$
\begin{array}{ccccc} &&1\\ &2&&2\\ 1&&4&&1\\&2&&2 \\ &&1 \end{array}
\longmapsto
\begin{array}{ccccc} &&1\\ &0&&0\\ 1&&20&&1\\&0&&0 \\ &&1 \end{array}
$$
and indeed produces a $K3$ surface $X$, because the automorphism
we modded out was algebraic.
We also obtain a rational map $\pi: T\rightarrow X$ of degree $l$ by
this procedure.
To fix a hyperk\a hler structure
we additionally need to pick the class of a K\a hler metric
on $X$. We will
consider orbifold limits of $K3$ surfaces, that is use the
\textsl{orbifold singular metric}
on $X$ which is induced from the flat metric on $T$ and
assigns volume zero to all
the exceptional divisors. The corresponding Einstein metric
is constructed by excising a
sphere around each singular point of $T/\Z_l$
and gluing in an Eguchi Hanson sphere $E_2$ instead for $l=2$, or a
generalized version $E_l$ with boundary
$\partial E_l=\S^3/\Z_l$ at infinity and nonvanishing Betti numbers
$b_0(E_l)=1, b_2(E_l)=b_2^-(E_l)=l-1$, i.e. $\chi(E_l) = l$.
The orbifold limit is the limit
these Eguchi Hanson type spheres have shrunk to zero size in.
The description \req{einsteinmod}
of the moduli space of Einstein metrics of volume 1
on $K3$  includes orbifold limits \cite{koto87},
and as was shown by Anderson \cite{an92} one can define an
\textsl{extrinsic $L^2$-metric} on the space ${\cal E}$
of regular Einstein metrics of volume 1 on $K3$ such that the completion
of ${\cal E}$ is contained in the set of regular and orbifold singular
Einstein metrics.

On the conformal field theory side the orbifold construction is in total
analogy to the geometric one described above. Assume we know the action
of $\Z_l$ on the space of states ${\cal H}$ of a conformal field theory
with geometric interpretation on the torus $T$ we had above. To construct
the orbifold conformal field theory, keep all the invariant states in
${\cal H}$ and then -- for the sake of modular invariance, if we
argue on the level of partition functions -- add twisted sectors. For
$\zeta\in\Z_l$, the $\zeta$-twisted sector consists
of states corresponding to fields $\phi$ which are only well defined
up to $\zeta$-action on the world sheet of the original theory,
that is $\phi:Z\rightarrow T, \phi(\sigma_0+1, \sigma_1) =
\zeta\phi(\sigma_0, \sigma_1)$.
$Z$ denotes the configuration space as mentioned in the introduction
and coordinates $(\sigma_0, \sigma_1),\sigma_0\sim\sigma_0+1,
(\sigma_0,\sigma_1)\sim(\sigma_0+\tau_0,\sigma_1+\tau_1)$
are chosen such that $\phi(0,0)$ is a fixed point.
In other words,
the constant mode in the Fourier expansion of $\phi$ is a fixed
point $p_\zeta$ of $\zeta$. The other modes are of non integral level, so
the ground state energy in the twisted sector
is shifted away from zero. More precisely, the ground state
$|\Sigma_{\zeta, p_\zeta}\rangle$
of the $\zeta$-twisted sector ${\cal H}_{\zeta,p_\zeta}$ belongs to the
Ramond sector and has   dimensions
$h=\qu{h}={c\over24}={1\over4}$. The corresponding field
$\Sigma_{\zeta, p_\zeta}$ introduces
a cut in $Z$ from $(0,0)$ to $(\tau_0,\tau_1)\sim(0,0)$
to establish the transformation property
$\phi(\sigma_0+1, \sigma_1) = \zeta\phi(\sigma_0, \sigma_1)$ for
$|\phi\rangle\in {\cal H}_{\zeta,p_\zeta}$, often referred to as
\textsl{boundary condition}. The field $\Sigma_{\zeta, p_\zeta}$
is called a \textsl{twist field}. For explicit formulae
of partition functions for $\Z_l$ orbifold
conformal field theories see \cite{eoty89}, for the special cases
$l=2$ and $l=4$ we are studying here see
\req{kummerpart} and \req{z4part}.

To summarize, we stress the analogy between orbifolds
in the geometric and the conformal field theory sense once again;
in particular,
the introduction of a twist field for each fixed point
and boundary condition corresponds
to the introduction of an exceptional divisor in the course of
blowing up the quotient singularity, if we use the metric which assigns
volume zero to all the exceptional divisors. 

By construction orbifold conformal field theories have a preferred
geometric interpretation in the sense of section \ref{k3mod}.
We will now investigate this geometric interpretation
for $\Z_2$ and $\Z_4$ orbifolds, particularly
taking advantage of their specific algebraic automorphisms.
A program for finding a stratification of the moduli space could even
be formulated as follows: Find all subspaces of theories having
a geometric interpretation $(\Sigma,V,B)$
with given algebraic automorphism group $G$.
Relations between such subspaces may be described by the modding out
of algebraic automorphisms. Any infinitesimal deformation of
$\Sigma$ by an element of $H^{1,1}(X,\R)^G$
will preserve  the symmetries in $G$, as well as volume deformations
and B-field deformations by elements in $H^2(X,\R)^G$.
The subspace of theories with given classical symmetry group $G$
in a geometric interpretation
therefore can maximally have real dimension
$3(\mu(G)-5) + 1 + \mu(G)-2 = 4(\mu(G)-4)$
in accord with \req{invdim}. In particular, for the minimal
value $\mu(G)=5$, the only deformations preserving the entire
symmetry are deformations of volume and those of the B-field by
elements of $\Sigma$.

Of course, the above program is far from utterly
realizable, even in the pure geometric context, but it might serve as
a useful line of thought.
$\Z_2$ Orbifolds actually yield the first item of this program:
We can map the entire torus moduli space into the $K3$ moduli space by
modding out the symmetry $z\mapsto -z$. The description is straightforward
if we make use of the geometric interpretation of torus theories given by
the triality automorphism \req{decomptor}, because the geometric data
then turn out to translate in a simple way into the
corresponding data on $K3$.
\subsection{$\Z_2$ Orbifolds in the moduli space}\label{z2orb}
Some comments on $\Z_2$ orbifold conformal
field theories as described at the beginning of the section are due,
before we can show where they are located within the moduli space
${\cal M}^{K3}$. We denote
the $\Z_2$ orbifold obtained from the nonlinear $\sigma$ model
${\cal T}(\Lambda,B_T)$ by ${\cal K}(\Lambda,B_T)$. If the theory
on the torus has an enhanced symmetry $G$ we frequently simply write
$G/\Z_2$, e.g. $SU(2)_1^{\,4}/\Z_2$ for ${\cal K}(\Z^4,0)$.

In the nonlinear $\sigma$ model on the torus $T=\R^4/\Lambda$
as described in section \ref{modtor}
the current $j_k$  generates translations
in direction of coordinate $x_k$. This induces
a natural correspondence between tangent vectors of $T$ and fields of
the nonlinear $\sigma$ model which is compatible with the $so(4)$ action
on the tangent spaces of $T$ and the moduli space, respectively.
After selection of an appropriate framing of $Q_l\otimes Q_r$ to identify
$su(2)^{susy}_{l,r}$ with $su(2)_{l,r}$ as described
in section \ref{modulispace}
the $\psi_k$ are the superpartners of the $j_k$.
Hence the choice of complex coordinates
$z_1 := {1\over\sqrt2}(x_1+ix_2),
z_2 := {1\over\sqrt2}(x_3+ix_4)$
corresponds to setting
\beq{compco}
\psi_\pm^{(1)} \;:=\; \inv{\sqrt2} (\psi_1\pm i\psi_2), \quad
\psi_\pm^{(2)} \;:=\; \inv{\sqrt2} (\psi_3\pm i\psi_4).
\eeq
The holomorphic
$W$-algebra of our theory has an $su(2)_1^{\,2}$-subalgebra
generated by
\beqn{su2h2curalg}
J \>:=\>
\psi_+^{(1)}\psi_-^{(1)}+\psi_+^{(2)}\psi_-^{(2)},
\quad J^+ \;:=\; \psi_+^{(1)}\psi_+^{(2)}, \quad
J^- \;:=\; \psi_-^{(2)}\psi_-^{(1)}; \e
A \>:=\>
\psi_+^{(1)}\psi_-^{(1)}-\psi_+^{(2)}\psi_-^{(2)},
\quad A^+ \;:=\; \psi_+^{(1)}\psi_-^{(2)}, \quad
A^- \;:=\; \psi_+^{(2)}\psi_-^{(1)}.
\eeqn
Its geometric counterpart on the torus is  the Clifford algebra generated
by the twoforms
$dz_1\wedge d\qu{z}_1+dz_2\wedge d\qu{z}_2,
dz_1\wedge dz_2, d\qu{z}_1\wedge d\qu{z}_2;
dz_1\wedge d\qu{z}_1-dz_2\wedge d\qu{z}_2,
dz_1\wedge d\qu{z}_2, dz_2\wedge d\qu{z}_1$ upon Clifford multiplication.

The nonlinear $\sigma$ model on the Kummer surface ${\cal K}(\Lambda)$
is the ``ordinary'' $\Z_2$ orbifold
of the above, where $\Z_2$ acts by $j_k\mapsto -j_k,
\psi_k\mapsto -\psi_k$, $k=1,\dots,4$. Note that the entire
$su(2)_1^{\,2}$-algebra \req{su2h2curalg} is invariant under this action,
thus any  nonlinear $\sigma$ model on a Kummer surface possesses an
$su(2)_1^{\,2}$-current algebra.
The $NS$-part of its partition function is
\beqn{kummerpart}
Z_{NS}(\tau,z)
\>=\> \inv{2}\left\{
Z_{\Lambda,B}(\tau) \left| {\theta_3(z)\over \eta}\right|^4
+ \left| {\theta_3\theta_4\over \eta^2}\right|^4
\left| {\theta_4(z)\over \eta}\right|^4 \right. \e
\>\> \hphantom{\inv{2}\inv{2}Z_{\Lambda,B}(\tau)}
\left.+ \left| {\theta_2\theta_3\over \eta^2}\right|^4
\left| {\theta_2(z)\over \eta}\right|^4
+ \left| {\theta_2\theta_4\over \eta^2}\right|^4
\left| {\theta_1(z)\over \eta}\right|^4
\right\}.
\eeqn
Here and in the following
we  decompose  partition functions into four parts
corresponding to the four sectors $NS, \wt{NS},$ $R, \wt{R}$, i.e.
with $y=exp(2\pi iz), \qu{y}=exp(-2\pi i\qu{z})$
\beq{sectors}
\begin{array}{rclcl}
Z
& = &
{1\over 2} \left( \vphantom{{1\over 2}}
Z_{NS} + Z_{\wt{NS}} + Z_{R} + Z_{\wt{R}} \right) ,
\e
Z_{NS}(\tau,z)
& = & \tr[NS]
\left[
q^{L_0-{c\over 24}}\qu{q}^{\qu{L}_0-{c\over 24}}  y^{J_0}\qu{y}^{\qu{J}_0}
\right],
\e
Z_{\wt{NS}}(\tau,z)
& = & \tr[NS]
\left[
(-1)^F
q^{L_0-{c\over 24}}\qu{q}^{\qu{L}_0-{c\over 24}} y^{J_0}\qu{y}^{\qu{J}_0}
\right]
& = & Z_{NS}(\tau,z+{1\over 2}),
\e
Z_{R}(\tau,z)
& = & \tr[R]
\left[
q^{L_0-{c\over 24}}\qu{q}^{\qu{L}_0-{c\over 24}} y^{J_0}\qu{y}^{\qu{J}_0}
\right]
& = &  (q\qu{q})^{c\over 24} (y\qu{y})^{c\over6}\,
Z_{NS}(\tau,z+{\tau\over 2}),
\e
Z_{\wt{R}}(\tau,z)
& = & \tr[R]
\left[
(-1)^F
q^{L_0-{c\over 24}}\qu{q}^{\qu{L}_0-{c\over 24}} y^{J_0}\qu{y}^{\qu{J}_0}
\right]
& = & Z_{R}(\tau,z+{1\over 2}).
\end{array}
\eeq
Given $Z_{NS}$ the entire partition function can be
determined by using the above flows to find $Z_{\wt{NS}}, Z_R$ and
$Z_{\wt{R}}$.

This orbifold model has an $N=(16,16)$ supersymmetry. We are interested
in deformations which conserve  $N=(4,4)$ subalgebras. As explained in 
section \ref{modulispace},
the latter are given by chiral and antichiral $({1\over2},{1\over2})$--fields.
Generically, the Neveu-Schwarz sector contains $144$ fields with
dimensions $(h,\qu{h})=({1\over2},{1\over2})$. Their quantum numbers
under $(J,A;\qu{J},\qu{A})$ are $(\eps_1,\eps_2;\eps_3,\eps_4), 
\eps_i\in\{\pm1\}$ ($16$ fields), $(\eps_1,0;\eps_3,0)$ ($64$ fields), and
$(0,\eps_2;0,\eps_4)$ ($64$ fields). The $80$ fields which are 
charged under $(J;\qu{J})$ yield the $N=(4,4)$ supersymmetric deformations
which conserve the superalgebra that contains the $J$ currents. The $80$ 
fields which are charged under $(A;\qu{A})$ yield deformations conserving
a different $N=(4,4)$ superalgebra. The latter corresponds to the opposite
torus orientation.

Let us now focus on the description of the resulting geometric objects,
namely \textsl{Kummer surfaces} denoted by ${\cal K}(\Lambda)$ if
obtained by the $\Z_2$ orbifold procedure
from the fourtorus
$T=\R^4/\Lambda$. Generators of the lattice $\Lambda$ are denoted by
$\lambda_1,\dots,\lambda_4$. From \req{decomptor} we obtain an associated
three--plane $\Sigma_T\subset H^2(T,\R)$, i.e. an Einstein metric on $T$, and
we must describe how the  Teich\-m\u l\-ler
space ${\cal T}^{3,3}$ of Einstein metrics of volume 1
on the torus is mapped into the corresponding space ${\cal T}^{3,19}$
for $K3$.
This is best understood in terms of the lattices
$H^2(T,\Z)\cong \Gamma^{3,3}$ and
$H^2(X,\Z)\cong \Gamma^{3,19}, X={\cal K}(\Lambda)$.
In our notation $H^2(T,\Z)$ is generated by
$\mu_j\wedge\mu_k, j,k\in\{1,\dots,4\}$ if
$(\mu_1,\dots,\mu_4)$ is the basis dual
to $(\lambda_1,\dots,\lambda_4)$.
$\Sigma_T$ is defined by its relative position to a reference lattice
$\Gamma^{3,3}\cong H^2(T,\Z)\subset H^2(T,\R)$.
Note that in order to simplify the
following argumentation we rather regard $\Sigma_T\subset H^2(T,\Z)$
as giving the  position of the lattice
$H^2(T,\Z)=\span_\Z(\mu_j\wedge\mu_k)$
relative to a fixed three--plane
$\span_\R(e_1\wedge e_2+e_3\wedge e_4,e_1\wedge e_3+e_4\wedge e_2,
e_1\wedge e_4+e_2\wedge e_3)$ with respect to the standard basis
$(e_1,\dots,e_4)$ of $\R^4$.

To make contact with the  theory of Kummer surfaces we
pick a complex structure $\Omega_T\subset\Sigma_T$.
The $\Z_2$ action on $T$ has $16$ fixed points
${1\over2}\sum_{k=1}^4 \eps_k\lambda_k, \eps\in\F_2^4$.
We can therefore choose indices in $\F_2^4$ to label the fixed 
points\footnote{$\F_2$ denotes the unique finite field with two elements.}.
Note that this is not only a labelling but the torus geometry indeed
induces a natural affine $\F_2^4$-structure on the set $I$ of
fixed points \cite[Cor.~5]{ni75}.
The twoforms corresponding to the $16$ exceptional
divisors obtained from blowing up the fixed points are denoted by
$\{ E_i\mid i\in I\}$. They are elements of $Pic(X)$ no matter what
complex structure we choose, because we are working in the orbifold limit,
i.e. $E_i\perp\Sigma\;\fa i\in I$.
Let $\Pi\subset Pic(X)$ denote the primitive
sublattice of the Picard lattice that contains $\{ E_i\mid i\in I\}$.
It is called \textsl{Kummer lattice} and by \cite[Th.~3]{ni75}:
\btheo{kummercriterion}
The Kummer lattice $\Pi$ is spanned by the exceptional divisors
$\{ E_i\mid i\in I\}$ and
$\{ {1\over2}\sum_{i\in H} E_i \mid H\subset I$ is a hyperplane$\}$.
On the other hand, a
$K3$ surface $X$ is a Kummer surface iff $Pic(X)$ contains
a primitive sublattice isomorphic to $\Pi$.
\etheo
%
Let $\pi:T\rightarrow X$ be the degree two map from the torus to
the orbifold singular Kummer surface. Using Poincar\'e duality, one gets
maps $\pi_\ast$ from the homology and cohomology groups of $T$ to those of
$X$, and $\pi^\ast$ in the other direction. In particular, this gives the
natural embedding $\pi_\ast:H^2(T,\Z)(2)\hookrightarrow H^2(X,\Z)$
(here $\Gamma(2)$ denotes $\Gamma$ with quadratic form scaled by $2$).
The image lattice will be called $K$.
We prefer to work with metric isomorphisms and therefore denote the
image in $K$ of an element $a\in H^2(T,\Z)$ by $\sqrt2 a$.
In particular, we write
$\sqrt2\mu_j\wedge\mu_k, j,k=1,\dots,4$ for the generators of $K$.
The lattice $H^2(X,\Z)$ contains $K\oplus\Pi$ and is contained in
the dual lattice $K^\ast\oplus\Pi^\ast$. The
three--plane $\Sigma\subset H^2(X,\R)$ which describes the location
of the singular Kummer surface
within the moduli space \req{einsteinmod} of Einstein metrics of volume
1 on $K3$ is given by $\Sigma=\pi_\ast\Sigma_T$.

A description of how the lattices
$K$ and $\Pi$ are embedded in $H^2(X,\Z)$ can be found in \cite{ni75}.
First notice
$K^\ast /K\cong (\Z_2)^6 \cong \Pi^\ast /\Pi$,
where $\Pi^\ast /\Pi$ is generated by
$\{ {1\over2}\sum_{i\in P} E_i \mid P\subset I$ is a plane$\}$.
The isomorphism $\gamma:K^\ast /K \longrightarrow \Pi^\ast /\Pi$
is most easily understood in terms of homology by assigning the
image in $X$ of a twocycle through four fixed points in a plane $P\subset I$
to  ${1\over2}\sum_{i\in P} E_i$. For example,
$\gamma(\inv{\sqrt2}\mu_j\wedge\mu_k)=\inv{2}\sum_{i\in P_{jk}}E_i$,
$P_{jk}=\span_{\F_2}(f_j,f_k)\subset\F_2^4$, $f_j\in\F_2^4$ the $j$th
standard basis vector. Note that $P_{jk}$ may be exchanged by any
of its translates $l+P_{jk}, l\in\F_2^4$.
Next check that the discriminant forms of $K^\ast /K$ and
$\Pi^\ast /\Pi$, i.e. the induced $\Q/2\Z$ valued quadratic forms, agree up
to a sign. Then
\beq{glue}
H^2(X,\Z) \cong \left\{ (x,y)\in K^\ast\oplus \Pi^\ast \mid
\gamma(\qu{x})=\qu{y} \right\},
\eeq
$\qu{x},\qu{y}$ denoting the images of $x,y$
under projection to $K^\ast /K$,
$\Pi^\ast /\Pi$. The isomorphism \req{glue} provides a  natural
primitive embedding $ K\perp\Pi \hookrightarrow H^2(X,\Z)$,
which is unique up to isomorphism \cite[Lemma 7]{ni75}.
Here, $H^2(X,\Z)\cong\Gamma^{3,19}$ is generated by
\beq{genh2}
M:=
\left\{ \inv{\sqrt2}\mu_j\wedge\mu_k
+ \inv{2}\smash{\sum_{i\in P_{jk}}} E_{i+l}, l\in I
\right\}\mbox{ and }
 \span_\Z\left(
E_i, i\in I\right).
\eeq
Hence
$\Gamma^{3,3}(2)\cong H^2(T,\Z)(2)\hookrightarrow H^2(X,\Z)\cong\Gamma^{3,19}$
is naturally embedded, and
in particular $\Sigma\subset H^2(X,\R)\cong H^2(X,\Z)\otimes\R$
is obtained directly by regarding
$\Sigma_T\subset H^2(T,\R)\cong H^2(T,\Z)\otimes\R
\hookrightarrow H^2(X,\Z)\otimes\R$ as three--plane in $H^2(X,\R)$.

To describe where the image ${\cal K}(\Lambda,B_T)$ of a
superconformal field theory ${\cal T}(\Lambda,B_T)$
under $\Z_2$ orbifold is located in ${\cal M}^{K3}$ we now
generalize the above construction to the quantum level. 
We have to lift $\pi_\ast$ to an embedding
$\wh{\pi}_\ast:H^{even}(T,\Z)(2)\hookrightarrow H^{even}(X,\Z)$.
The image will be denoted by $\wh{K}$. Apart from $\mu_j\wedge\mu_k$
the lattice $H^{even}(T,\Z)$ has generators $\upsilon,\upsilon^0$ as defined
in \req{expdecomp}. Note that
$\wh{K}$ cannot be embedded
as primitive sublattice in $\Gamma^{4,20}$ such that $\wh{K}\perp\Pi$
because $\wh{K}^\ast/\wh{K}\cong(\Z_2)^8\not\cong(\Z_2)^6\cong\Pi^\ast/\Pi$.
This means that the B-field of the orbifold theory must have components
in the Picard lattice.

The torus model is given by a four--plane $x_T\subset H^{even}(T,\R)$,
the corresponding orbifold model by its image $x=\wh{\pi}_\ast x_T$ in
$H^{even}(X,\Z)\otimes \R$. To arrive at a complete
description, we must find the
embedding of $H^{even}(X,\Z)$ in $\wh{K}\otimes\R + H^2(X,\R)$.
Since scalar products with elements of $\wh{K}$ must be integral
and $\sqrt2\upsilon^0\in\wh{K}$, 
every $a\in \Pi$ must have a lift ${1\over\sqrt2}\upsilon+a$
or $0+a$ in $H^{even}(X,\Z)$. Those elements
for which the lift has the form
$0+a$ must form an $O^+(H^{even}(T,\Z))$ invariant sublattice of $\Pi$.
One may easily check that this sublattice cannot contain the 
exceptional divisors $E_i, i\in I$. Moreover, as unimodular lattice
$H^{even}(X,\Z)$ must contain an element of the form
${1\over\sqrt2}\upsilon^0+a$ with $a\in \Pi^\ast$. One finds 
that $H^{even}(X,\Z)$ must contain the set of elements
\beq{mroof}
\wh{M} := M\cup
\left\{ \inv{\sqrt2}\upsilon^0 - \inv{4}\smash{\sum_{i\in I}E_i};\;
-\inv{\sqrt2}\upsilon +E_i , i\in I \right\}.
\eeq

In analogy to
Nikulin's description \req{glue} and \req{genh2}
of  $H^2(X,\Z)\cong \Gamma^{3,19}$
we now find
\blem{gamma420}
The lattice $\Gamma$ spanned by $\wh{M}$ and
$\{\pi\in\Pi \mid \fa m\in\wh{M}:\, \langle\pi,m\rangle\in\Z \}$
is isomorphic to $\Gamma^{4,20}$.
\elem
\bpr
Define
\beq{newlattice}
\wh{\upsilon} := \sqrt2\upsilon,\quad
\wh{\upsilon}^0
:= \inv{\sqrt2}\upsilon^0 -\inv{4}\sum_{i\in I}E_i +\sqrt2\upsilon, \quad
\wh{E_i} := -\inv{\sqrt2}\upsilon + E_i.
\eeq
Then $\Gamma$ is generated by $\wh{\upsilon},\wh{\upsilon}^0$ and the
lattice
$$
\wh{\Gamma}:=\span_\Z\left(
\inv{\sqrt2}\mu_j\wedge\mu_k + \inv{2}\smash{\sum_{i\in P_{jk}}}
\wh{E\hphantom{_i}}_{\!\!i+l},
l\in I;\;\;
\wh{E_i}, i\in I
\right).
$$
Because $\langle\wh{E_i},\wh{E_j}\rangle = -2\delta_{ij}$
and upon comparison to \req{genh2}
it is now easy to see that $\wh{\Gamma}\cong\Gamma^{3,19}$.
Moreover, $\wh{\upsilon},\wh{\upsilon}^0\perp\wh{\Gamma}$
and $\span_{\Z}(\wh{\upsilon},\wh{\upsilon}^0)\cong U$ completes
the proof.
\epr
In particular, lemma \ref{gamma420} describes a natural embedding
$\Gamma^{4,4}(2)\cong H^{even}(T,\Z)(2)$ $\hookrightarrow
H^{even}(X,\Z)\cong\Gamma^{4,20}$. As in the case of embedding
the Teichm\u ller spaces ${\cal T}^{3,3}\hookrightarrow{\cal T}^{3,19}$
this enables us to directly locate
the image under $\Z_2$ orbifold
of a conformal field theory corresponding to
a four--plane $x\subset H^{even}(T,\R)\cong\Gamma^{4,4}\otimes\R$
within ${\cal M}^{K3}$ by regarding $x$ as four--plane
in $H^{even}(X,\R)\cong\Gamma^{4,20}\otimes\R$. Note that in this geometric
interpretation
$\wh{\upsilon},\wh{\upsilon}^0$ are the generators of
$H^4(X,\Z)$ and $H^0(X,\Z)$.
\btheo{toremb}
Let $(\Sigma_T,V_T,B_T)$ denote a geometric interpretation of the nonlinear
$\sigma$ model ${\cal T}(\Lambda,B_T)$ as given by \req{decomptor}.
Then the corresponding orbifold conformal field theory ${\cal K}(\Lambda,B_T)$
associated to the Kummer surface $X={\cal K}(\Lambda)$ has
geometric interpretation $(\Sigma,V,B)$ where $\Sigma\in{\cal T}^{3,19}$
as described after theorem \ref{kummercriterion}, $V={V_T\over2}$
and $B={1\over\sqrt2}B_T+{1\over2}B_\Z^{(2)},
B_\Z^{(2)}={1\over2}\sum_{i\in I}\wh{E_i}\in H^{even}(X,\Z)$
with  $\wh{E_i}\in H^{even}(X,\Z)$ of length squared -2 given in
\req{newlattice}.

In particular, the $\Z_2$ orbifold procedure induces an embedding
${\cal M}^{tori}\hookrightarrow{\cal M}^{K3}$ as quaternionic submanifold.
\etheo
\bpr
Pick a basis $\sigma_i, i\in\{1,2,3\}$ of $\Sigma_T$. Then by \req{expdecomp}
the nonlinear $\sigma$ model ${\cal T}(\Lambda,B_T)$ is given by the
four--plane $x$ with generators
$\xi_i=\sigma_i - \langle\sigma_i,B_T\rangle\upsilon, i\in\{1,2,3\}$
and
$\xi_4=\upsilon^0+B_T+\left(V_T-\inv[\|B_T\|^2]{2}\right)\upsilon$.
By the embedding
$\Gamma^{4,4}\otimes\R \cong H^{even}(T,\R)\hookrightarrow
H^{even}(X,\R)\cong \Gamma^{4,20}\otimes\R$ given in lemma \ref{gamma420}
it is now a simple task to reexpress the generators of $x$ using the
generators $\wh{\upsilon},\wh{\upsilon}^0$  of
$H^4(X,\Z)$ and $H^0(X,\Z)$:
\begin{eqnarray*}
\sqrt2\left(\sigma_i-\langle\sigma_i,B_T\rangle\upsilon\right)
&=& \sqrt2\sigma_i-\langle\sqrt2\sigma_i,\inv{\sqrt2}B_T\rangle\wh{\upsilon}\\
\inv{\sqrt2}\left(\upsilon^0+B_T+\left(V_T-\inv[\|B_T\|^2]{2}\right)
\upsilon\right)
&=& \wh{\upsilon}^0+\inv{\sqrt2}B_T+\inv{2}B_\Z^{(2)}\\
&&\quad
+\left(\inv[V_T]{2}-\inv{2}\left\|\inv{\sqrt2}B_T+\inv{2}B_\Z^{(2)}\right\|^2
\right)\wh{\upsilon}.
\end{eqnarray*}
Comparison with \req{expdecomp} directly gives the assertion of
the theorem.
\epr
Theorem \ref{toremb} makes precise how the statement
that orbifold conformal field theories tend to give value $B={1\over2}$
to the B-field in direction of exceptional divisors
\cite[\para4]{as96} is to be understood.
Note that $x^\perp\cap\Gamma^{4,20}$ does not contain vectors of length
squared $-2$, namely $E_i\in x^\perp, \|E_i\|^2=-2$ but
$E_i\not\in H^{even}(X,\Z)$.
In the context of compactifactions of the type IIA
string on $K3$ this  proves that $\Z_2$ orbifold conformal
field theories do not have enhanced gauge symmetry.
A similar statement was made
in \cite{as95} and widely spread in the literature, but we were unable
to follow the argument up to our result of theorem \ref{toremb}.
\subsection{T--duality and Fourier--Mukai transform}\label{foumuk}
By theorem \ref{toremb}  any automorphism on the Teichm\u ller space
${\cal T}^{4,4}$ of ${\cal M}^{tori}$ is conjugate to
an automorphism on the Teichm\u ller space ${\cal T}^{4,20}$ of
${\cal M}^{K3}$. In particular, nonlinear
$\sigma$ models on tori  related by T--duality must give
isomorphic theories on $K3$ under $\Z_2$ orbifolding.
To show this explicitly and discuss the
duality transformation on ${\cal M}^{K3}$ obtained
this way is the object of this subsection.

For simplicity first assume that our $\sigma$ model on
the torus $T=\R^4/\Lambda$ has vanishing B-field,
where we have chosen a geometric interpretation $(\Sigma_T,V_T,0)$.
Then T--duality acts by $(\Sigma_T,V_T,0)\mapsto (\Sigma_T,1/{V_T},0)$.
By theorem \ref{toremb} the corresponding $\Z_2$ orbifold theories
have geometric interpretations
$(\Sigma,{V_T}/2,B)$ and $(\Sigma,1/{2V_T},B)$, respectively, where
$\Sigma$ is obtained as image of the embedding
$\Sigma_T\subset H^2 (T,\R)\hookrightarrow H^2(X,\R)$ and
$B={1\over2}B_\Z^{(2)}={1\over4}\sum_{i\in I}\wh{E_i}$.
We will now construct an automorphism $\Theta$
of the lattice $H^{even}(X,\Z)$
which fixes the four--plane $x$ corresponding to the model with geometric
interpretation $(\Sigma,{V_T}/2,B)$ and acts by
${V_T}/2\mapsto1/{2V_T}$.
In other words, we will explicitly construct the duality transformation
induced by torus T--duality on ${\cal M}^{K3}$.
Our transformation $\Theta$ below was already given in \cite{rawa98} but
not with complete proof. Within the context of boundary conformal
field theories,
in \cite{ber99} it was shown that $\Theta$ induces an isomorphism
on the corresponding conformal field theories. The relation to the
Fourier--Mukai transform which we will show in theorem \ref{foumukt} has
not been clarified up to now.

By \req{expdecomp}, the four--plane $x\subset H^{even}(X,\Z)$ is spanned by
$\wt\Sigma=\xi(\Sigma)$ and the vector
$\xi_4=\wh{\upsilon}^0+B+({V_T\over2}+1)\wh{\upsilon}$
(notations as in theorem \ref{toremb}).
Because by the above $\Theta$  fixes $x$ and $\wt\Sigma$ pointwise,
the unit vector ${\xi_4}/\sqrt{V_T}\in\Sigma^\perp\cap x$ must be invariant,
too, i.e. invariant under the
transformation $V_T\mapsto1/{V_T}$. Hence
$$
\inv{\sqrt{V_T}}\wh{\upsilon}^0 + \inv{\sqrt{V_T}}B
+ \left( \inv{2}\sqrt{V_T}+\inv{\sqrt{V_T}}\right)\wh{\upsilon}
= \sqrt{V_T}\wt{\upsilon}^0 + \sqrt{V_T}\wt{B}
+ \left( \inv{2\sqrt{V_T}}+\sqrt{V_T}\right)\wt{\upsilon}
$$
for any value of $V_T$. We set
$\wt{\upsilon}:=\Theta({\wh{\upsilon}}),
\wt{\upsilon}^0:=\Theta({\wh{\upsilon}^0})$ etc. and deduce
\beq{involution}
\wt{\upsilon}^0 + \wt{B} + \wt{\upsilon} = \inv{2}\wh{\upsilon}, \quad
\wh{\upsilon}^0 + {B} + \wh{\upsilon} = \inv{2}\wt{\upsilon} .
\eeq
The first equation together with
$\langle \wt{B}, \wt{\upsilon}\rangle
=\langle \wt{B}, \wt{\upsilon}^0\rangle = 0,
\|\wt{B}\|^2=-2$
implies $\langle \wt{B},\wh{\upsilon}\rangle = -4$
and justifies the ansatz
$$
\wt{B} = -4\wh{\upsilon}^0 - \sum_{i\in I} \alpha_i \wh{E_i} +a\wh{\upsilon}
\;\Longrightarrow\;
\sum_{i\in I}(\alpha_i-1)^2=1, \quad \sum_{i\in I}\alpha_i =8-2a.
$$
The only solutions satisfying
$\sum_{i\in I} \alpha_i \wh{E_i}\in H^{even}(X,\Z)$, which must be true by
\req{involution}, are $\alpha_{i_0}\in\{0,2\}$ for some $i_0\in I$ and
$\alpha_{i}=1$ for $i\neq i_0$, correspondingly
$a\in\{ -{7\over2}, -{9\over2}\}$. We conclude that if the automorphism
$\Theta$ exists, then it is already
uniquely determined up to the choice of $a$ and of one point $i_0\in I$.
The two possible choices of $a$ turn out to be related by the B-field shift
$\wt{B}\mapsto \wt{B}-2\wt{B}=-\wt{B}$ and yield equivalent results.
In the following we pick $a=-{7\over2}$ and find
\beq{fouriermukai}
\wt{\upsilon} = 2(\wh{\upsilon}+\wh{\upsilon}^0)
+ \inv{2}\sum_{i\in I} \wh{E_i},\quad
\wt{\upsilon}^0 = 2(\wh{\upsilon}+\wh{\upsilon}^0)
+ \inv{2}\sum_{i\in I} \wh{E_i} - \wh{E\hphantom{_i}}_{\!\!i_0}.
\eeq
One easily checks that
$\wt{U}:=\span_\Z(\wt{\upsilon},\wt{\upsilon}^0)\cong U$. By $\wt{\Pi}$
we denote the orthogonal complement of $\wt{U}$ in
$\span_\Z(\wh{\upsilon},\wh{\upsilon}^0)\perp\Pi\cong U\perp\Pi$, where
$\Pi$ is the Kummer lattice of $X$ as introduced in theorem
\ref{kummercriterion}. Note that in $I$ there are 15 hyperplanes
$H_i, i\in I_0=I-\{i_0\}$ which do not contain $i_0$.
The label $i\in I_0$ is understood as the vector dual to the hyperplane $H_i$.
Since the choice of $i_0$ can be seen as the choice of
an origin in the affine space $\F_2^4$, the latter can be
regarded as a vector space, and we have a unique natural isomorphism
$(\F_2^4)^\ast\cong\F_2^4$. 
One now checks that $\wt{\Pi}$ is spanned by $\wt{E_i}, i\in I$, with
\beqn[rcll]{foumupi}
\wt{E\hphantom{_i}}_{\!\!i_0}
\>:=\>\wh{\upsilon}-\wh{\upsilon}^0, \quad
\wt{E_{i}} \,:=\, -\inv{2}\sum_{j\in H_i} \wh{E_j} -
\wh{\upsilon}-\wh{\upsilon}^0 \>\quad (i\neq i_0)
\eeqn
as well as ${1\over2}\sum_{i\in H}\wt{E_{i}}$ for any hyperplane
$H\subset  I$. The signs of the $\wt{E_{i}}$ have been chosen such that
$\wt{B}={1\over2}\wt{B}_\Z^{(2)}={1\over4}\sum_{i\in I}\wt{E_i}$.

Since $\langle \wt{E_i},\wt{E_j}\rangle = -2\delta_{ij}$, 
one has $\wt{\Pi}\cong\Pi$. Hence
$\Theta(\wh{E_i})=\wt{E_i}$ is a continuation of \req{fouriermukai} to
an automorphism of lattices $U\perp\Pi\cong\wt{U}\perp\wt{\Pi}$, and
we find $\Theta^2=\id$.
Note that the action of $\Theta$ can be viewed as
a duality transformation exchanging vectors $i\in I$ with
hyperplanes $H_i, i\in I$. Two--planes $P\subset I$ are exchanged with
their duals $P^\ast$ which shows that $\Theta$ can be continued to a map
on the entire lattice $H^\ast(X,\Z)$ consistently with \req{glue}. The induced
action on $K=\pi_\ast H^2(T,\Z)$ leaves $\Sigma$ invariant.
We also see that the above procedure is easily generalized to
arbitrary nonlinear $\sigma$ models ${\cal T}(\Lambda,B_T)$.

Let $S$ denote the classical symmetry which changes the sign of
$\wh{E\hphantom{_i}}_{\!\!i_0}$ and leaves the other lattice generators
$\wh{E_i}$, $i\neq i_0$, $\wh{\upsilon}$, $\wh{\upsilon}^0$,
$\mu_j\wedge \mu_k$ invariant. By \req{fouriermukai} and \req{foumupi}
one has $\Theta S=T_{FM} \Theta$, where $T_{FM}$ is the Fourier-Mukai
transformation which exchanges $\wh{\upsilon}$ with $\wh{\upsilon}^0$. 
Since $T_{FM}=\Theta S \Theta$,
all in all we have
\btheo{foumukt}
Torus T--duality induces a duality transformation 
$\Theta$ as given by \req{fouriermukai} and \req{foumupi}
on the subspace of
${\cal M}^{K3}$ of theories associated to Kummer surfaces in the orbifold
limit (see also \cite{rawa98}). The Fourier--Mukai transform $T_{FM}$ which
exchanges $\wh{\upsilon}$ with $\wh{\upsilon}^0$ is conjugate to 
a classical symmetry $S$ by the image $\Theta$ of the T-duality map
on theories associated to the torus.
\etheo
Note that by theorem \ref{foumukt} we can prove
Aspinwall's and Morrison's description \req{modallgglobal}
of the moduli space ${\cal M}^{K3}$
purely within conformal field theory without
recourse to Landau--Ginzburg arguments. Namely, as explained in section
\ref{modulispace}, the group ${\cal G}^{(16)}$ needed to project from the
Teichm\u ller
space \req{modallg} to the component ${\cal M}^{K3}$ of the moduli space
contains the group $O^+(H^2(X,\Z))$ of classical symmetries which fix
the vectors $\wh{\upsilon}, \wh{\upsilon}^0$ determining our geometric
interpretation. Moreover, for any primitive
nullvector $\wt{\upsilon}^0$ with 
$\langle\wh{\upsilon},\wt{\upsilon}^0\rangle=1$
there exists an element
$\wt{g}\in{\cal G}^{(16)}$ such that $\wt{g}\wh{\upsilon}=\wh{\upsilon}$
and $\wt{g}\wh{\upsilon}^0=\wt{\upsilon}^0$. By theorem \ref{foumukt}
the symmetry $T_{FM}\in O^+(H^{even}(X,\Z))$ which
exchanges $\wh{\upsilon}$ and $\wh{\upsilon}^0$ and leaves $x$
invariant also is an
element  of ${\cal G}^{(16)}$, thus
$O^+(H^{even}(X,\Z))\subset{\cal G}^{(16)}$
and $O^+(H^{even}(X,\Z))={\cal G}^{(16)}$ under the assumption
that ${\cal M}^{K3}$ is Hausdorff, as argued in section \ref{modulispace}.
\subsection{Algebraic automorphisms of Kummer surfaces}\label{kummerautos}
To describe strata of the moduli space ${\cal M}^{K3}$
we will study subspaces of the \textsl{Kummer stratum} found above which
consist of theories with enhanced
classical symmetry groups in the geometric interpretation given there.
Concentrating on the geometric objects first, in this subsection
we investigate algebraic automorphisms of Kummer surfaces
which fix the orbifold singular metric. Such an automorphism
induces an automorphism of the Kummer lattice $\Pi$ because by
$K\cong H^2(T,\Z)(2)$ and \req{glue}
all the lattice vectors of length squared $-2$ in
$\Sigma^\perp$ belong to $\Pi$, and $\Pi\otimes\R$
by theorem \ref{kummercriterion}
is spanned by the lattice vectors $E_i,i\in I$ of length squared
$-2$. Vice versa,

\blem{kummerautpi}
The action of an algebraic automorphism $\alpha$ which fixes the orbifold
singular metric on a Kummer surface $X$ is uniquely determined by
its action on the set $\{E_i\mid i\in I\}$ of forms corresponding to
exceptional divisors,
i.e. by an affine transformation $A_\alpha\in \Aff(I)$.
\elem
\bpr
Let $\alpha^\ast$ denote the induced automorphism on the Kummer lattice
$\Pi$.
By theorem \ref{kummercriterion} and \req{glue} the intersection form
on $\Pi$ is negative definite
and the $\pm E_i, i\in I$ are the only lattice vectors of length squared $-2$.
Therefore, $\alpha^\ast$ is uniquely determined by
$\alpha^\ast(E_i) = \eps_i(\alpha) e_{A_\alpha(i)}$ for $i\in I$, where
$\eps_i(\alpha)\in\{\pm1\}$ and
$A_\alpha\in \Aff(I)$. Actually, $\eps_i(\alpha)=\eps_i(A_\alpha)$,
because $A_\alpha(i)=\id\Longrightarrow \eps_i( \alpha)=1$ for
otherwise $E_i\in (H^2(X,\Z)^{\alpha^\ast})^\perp$ with length squared $-2$
contradicting theorem \ref{algautocriterion}.
Assume $A_\alpha=A_{\alpha^\prime}$ for another algebraic automorphism
$\alpha^\prime$ fixing the metric. Then
$g:=(\alpha^{-1}\circ\alpha^\prime)^\ast$ acts trivially on $\Pi$, and because
$\Sigma$ is fixed by $g$ as well,
for the group $G$ generated by $\alpha^{-1}\circ\alpha^\prime$ we find
$\mu(G)\geq 2+3+16 = 21$. Now \req{mukabsch} shows that $G$ is trivial,
proving $\alpha=\alpha^\prime$.
\epr
By abuse of language in the following we will frequently use the
induced action of an algebraic automorphism on $\Pi$ or in $\Aff(I)$
as a shorthand for the entire action.
\btheo{generickummerautos}
For every Kummer surface $X$ the group of algebraic automorphisms
fixing the orbifold singular metric contains $\F_2^4\subset \Aff(I)$,
which acts by translations on $I$.
\etheo
\bpr
Any translation $t_i\in \Aff(I)$ by $i\in I$ acts trivially
on $\Pi^\ast /\Pi$. Thus $t_i$ can be continued trivially to
$H^2(X,\Z)$ by \req{glue}. One now easily checks that the resulting
automorphism of $H^2(X,\C)$ satisfies the criteria of theorem
\ref{algautocriterion}.
\epr
Next we will determine the group of algebraic automorphisms for
the Kummer surface associated to a torus with enhanced symmetry:
\btheo{su2h4autos}
The group of algebraic automorphisms fixing the orbifold singular
metric of $X={\cal K}(\Lambda), \Lambda\sim\Z^4$ is
${\cal G}^+_{Kummer}=\Z_2^2\ltimes\F_2^4$.
Here, $\Z_2^2\ltimes\F_2^4\subset GL(\F_2^4)\ltimes\F_2^4=\Aff(I)$
is equipped with the standard semidirect product.

For $\wt{X}={\cal K}(\wt{\Lambda})$, where $\wt{\Lambda}$ is generated
by $\Lambda_i\cong R_i\Z^2, R_i\in\R, i=1,2$,
the group of algebraic automorphisms fixing the orbifold singular
metric generically is $\wt{\cal G}^+_{Kummer}=\Z_2\ltimes\F_2^4$.
\etheo
\bpr
To demonstrate $\Z_2^2\ltimes\F_2^4\subset{\cal G}^+_{Kummer}$
we will show that certain algebraic automorphisms on the
underlying torus $T=\R^4/\Lambda$ can be pushed to $X$ and generate
an additional group of automorphisms $\Z_2^2\subset GL(\F_2^4)$
on $\Pi$. Namely, in terms of standard coordinates $(x_1,\dots,x_4)$
on $T$, we are looking for automorphisms which leave the
forms
\beq{realforms}
dx_1\wedge dx_3 + dx_4\wedge dx_2, \quad
dx_1\wedge dx_4 + dx_2\wedge dx_3, \quad
dx_1\wedge dx_2 + dx_3\wedge dx_4
\eeq
invariant. This is true for
\beqn{klein}
r_{12}: \>\> (x_1, x_2, x_3, x_4) \mapsto (-x_2, x_1, x_4, -x_3),\e
r_{13}: \>\> (x_1, x_2, x_3, x_4) \mapsto (-x_3, -x_4, x_1, x_2),\e
r_{14}=r_{12}\circ r_{13}:
\>\> (x_1, x_2, x_3, x_4) \mapsto (x_4, -x_3, x_2, -x_1).
\eeqn
The induced action on $\Pi$ is described by permutations $A_{kl}\in \Aff(I)$
of the $\F_2^4$-coordinates, namely
$r_{12}\;\wh{=}\; A_{12}=(12)(34),r_{13}\;\wh{=}\; A_{13}=(13)(24)$.
To visualize this action we introduce the following
helpful pictures first used by H. Inose \cite{in76}:
\begin{figure}[ht]
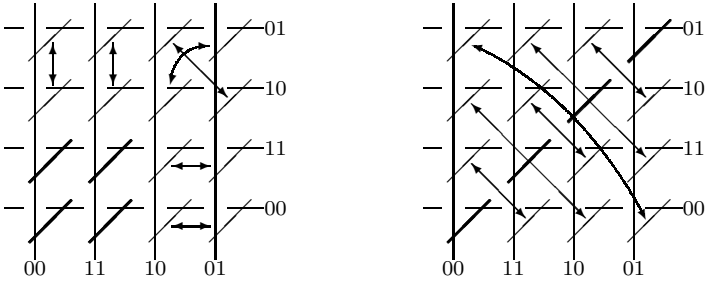

\hbox to \hsize{\hspace*{\fill}
\zzwei[4mm]{
\multiput(0.8,0.9)(0.01,-0.01){5}{\line(1,1){1.4}}
\multiput(2.8,0.9)(0.01,-0.01){5}{\line(1,1){1.4}}
\multiput(0.8,2.9)(0.01,-0.01){5}{\line(1,1){1.4}}
\multiput(2.8,2.9)(0.01,-0.01){5}{\line(1,1){1.4}}
\multiput(1.6,6.8)(2,0){2}{\vector(0,1){0.7}}
\multiput(1.6,6.8)(2,0){2}{\vector(0,-1){0.7}}
\multiput(6.2,1.4)(0,2){2}{\vector(1,0){0.65}}
\multiput(6.2,1.4)(0,2){2}{\vector(-1,0){0.65}}
\qbezier(5.5,6.4)(5.8,7.4)(6.7,7.4)
\put(5.6,6.5){\vector(-1,-4){0.1}}
\put(6.7,7.4){\vector(1,0){0.1}}
\put(6.5,6.6){\vector(1,-1){0.9}}
\put(6.5,6.6){\vector(-1,1){0.9}}
}\hspace*{\fill}
\zzwei[4mm]{
\multiput(0.8,0.9)(0.01,-0.01){5}{\line(1,1){1.4}}
\multiput(2.8,2.9)(0.01,-0.01){5}{\line(1,1){1.4}}
\multiput(4.8,4.9)(0.01,-0.01){5}{\line(1,1){1.4}}
\multiput(6.8,6.9)(0.01,-0.01){5}{\line(1,1){1.4}}
\multiput(2.5,2.6)(2,2){3}{\vector(-1,1){0.9}}
\multiput(2.5,2.6)(2,2){3}{\vector(1,-1){0.9}}
\multiput(3.5,3.6)(2,2){2}{\vector(-1,1){1.9}}
\multiput(3.5,3.6)(2,2){2}{\vector(1,-1){1.9}}
\qbezier(1.7,7.4)(5.1,5.8)(7.3,1.8)
\put(1.7,7.4){\vector(-3,1){0.1}}
\put(7.3,1.8){\vector(1,-2){0.1}}
}\hspace*{\fill}}
\caption{Action of the algebraic automorphisms
$r_{12}$ (left) and $r_{13}$ (right) on $\Pi$.}
\label{z4gen}
\end{figure}
The vertical line labelled by $j\in\F_2^{\,2}$ symbolizes
the image of the twocycle
$\{ x\in T\mid (x_1,x_2)={1\over2} j\}$ in $X$, and analogously for the
horizontal line labelled by $j^\prime\in\F_2^{\,2}$ we have
$\{ x\in T\mid (x_3,x_4)={1\over2} j^\prime\}$. Then the diagonal lines
from cycle $j$ to cycle $j^\prime$ symbolize the exceptional divisor
obtained from blowing up the fixed point labelled $(j,j^\prime)\in I$.
Fat diagonal lines mark those exceptional divisors which are fixed by
the respective automorphism.

One may now easily check that the automorphisms \req{klein},
viewed as automorphisms on $H^2(X,\C)$, satisfy the
criteria of theorem \ref{algautocriterion} and thus indeed are induced
by algebraic automorphisms of $X$.

To see that ${\cal G}^+_{Kummer}$ does not contain any further elements,
by lemma \ref{kummerautpi} it will suffice to show that no other element
of $Aut(\Pi)$ can be continued to $H^2(X,\Z)$ consistently such that
it satisfies the criteria of theorem \ref{algautocriterion}. Because all
the translations of $I$ are already contained in ${\cal G}^+_{Kummer}$
we can restrict our investigation to those elements
$A\in GL(\F_2^4)\subset \Aff(I)$
which can be continued to $H^2(X,\Z)$ preserving the symplectic forms
on $\F_2^4$ that correspond to \req{realforms}.
After some calculation one finds that $A$ must commute with all the
transformations listed in \req{klein}. This means that $A$ acts on $I$
by $A_{kl}^\prime(i)=A_{kl}(i) + |i|(1,1,1,1), |i|=\sum_k i_k\in\F_2$.
But if any such  $A_{kl}^\prime\in{\cal G}^+_{Kummer}$, then also
$A^\prime\in{\cal G}^+_{Kummer}$, where
$A^\prime(i)=i + |i|(1,1,1,1)$. $A^\prime$
leaves invariant a  sublattice  of $\Pi$ of rank 12.
But then, because of \req{mukabsch} and
from \req{invdim} 
$A^\prime$  cannot be induced by an algebraic automorphism
fixing the orbifold singular metric of $X$.

The result for $\wt{\cal G}^+_{Kummer}$ follows from the above proof.
Namely, if $(x_1,x_2)$ are standard coordinates on $\Lambda_1\otimes\R$
and $(x_3,x_4)$ on $\Lambda_2\otimes\R$, then
among the automorphisms \req{klein} only $r_{12}$ is  generically defined on
$\wt{\Lambda}$.
\epr
\subsection{$\Z_4$ Orbifolds in the moduli space}\label{z4orbifold}
This subsection is devoted to the study of $\Z_4$ orbifolds in the moduli space
${\cal M}^{K3}$.
We first turn to some features of the $\Z_4$ orbifold construction on
the conformal field theory side which need further discussion. From
what was said at the beginning of the section,
in terms of complex coordinates \req{compco} on $T=\R^4/\Lambda$
the $\Z_4$ action
on the nonlinear $\sigma$ model is given by
$(\psi_\pm^{(1)}, \psi_\pm^{(2)}) \mapsto
(\pm i\psi_\pm^{(1)}, \mp i\psi_\pm^{(2)})$.
From \req{su2h2curalg} we readily read off that there always
is a surviving
$su(2)_1\oplus u(1)$ subalgebra of the holomorphic W-algebra generated
by $J,J^\pm, A$. To have a $\Z_4$ symmetry on the entire space of states
of the torus theory, the charge lattice \req{chargelattice}
must obey this symmetry. So in addition to picking a
$\Z_4$ symmetric torus, i.e. a lattice $\Lambda$ generated by
$\Lambda_i\cong R_i\Z^2, R_i\in\R, i=1,2$,
we must have an appropriate B-field $B_T$ in
the nonlinear $\sigma$ model on $T$ which preserves this symmetry. In
terms of cohomology we need
$B_T\in H^2(T,\R)^{\Z_4}=\span_\R\left( \mu_1\wedge\mu_2, \mu_3\wedge\mu_4,
\mu_1\wedge\mu_3+\mu_4\wedge\mu_2, \mu_1\wedge\mu_4+\mu_2\wedge\mu_3 \right)$.
As in section \ref{z2orb} $(\mu_1,\dots,\mu_4)$ denotes a basis
dual to $(\lambda_1,\dots,\lambda_4)$, $\lambda_i$ being generators of
$\Lambda$ and $\Sigma_T\subset H^2(T,\R)$ is regarded as giving
the position of $H^2(T,\Z)$ relative to a fixed three--plane
$\span_\R(e_1\wedge e_2+e_3\wedge e_4,e_1\wedge e_3+e_4\wedge e_2,
e_1\wedge e_4+e_2\wedge e_3)$.

To determine the partition function, a lengthy but straightforward calculation
using \cite[(5.2)-(5.5)]{eoty89} shows
\begin{equation}\label{z4part}
\begin{array}{rcl} 
\noalign{$\ds Z_{NS}(\tau,z)\vphantom{\sum_{2}}$}
\hphantom{\Sigma}\>=\> \inv{2}\left[
\left\{\left(
\inv{2}Z_{\Lambda,B_T}(\tau)
+ \inv{2}\left| {\theta_3\theta_4\over \eta^2}\right|^4
+ \inv{2}\left| {\theta_2\theta_3\over \eta^2}\right|^4
+ \inv{2}\left| {\theta_2\theta_4\over \eta^2}\right|^4
\right)  \left| {\theta_3(z)\over \eta}\right|^4 \right\}\right. \edd
\>\> \;\left.\hphantom{Z(\tau)}
+ \left| {\theta_3\theta_4\over \eta^2}\right|^4
\left| {\theta_4(z)\over \eta}\right|^4
+ \left| {\theta_2\theta_3\over \eta^2}\right|^4
\left| {\theta_2(z)\over \eta}\right|^4
+ \left| {\theta_2\theta_4\over \eta^2}\right|^4
\left| {\theta_1(z)\over \eta}\right|^4
\right],
\end{array}
\end{equation}
where for $Z_{\Lambda,B_T}(\tau)$ one has to insert the expression
for the specific torus $T$ as obtained from \req{torpartition}.
Comparing to \req{kummerpart} the partition function \req{z4part}
coincides with that
of the $\Z_2$ orbifold of a theory
whose NS-partition function
is  the expression in curly brackets in \req{z4part}.
Indeed, the partition function of $SU(2)_1^{\,4}/\Z_4$, i.e.  of the
$\Z_4$ orbifold of $T=\R^4/\Z^4$ with $B_T=0$, agrees with that of
the $\Z_2$ orbifold ${\cal K}(D_4,0)$ \cite{eoty89}.
In section \ref{z2orb} we showed that every $\Z_2$ orbifold conformal
field theory has an $su(2)_1^{\,2}$
subalgebra of the holomorphic W-algebra.
On the other hand, as demonstrated above, the $\Z_4$ orbifold generically only
possesses an $su(2)_1\oplus u(1)$ current algebra. For $SU(2)_1^{\,4}/\Z_4$
this is enhanced to $su(2)_1\oplus u(1)^3$ which still does not
agree with the one for Kummer surfaces. Hence  although the
theories have the same partition function, they are not isomorphic.

Similarly, the partition function of the $\Z_4$ orbifold of the torus model
with $SO(8)_1$ symmetry agrees with that of ${\cal K}(\Z^4,0)$
as can be seen from \req{su2h4part}. In this
case the theories indeed are the same as will be shown in theorem
\ref{z2andz4}.

To have a better understanding of their location
within the moduli space and their geometric properties
we now construct  $\Z_4$ orbifolds by applying
another orbifold procedure to theories with enhanced symmetries
which have already been located in moduli  space.
\btheo{z4production}
Let $\wt{\Lambda}$ denote a lattice generated by
$\Lambda_i\cong R_i\Z^2, R_i\in\R, i=1,2$. Consider
the $K3$ surface $X$ obtained from the Kummer surface
${\cal K}(\wt{\Lambda})$
by modding out the algebraic automorphism
$r_{12}\in \wt{\cal G}^+_{Kummer}$, blowing up the singularities and
using the induced orbifold singular metric. Then $X$ is the $\Z_4$
orbifold of $T=\R^4/\wt{\Lambda}$.
\etheo
\bpr
By construction \req{klein}, $r_{12}$ is induced by the automorphism
$(x_1,x_2,x_3,x_4)$ $\mapsto (-x_2,x_1,x_4,-x_3)$ with respect to
standard coordinates on $T$. In terms of complex coordinates
as in \req{compco} this is just the action
$\rho: (z_1,z_2)\mapsto (iz_1, -iz_2)$, and because
${\cal K}(\Lambda)=\wt{T/\rho^2}$, the assertion is clear.
\epr
\textsl{Remark:}\hspace*{\fill}\\
Study figure \ref{z4gen} to see how the structure $A_1^6\oplus A_3^4$
of the exceptional divisors in the $\Z_4$ orbifold comes about:
Twelve of the fixed points in ${\cal K}(\wt{\Lambda})$
are identified pairwise to yield six $\Z_2$ fixed points in the
$\Z_4$ orbifold, that is $A_1^6$. The four points labelled
$i\in\{ (0,0,0,0), (1,1,0,0),$ $(0,0,1,1), (1,1,1,1) \}$ are true
$\Z_4$ fixed points. The induced action of $r_{12}$ on the corresponding
exceptional divisor $\C\Pn^1\cong\S^2$ is just a $180^\circ$ rotation
about the north-south axis, and north and south poles are fixed points.
Blow up the resulting singularities in ${\cal K}(\Lambda)/r_{12}$
to see how an $A_3$ arises from the $A_1$ over each true
$\Z_4$ fixed point.\hspace*{\fill}\\[\myskip]
For a $\Z_4$ orbifold $X$ there is an analog $\bipi$ of the Kummer
lattice $\Pi$ described in theorem \ref{kummercriterion}, the primitive
sublattice of $Pic(X)$ containing all the twoforms which correspond to
exceptional divisors by Poincar\'e duality. We will give
an analogous description of $\bipi$ as for $\Pi$ in lemma \ref{bipiinh2} below.
The embedding of the moduli
space of $\Z_4$ orbifolds in ${\cal M}^{K3}$ then works 
analogously to that of $\Z_2$ orbifold conformal field theories
as described in subsection \ref{z2orb}. 

Let us fix some notations. Let $\pi:T\rightarrow X$ denote the rational
map of degree four. Then $K:=\pi_\ast H^2(T,\Z)^{\Z_4}
=\span_\Z\left( 2\mu_1\wedge\mu_2, 2\mu_3\wedge\mu_4,
\mu_1\wedge\mu_3+\mu_4\wedge\mu_2,\right.$
$\left.\mu_1\wedge\mu_4+\mu_2\wedge\mu_3\right)$. 
For the twoforms corresponding to
the exceptional divisors of the $\Z_4$ orbifold we adopt the labelling
of fixed points by $I\cong\F_2^4$ as used in the $\Z_2$ orbifold case.
Here, we have six $\Z_2$ fixed points labelled by
$i\in I^{(2)}:=\{ (j_1,j_2,1,0), (1,0,j_3,j_4)\mid j_k\in\F_2\}$.
The four true $\Z_4$ fixed
points are labelled by $i\in I^{(4)}:=\{ (i,i,j,j)\mid i,j\in\F_2\}$.
The corresponding twoforms are denoted by $E_i$ for $i\in I^{(2)}$,
and for each $\Z_4$ fixed point $i\in I^{(4)}$ we have
three exceptional divisors Poincar\'e dual to $E_i^{(\pm)}, E_i^{(0)}$
such that
$\langle E_i^{(\pm)}, E_i^{(0)}\rangle =1,
\langle E_i^{(+)}, E_i^{(-)}\rangle =0$. For ease of notation
we also use the combination $E_i:=3E_i^{(+)}+2E_i^{(0)}+E_i^{(-)}$
if $i\in I^{(4)}$. 

As a first step we determine the analogs of \req{glue} and \req{genh2}
in order to describe the primitive embedding
$K\perp\bipi\hookrightarrow H^2(X,\Z)$. By \req{glue} images
$\kappa\in K^\ast$ of forms corresponding to torus cycles do not
necessarily correspond to cycles in $H_2(X,\Z)$.
Namely, the Poincar\'e dual of
a  representative $\kappa$ of 
$\qu\kappa\in K^\ast/K$ built from combinations of 
${1\over2}\mu_j\wedge\mu_k$
can be interpreted as the $\pi_\ast$ image
of a torus cycle which contains $\Z_4$ fixed points. It is not a cycle on $X$,
since it has boundaries where the exceptional divisors were glued in 
instead of the fixed points by the blow up procedure. 
Since the discriminant forms of $K^\ast /K$ 
and $\bipi^\ast /\bipi$ agree up
to a sign, there is a representative $\eta$ of $\qu \eta\in\bipi^\ast /\bipi$ 
in the image of $\qu\kappa$ whose Poincar\'e dual has the
same boundary as $\kappa$ but orientation reversed. We can glue
a part $\eta$ of a rational sphere 
into the boundary of $\kappa$ to obtain a cycle 
$\kappa+\eta\in H^2(X,\Z)$, where up  to a sign
$\kappa$ has the same intersection number as $\eta$  
with every exceptional divisor.

We again adopt the notation
$P_{jk}=\span_{\F_2}(f_j,f_k)$ used in subsection \ref{z2orb}. Remember to
count $\Z_2$ fixed points only once, e.g.
$P_{12}=\{ (0,0,0,0), (1,0,0,0),$ $(1,1,0,0) \}$. We then have
\blem{bipiinh2}
The lattice generated by the set $M$ which consists of
$$
\begin{array}{l}\ds
\inv{2}\mu_1\wedge\mu_2- \inv{2}E_{(0,1,0,0)+\eps(0,0,1,1)}
- \inv{4} \sum_{i\in P_{12}\cap I^{(4)}} E_{i+\eps(0,0,1,1)}
,\quad \eps\in\{0,1\}; \edd
\inv{2}\mu_3\wedge\mu_4+ \inv{2} E_{(0,0,0,1)+\eps(1,1,0,0)}
+ \inv{4} \sum_{i\in P_{34}\cap I^{(4)}} E_{i+\eps(1,1,0,0)},
\quad \eps\in\{0,1\};\edd
\inv{2}\left(\mu_1\wedge\mu_3+\mu_4\wedge\mu_2\right)
- \inv{2}\sum_{i\in P_{13}} E_{i+j},\quad j\in I^{(4)};\edd
\inv{2}\left(\mu_1\wedge\mu_4+\mu_2\wedge\mu_3\right)
- \inv{2}\sum_{i\in P_{14}} E_{i+j},\quad j\in I^{(4)};
\end{array}
$$
and by $\mathcal{E}:=
\{ E_i^{(\pm)}, E_i^{(0)}, i\in I^{(4)}; E_i, i\in I^{(2)}\}$
is isomorphic to $\Gamma^{3,19}$.
In particular, $\bipi$ is generated by $\mathcal{E}$ and
$$
\begin{array}{l}\ds
\inv{4}\left(
E_{(0,0,0,0)} + E_{(1,1,1,1)} - E_{(0,0,1,1)} - E_{(1,1,0,0)}\right) 
+\inv{2}\left( E_{(0,1,0,1)} + E_{(0,1,1,0)} \right),\edd
\inv{2}\left(E_{(0,0,0,0)} +  E_{(0,0,1,1)} 
+ E_{(0,1,0,0)} + E_{(0,1,1,1)} + E_{(0,1,0,1)} + E_{(0,1,1,0)} \right) ,\edd
\inv{2}\left(E_{(1,1,0,0)} +  E_{(0,0,1,1)} 
+ E_{(0,0,0,1)} + E_{(0,1,0,0)} + E_{(1,1,0,1)} + E_{(0,1,1,1)}\right) 
.
\end{array}
$$
This gives a natural embedding $K\perp\bipi\hookrightarrow H^2(X,\Z)$,
and
$(H^2(T,\Z))^{\Z_4}$ $\hookrightarrow H^2(X,\Z)\cong\Gamma^{3,19}$. Given a
K\a hler--Einstein metric in ${\cal T}^{3,3}$ defined by
$\Sigma_T\subset H^2(T,\R)^{\Z_4}$, its image $\Sigma$ under
the $\Z_4$ orbifold procedure is read off from 
$\Sigma_T\subset H^2(T,\R)^{\Z_4}\cong(H^2(T,\Z))^{\Z_4}\otimes\R
\hookrightarrow H^2(X,\Z)\otimes\R \cong H^2(X,\R)$.
\elem
In order to prove lemma \ref{bipiinh2} one has to show
that the lattice under inspection
has signature $(3,19)$ and is self dual. We omit the
tedious calculation. The construction will be described in
more detail in \cite{we00}.

To give the location in ${\cal M}^{K3}$
of the image of ${\cal T}(\Lambda,B_T)$ under the
$\Z_4$ orbifold we have to lift the above picture to the quantum level.
As before, $H^{even}(T,\Z)\cong \Gamma^{4,4}$ is generated by
$\mu_j\wedge\mu_k$ and $\upsilon,\upsilon^0$ defined in \req{expdecomp}.
As in \req{mroof} we extend the set $M$ of lemma \ref{bipiinh2}
to $\wh{M}:=M\cup\{ \wh{\upsilon},\wh{\upsilon}^0\}$ by
$$
\wh{\upsilon} := 2\upsilon , \quad
\wh{\upsilon}^0 :=
\inv{2}\upsilon^0 -\inv{4}\sum_{i\in I^{(2)}} E_i
-\inv{8}\sum_{i\in I^{(4)}}\left(
3E_i^{(+)}+4E_i^{(0)}+3E_i^{(-)}\right)
+ 2\upsilon .
$$
Defining
\beqn[ll]{hatforms}
\mbox{for } i\in I^{(4)}: \;
\> \wh{E}_i^{(\pm)}:= -\inv{2}\upsilon+E_i^{(\pm)},\quad
\wh{E}_i^{(0)}:= -\inv{2}\upsilon+E_i^{(0)},\e
\mbox{for } i\in I^{(2)}: 
\> \wh{E}_i:= -\upsilon+E_i
\eeqn
one now  checks in exactly the same fashion as in lemma
\ref{gamma420}
\blem{gamma420z4}
The lattice generated by $\wh{M}$ and
$\{\pi\in\span_\Z(\wh{E}_i^{(\pm)}, \wh{E}_i^{(0)}, i\in I^{(4)};
\wh{E}_i, i\in I^{(2)}) \mid
\fa m\in\wh{M}:\, \langle\pi,m\rangle\in\Z \}$
is isomorphic to $\Gamma^{4,20}$.
\elem
The embedding $H^{even}(T,\Z)^{\Z_4}\hookrightarrow H^{even}(X,\Z)$ that is now
established actually is the unique one up to lattice automorphisms
(see \cite{we00}, where also the
other $\Z_M$ orbifold conformal field theories, $M\in\{3,6\}$, will
be treated).
Now use
\beq{bz4}
B_\Z^{(4)}
:= \sum_{i\in I^{(2)}} \wh{E}_i
+ \inv{2} \sum_{i\in I^{(4)}} \left(3\wh E_i^{(+)}+4\wh E_i^{(0)}
+3\wh E_i^{(-)}\right)
\;\in H^{even}(X,\Z)
\eeq
to find
\begin{eqnarray*}
2\left(\sigma_i-\langle\sigma_i,B_T\rangle\upsilon\right)
&=& 2\sigma_i-\langle2\sigma_i,\inv{2}B_T\rangle\wh{\upsilon}\\
\inv{2}\left(\upsilon^0+B_T+\left(V-\inv[\|B_T\|^2]{2}\right)
\upsilon\right)
&=& \wh{\upsilon}^0+\inv{2}B_T+\inv{4}B_\Z^{(4)}\\
&&\quad 
+\left(\inv[V_T]{4}-\inv{2}\left\|\inv{2}B_T+\inv{4}B_\Z^{(4)}\right\|^2
\right)\wh{\upsilon},
\end{eqnarray*}
hence
\btheo{z4emb}
Let $(\Sigma_T,V_T,B_T)$ denote a geometric interpretation of the nonlinear
$\sigma$ model ${\cal T}(\Lambda,B_T)$ as given by \req{decomptor}.
Assume that $\Lambda$ is generated by $\Lambda_i\cong R_i\Z^2, R_i\in\R,
i=1,2$, and $B_T\in H^2(T,\Z)^{\Z_4}$ such that a $\Z_4$ action is
well defined on ${\cal T}(\Lambda,B_T)$. Then the image
$x\in{\cal T}^{4,20}$ under the
$\Z_4$ orbifold procedure has geometric interpretation
$(\Sigma,V,B)$ where $\Sigma\in{\cal T}^{3,19}$ is found as described in
lemma \ref{bipiinh2},  $V={V_T\over4}$,
and $B={1\over2}B_T+{1\over4}B_\Z^{(4)},
B_\Z^{(4)}\in H^{even}(X,\Z)$ as in \req{bz4}.

In particular, the moduli space of superconformal field theories
admitting an interpetation as
$\Z_4$ orbifold is a quaternionic submanifold of ${\cal M}^{K3}$.
Moreover, $x^\perp\cap H^{even}(X,\Z)$ does not contain
vectors of length squared $-2$.
\etheo
Note that from
\req{bz4} it is easy to read off the flow of the B-field
obtained from the orbifold procedure through an
$A_3$ divisor over one of the true $\Z_4$ fixed points of $X$:
On integration over any of the divisors that correspond to a
$\Z_m$ fixed point, we get B--field flux ${1\over m}$. This is also true 
for the other
$\Z_M$ orbifold conformal field theories 
and confirms earlier results \cite{do97,blin97}
obtained in the context of brane physics.

Theorem \ref{z4emb} proves that $\Z_4$ orbifold conformal field theories
do not correspond to string compactifications of the type IIA string on
$K3$ with enhanced gauge symmetry.
Concerning the algebraic automorphism group of $\Z_4$ orbifolds we can prove
\btheo{z4autos}
Let $X$ denote the $\Z_4$ orbifold of $T=\R^4/\Lambda$.
Then the group ${\cal G}$ of algebraic automorphisms
fixing the orbifold singular
metric of $X$  consists of all the residual symmetries induced
by algebraic automorphisms of ${\cal K}(\Lambda)$ which commute with
$r_{12}$.
Thus, generically ${\cal G}\cong\F_2^2$ is generated by the induced actions of
$t_{1100}$ and $t_{0011}$. If $\Lambda\sim\Z^4$,
${\cal G}\cong D_4$ is generated by the induced actions of $t_{1100}$
and $r_{13}$.
\etheo
If we want invariance
of the conformal field theory under the entire group ${\cal G}\cong D_4$
of algebraic automorphisms found in theorem \ref{z4autos} we must restrict
$B_T$ to values such that $B_T\in$
$H^2(T,\R)^{\Z_4}\cap H^2(X,\R)^{D_4} = \Sigma$
where we regard $H^2(T,\R)^{\Z_4}\hookrightarrow H^2(X,\R)$ as described
in lemma \ref{bipiinh2}.
If $B_T$ is viewed as element of $Skew(4)$ acting on $\R^4$ this condition
is equivalent to $B_T$ commuting with the automorphisms listed in
\req{klein}.
\subsection{Application: Fermat's description for $SU(2)_1^{\,4}/\Z_4$}
\label{application}
\hspace*{\fill}\\[-1em]
\btheo{quartic}
The $\Z_4$ orbifold of ${\cal T}(\Z^4,0)$ admits a geometric interpretation
on the Fermat quartic
\beq{kuqua}
\vphantom{\sum^3}
{\cal Q} =
\left\{(x_0,x_1,x_2,x_3)\in \C\Pn^3 \left|
\smash{\sum_{i=0}^3} x_i^4 = 0\right.
\right\}
\eeq
in $\C\Pn^3$ with volume $V_{\mathcal Q}={1\over2}$ and B--field 
$B_{\mathcal Q}=-{1\over2}\sigma_1^{(\mathcal Q)}$ up to a shift in 
$H^2(X,\Z)$, where $\sigma_1^{(\mathcal Q)}$ denotes the K\a hler
class of $\mathcal Q$.
\etheo
\bpr
Let $e_1,\dots,e_4$ denote the standard basis of $\Z^4$. Then
$\mu_i=e_i$, and by theorem \ref{z4emb} with $\|B_{\Z}^{(4)}\|^2=-32$
the $\Z_4$ orbifold of ${\cal T}(\Z^4,0)$
is described by the four--plane $x\in{\cal T}^{4,20}$ spanned by
$$
\begin{array}{rclrcl}\ds
\xi_1\>=\>\mu_1\wedge\mu_3 + \mu_4\wedge\mu_2,\quad\>
\xi_2\>=\>\mu_1\wedge\mu_4 + \mu_2\wedge\mu_3,\e
\xi_3\>=\>2(\mu_1\wedge\mu_2 + \mu_3\wedge\mu_4),\quad\>
\xi_4\>=\>4\wh{\upsilon}^0+B_\Z^{(4)} + 5\wh{\upsilon}.
\end{array}
$$
To read off a different geometric interpretation, we define
\beqn{nullvectors}
\upsilon_{\mathcal Q} \>:=\>
\inv{2} \left(\mu_1\wedge\mu_3 + \mu_4\wedge\mu_2
-\mu_1\wedge\mu_4 - \mu_2\wedge\mu_3\right)\e
\>\>\quad
+ \inv{2}\left( \wh{E}_{(0,1,1,0)}-\wh{E}_{(1,0,1,0)}\right) ,\e
\upsilon_{\mathcal Q}^0 \>:=\>
\mu_1\wedge\mu_3 + \mu_4\wedge\mu_2 +\mu_1\wedge\mu_2\e
\>\>\quad
+ \inv{2}\left( \wh{E}_{(0,0,0,1)}+\wh{E}_{(1,1,0,1)}
-\wh{E}_{(0,1,1,0)}-\wh{E}_{(1,0,1,0)}\right).
\eeqn
One  checks $\upsilon_{\mathcal Q},\upsilon_{\mathcal Q}^0\in H^{even}(X,\Z)$
as given in lemma \ref{gamma420z4},
$\|\upsilon_{\mathcal Q}\|^2=\|\upsilon_{\mathcal Q}^0\|^2=0$
and $\langle\upsilon_{\mathcal Q},\upsilon_{\mathcal Q}^0\rangle = 1$ to show that
$\upsilon_{\mathcal Q},\upsilon_{\mathcal Q}^0$ is an admissible choice for nullvectors in
\req{expdecomp}. For the corresponding geometric interpretation
$(\Sigma_{\mathcal Q},V_{\mathcal Q},B_{\mathcal Q})$ we find that $\Sigma_{\mathcal Q}$  is spanned by
\begin{eqnarray*}
\sigma_1^{({\mathcal Q})} &=& \mu_1\wedge\mu_3 + \mu_4\wedge\mu_2
+ \mu_1\wedge\mu_4 + \mu_2\wedge\mu_3 -2 \upsilon_{\mathcal Q}, \\
\sigma_2^{({\mathcal Q})} &=& 2(\mu_1\wedge\mu_2 + \mu_3\wedge\mu_4)-2 \upsilon_{\mathcal Q}, \\
\sigma_3^{({\mathcal Q})} &=& 4 \wh{\upsilon}^0+ B_\Z^{(4)} + 5\wh{\upsilon}.
\end{eqnarray*}
As complex structure $\Omega_{\mathcal Q}\subset\Sigma_{\mathcal Q}$ we pick the two--plane spanned
by $\sigma_2^{({\mathcal Q})}$ and $\sigma_3^{({\mathcal Q})}$. Note that this plane is generated
by lattice vectors, so the \textsl{Picard number}
$\rho(X):=\rk Pic(X) = \rk(\Omega^\perp\cap H^2(X,\Z))$
of the corresponding geometric interpretation $X$ is 20,
the maximal possible value. $K3$ surfaces with Picard number 20  are called
\textsl{singular} and  are classified by the quadratic form on their
\textsl{transcendental lattice}
$Pic(X)^\perp\cap H^2(X,\Z)$. In other words there is a
one to one correspondence between singular $K3$ surfaces and even
quadratic positive definite forms modulo $SL(2,\Z)$ equivalence
\cite{shin77}\footnote{
We thank Noriko Yui and Yasuhiro Goto for drawing our attention to the
relevant literature concerning singular $K3$ surfaces.}.
Because $\sigma_2^{({\mathcal Q})},\sigma_3^{({\mathcal Q})}$ are primitive lattice vectors,
one now easily checks that $X$ equipped with the complex structure
given by $\Omega_{\mathcal Q}$ has quadratic form $\diag(8,8)$ on the transcendental
lattice. By \cite{in76} this means that our variety indeed is the
Fermat quartic \req{kuqua} in $\C\Pn^3$.

Volume and B-field can now be read off using \req{expdecomp} and noting
that in our geometric interpretation
$$
\mu_1\wedge\mu_3 + \mu_4\wedge\mu_2-\mu_1\wedge\mu_4 - \mu_2\wedge\mu_3
=
\xi_4^{({\mathcal Q})} \sim
{\upsilon}^0_{\mathcal Q}+B_{\mathcal Q}
+\left(V_{\mathcal Q}-\inv{2}\left\|B_{\mathcal Q}\right\|^2
\right){\upsilon}_{\mathcal Q}.
$$
\epr

\section{Special points in  moduli space:
Gepner and Gepner type models}
\label{gepinmodsp}
Finally we discuss the probably best understood models of 
superconformal field theories associated to $K3$ surfaces, namely 
Gepner models \cite{ge87,ge88}.
The latter are rational conformal field theories and thus exactly solvable. 
For a short account on the Gepner construction 
and its most important features in the context of our investigations
see appendix \ref{gepner}. 
In this section, we explicitly locate the Gepner model
$(2)^4$ and some of its orbifolds within the moduli space ${\cal M}^{K3}$.
This is achieved by giving $\sigma$ model descriptions of these models
in terms of $\Z_2$ and $\Z_4$ orbifolds which we know how to locate in
moduli space by the results of section \ref{orbifolds}.
\subsection{Discrete symmetries of Gepner models and algebraic automorphisms
of $K3$ surfaces}
\label{gepsym}
As argued before, a basic tool to characterize a given
conformal field theory is the study of its discrete symmetry group.
We will first discuss the abelian group given by phase symmetries
of a Gepner model 
$\prod_{j=1}^r (k_j)$ with central charge $c=6$ and $r$ even \cite{ge87}.
Recall that this theory is obtained from the fermionic tensor product of the 
$N=2$ superconformal minimal
models $(k_j), j=1,\dots,r$, by modding out a cyclic group
${\cal Z} \cong \Z_n$, $n=\lcm\{2; k_i+2, i=1,\dots, r\}$.
The model therefore inherits a $\Z_{k_j+2}$ symmetry
from the parafermionic subtheories of 
each minimal model factor $(k_j)$ whose generator in the bosonic sector
acts by
\beq{phasesymmetries}
\Phi^{l_j}_{m_j,s_j;\qu{m}_j,\qu{s}_j}
\longmapsto 
e^{{2\pi i \over 2(k_j+2)}(m_j+\qu{m}_j)} \; 
\Phi^{l_j}_{m_j,s_j;\qu{m}_j,\qu{s}_j}
\eeq
on the $j$th factor.
The resulting abelian symmetry group of $\prod_{j=1}^r (k_j)$ is
$\Z_2\times{\cal G}_{ab}$, where $\Z_2$ denotes charge conjugation and
${\cal G}_{ab} = ( \prod_{j=1}^r \Z_{k_j+2} )/ \Z_m, 
m=\lcm\{k_i+2, i=1,\dots, r\}$. 
Here,  $\Z_m$ acts by
$$
\prod_{j=1}^r \Z_{k_j+2} \longrightarrow
\prod_{j=1}^r \Z_{k_j+2} , \quad
[a_1,\dots, a_r]\longmapsto [a_1+1,\dots, a_r+1] 
$$
(see also \cite{grpl90}). 
Note that only elements of the subgroup
\beq{gepcondalg}
{\cal G}_{ab}^{alg} :=
\left\{ [a_1,\dots, a_r]\in{\cal G}_{ab} \left|\;
\smash{\sum_{j=1}^r} {a_j\over k_j+2} \right. \in \Z \right\}
\subset {\cal G}_{ab}
\eeq 
commute with spacetime
supersymmetry, elements of ${\cal G}_{ab}-{\cal G}_{ab}^{alg}$ 
describe R-symmetries \cite{ge87}.

Assume we can locate our Gepner model within ${\cal M}^{K3}$,
that is we explicitly know the corresponding
four--plane $x\subset H^{even}(X,\R)$ as 
described in section \ref{modulispace}. Furthermore assume that by
picking a primitive
nullvector $\upsilon\in H^{even}(X,\Z)$ we have chosen a specific
geometric interpretation $(\Sigma, V, B)$.
By construction, a Gepner model comes with a specific choice of the
$N=(2,2)$ subalgebra corresponding to  a specific twoplane
$\Omega\subset\Sigma$.
We stress that this is true for any geometric interpretation of 
$\prod_j(k_j)$: The
choice of the $N=(2,2)$ subalgebra does \textsl{not} fix a complex structure
\textsl{a priori}, it fixes a choice of complex structure \textsl{in every
geometric interpretation} of our model, as was explained in section
\ref{modulispace}. Still,
we now assume our $K3$ surface $X$ to be equipped with complex 
structure and K\a hler metric.  By our discussion in section \ref{k3mod}
we know that 
any symmetry of the Gepner model which leaves the 
$su(2)_l^{susy}\oplus su(2)_r^{susy}$ currents $J,J^\pm, \qu{J}, \qu{J}^\pm$ 
and the vector $\upsilon$ invariant may act as an algebraic automorphism
on $X$. Because $J^\pm=\left(\Phi^0_{\mp 2, 2;0,0}\right)^{\otimes r}$ 
and $\qu{J}^\pm=\left(\Phi^0_{0,0;\mp 2, 2}\right)^{\otimes r}$
(see appendix \ref{gepner}) we conclude from \req{gepcondalg} that
elements of  ${\cal G}_{ab}^{alg}$
can act as algebraic automorphisms on $X$ fixing the B-field
$B\in H^2(X,\R)$, and vice versa. More explicitly by what was said in 
section \ref{k3mod},
the action of such a Gepner-symmetry on the 
$\left({1\over 2},{1\over 2}\right)$-fields 
with charges, say, $Q=\qu{Q}=1$ should be identified with the induced action
of an algebraic automorphism of $X$ on $H^{1,1}(X,\R)$. With 
reference to its possible geometric 
interpretation we call ${\cal G}_{ab}^{alg}$ the
\textsl{abelian algebraic symmetry group of the Gepner model}.

In the following subsections we will investigate where in the moduli space 
of superconformal field theories associated to $K3$ surfaces
to locate the Gepner
model $(2)^4$ and some of its orbifolds by elements of 
${\cal G}_{ab}^{alg} \cong (\Z_4)^2$.
From the above discussion it is clear that given a definite 
geometric interpretation for $(2)^4$ 
the geometric interpretation of its orbifold models is obtained by modding 
out the corresponding algebraic automorphisms.

Apart from symmetries in $\Z_2\times{\cal G}_{ab}$ our Gepner model
will possess permutation symmetries involving identical factor theories.
Their discussion is a bit more subtle, because 
as noted in \cite{fks92} permuting
fermionic fields will involve additional signs
\req{gepferm}. This in particular applies to 
$J^\pm=\left(\Phi^0_{\mp 2, 2;0,0}\right)^{\otimes r}$, meaning that odd 
permutations can only act algebraically when accompanied by a phase symmetry
\beq{phasecond}
[a_1,\dots, a_r]\in{\cal G}_{ab}:\quad
\sum_{j=1}^r {a_j\over k_j+2}  \in \Z+\inv{2}  .
\eeq
We will discuss this phenomenon in detail for the example of prime 
interest to us, namely the Gepner model $(2)^4$. Here 
${\cal G}_{ab}^{alg} \cong (\Z_4)^2$,
and the entire algebraic symmetry group is
generally believed to be 
${\cal G}^{alg} \cong (\Z_4)^2\rtimes {\cal S}_4$ \cite{as95}. 
Moreover, based on Landau-Ginzburg computations and
comparison of symmetries \cite{gvw89,grpl90,fkss90,as95}
it is generally believed that $(2)^4$ has a geometric
interpretation $(\Sigma_{\cal Q}, V_{\cal Q}, B_{\cal Q})$ 
given by the Fermat quartic \req{kuqua} in $\C\Pn^3$.
Indeed, ${\cal Q}$ is a $K3$ surface  with
algebraic automorphism group $(\Z_4)^2\rtimes {\cal S}_4$ \cite{mu88},
and arguments in favour of the viewpoint that it yields a geometric
interpretation of $(2)^4$  
will arise from the following discussion.
It is proved in corollary \ref{gepquart}.

To give the action of the two
generators $[1,3,0,0]$ and $[1,0,3,0]$ of 
${\cal G}_{ab}^{alg} \cong (\Z_4)^2$ on the 
$\left({1\over 2},{1\over 2}\right)$-fields with charges $Q=\qu{Q}=1$
we use the shorthand notation
\beqn{shorthand}
X\>:=\>(\Phi^1_{1,0;-3,2})^{\otimes 4} ,\e
Y(n_1,n_2,n_3,n_4) 
\>:=\>
\Phi^{n_1}_{n_1,0;n_1,0} \otimes\Phi^{n_2}_{n_2,0;n_2,0} \otimes
\Phi^{n_3}_{n_3,0;n_3,0} \otimes\Phi^{n_4}_{n_4,0;n_4,0} 
\eeqn
($n_i\in\N$) and find
\begin{equation}\label{z4on22the4}
\begin{array}{|c||c|c|c|c|}
\hline
\vphantom{\int^{C^\infty}}\ds [1,3,0,0]\rightarrow & 1  & -1 &  i & -i \\[3pt]
\vphantom{\int_\infty}\ds\downarrow [1,0,3,0] &&&&\\
\hline\hline
\hoch
1 
& \hoch Y(1,1,1,1),\; X
& \hoch Y(0,2,0,2),&&\\
&&Y(2,0,2,0)
&  Y(1,0,1,2) 
&  Y(1,2,1,0)\\
\hline
-1
& \hoch Y(2,2,0,0),
& \hoch Y(2,0,0,2) , &&\\
&Y(0,0,2,2)
&Y(0,2,2,0)
& Y(2,1,0,1)  
&  Y(0,1,2,1)   \\
\hline
i
& \hoch Y(1,1,0,2)
& \hoch  Y(2,0,1,1) 
& \hoch Y(2,1,1,0)
& \hoch Y(1,2,0,1)\\
\hline
-i
& \hoch Y(1,1,2,0)
& \hoch  Y(0,2,1,1) 
& \hoch Y(1,0,2,1) 
& \hoch Y(0,1,1,2) \\
\hline
\end{array}
\end{equation}
Note first that by 
\req{mukainum} we have $\mu(\Z_4\times\Z_4)=6$, in 
accordance with \req{invdim} and $2=6-4$ invariant fields in the above table.
One moreover easily checks that the spectrum of every element
$g\in{\cal G}_{ab}^{alg}$ of order four agrees with the one computed
in  \req{fourspec} for algebraic automorphisms of order four on 
$K3$ surfaces. This is a strong and highly non-trivial evidence for 
the fact that one possible geometric
interpretation of $(2)^4$ is given by a $K3$ surface
whose algebraic automorphism group contains $(\Z_4)^2$.

As stated above, further discussion is due 
concerning the action of ${\cal S}_4$ because transpositions
of fermionic modes introduce sign flips \req{gepferm}. 
In particular, odd elements
of ${\cal S}_4$ do not leave $J^\pm$ invariant. To have an algebraic
action of the entire group ${\cal S}_4$ we must therefore accompany 
$\sigma\in{\cal S}_4$ by a phase symmetry 
$a_\sigma = [a_1(\sigma), a_2(\sigma),a_3(\sigma),a_4(\sigma)]
\in{\cal G}_{ab}$ which for odd $\sigma$ satisfies \req{phasecond}.
Thus a transposition $(\alpha,\omega)\in{\cal S}_4$ must be represented by 
$\rho( (\alpha,\omega) )=(\alpha,\omega)\circ a_{(\alpha,\omega)} 
= a_{(\alpha,\omega)}\circ (\alpha,\omega)$ in order to have
$\rho( (\alpha,\omega) )^2 = \id$. With any such choice of $\rho$ on generators
$(\alpha_j,\omega_j)$ of ${\cal S}_4$ one may then check explicitly that $\rho$
defines an algebraic action of ${\cal S}_4$, i.e. its spectrum on the 
$\left( {1\over 2}, {1\over 2}\right)$-fields  coincides with the spectrum
of the algebraic automorphism group ${\cal S}_4$. Namely,
any element of order two (or three, four) in ${\cal S}_4$ 
leaves $\mu(\Z_2)-4=12$ (or $\mu(\Z_3)-4=8, \mu(\Z_4)-4=6$) states invariant,
and elements of order four have the spectrum given in \req{fourspec}.
Note in particular that by \req{phasecond} with any consistent choice
of $\sigma\mapsto a_\sigma$ the group ${\cal S}_4$ acts by 
$\sigma\mapsto\sign(\sigma)$ on $Y(1,1,1,1)$ and trivially on $X$. This 
leaves $X=(\Phi^1_{1,0;3,2})^{\otimes 4}$ as the unique invariant 
state upon the action of $(\Z_4)^2\rtimes {\cal S}_4$ in accordance with
$\mu((\Z_4)^2\rtimes {\cal S}_4) = 5$ and \req{invdim}.

Summarizing, we have shown that the action of the entire algebraic 
symmetry group 
${\cal G}^{alg} = (\Z_4)^2\rtimes {\cal S}_4$ of $(2)^4$ 
as described above exhibits a spectrum
consistent with its interpretation as group of algebraic automorphisms
of a $K3$ surface, e.g. the Fermat quartic with geometric interpretation
$(\Sigma_{\cal Q},V_{\cal Q},B_{\cal Q})$. Remember that 
$\mu\left({\cal G}^{alg}\right)=5$ is the minimal possible value of 
$\mu$ by the discussion in section \ref{k3mod}.
Thus by what was said in section \ref{orbifolds} the only four
invariant $({1\over 2},{1\over 2})$-fields 
$(\Phi^1_{\pm 1,0;\pm 3,2})^{\otimes 4}$, 
$(\Phi^1_{\pm 1,0;\pm 1,0})^{\otimes 4}$
are those corresponding to moduli
of volume deformation and of B-field deformation in direction of 
$\Sigma_{\cal Q}$.
\subsection{Ideas of proof: An example with $c=3$}\label{protoproof}
In this subsection  we
give a survey on the steps of proof we will perform to show equivalences 
between Gepner or 
Gepner type models and nonlinear $\sigma$ models. As an illustration
we then prove the \textsl{well known} fact that Gepner's model $(2)^2$ admits a
nonlinear $\sigma$ model description on the torus associated to the
$\Z^2$ lattice.

Given two $N=2$ superconformal field theories ${\cal C}^1, {\cal C}^2$
with central charge $c=3d/2$ ($d=2$ or $d=4$) 
and  spaces of states ${\cal H}^1, {\cal H}^2$, to prove their equivalence
we show the following:
\begin{enumerate}
\item
The partition functions of the two theories agree sector by sector
in the sense of \req{sectors}.
\item
The fields of dimensions $(h,\qu{h})=(1,0)$ 
in the two theories generate the same
algebra ${\cal A}={\cal A}_f\oplus{\cal A}_b$, where
${\cal A}_f=u(1)$ for $d=2$, ${\cal A}_f=su(2)_1^{\,2}$ for $d=4$, 
and $u(1)^{d}\subset {\cal A}_b$. 
In particular, $u(1)^c\subset{\cal A}$. ${\cal A}_f$ 
contains the $U(1)$-current $J^{(1)}=J$ of the $N=2$ superconformal 
algebra, and a second $U(1)$-generator $J^{(2)}$ if $d=4$. 
Furthermore, the fields of dimensions $(h,\qu{h})=(0,1)$ 
in both theories generate  algebras isomorphic to ${\cal A}$
as well, such that each of the left moving $U(1)$-currents $j$ has a
right moving partner $\qu{\jmath}$.
\item
For $i=1,2$ define
$$
{\cal H}^i_b := \left\{ |\phi\rangle\in{\cal H}^i
\,\left| \, J^{(k)} |\phi\rangle =   0,\quad
k\in\{1,\inv[d]{2}\} \right. \right\}
$$
and denote the $U(1)$-currents in 
$u(1)^{d}\subset{\cal A}_b$ by $j^1,\dots,j^{d}$. We normalize them to
\beq{norm}
j^k(z)\;j^l(w) \sim {\delta_{kl}\over (z-w)^2}.
\eeq
Let $j^{d+k}\sim J^{(k)}, k\in\{1,{d\over 2}\}$ denote the 
remaining $U(1)$-currents when normalized to \req{norm}, too, and set
${\cal J}:=(j^1,\dots,j^d;\qu{\jmath}^1,\dots,\qu{\jmath}^{d})$.
The charge lattices
$$
\Gamma_b^i := \left\{ \gamma\in\R^{d;d} \left| \,\exists\;
|\phi\rangle\in {\cal H}^i_b: {\cal J}|\phi\rangle = \gamma|\phi\rangle
\right.\right\}
$$ 
of ${\cal H}^1_b$ and ${\cal H}^2_b$ with respect to 
${\cal J}$ 
are isomorphic to the same self dual lattice $\Gamma_b\subset\R^{d;d}$; 
because the states in ${\cal H}^i_b$ are
pairwise local, in order to prove this it suffices to show agreement of the 
${\cal J}$-action on a set of states whose
charge vectors generate a self dual lattice $\Gamma_b$.
\end{enumerate}
\btheo{idproof}
If i.-iii. are true then  theories ${\cal C}^1$ and ${\cal C}^2$
are isomorphic (the converse generically is wrong, of course).
\etheo
\bpr
Using i.-iii. we first show 
${\cal H}^1_b\cong {\cal H}^2_b=: {\cal H}_b$. 
Denote by $V^i[\gamma]$ the primary field corresponding to a
state in ${\cal H}^i_b$ with charge 
$\gamma=(\gamma_l;\gamma_r)\in\Gamma_b$.
Notice that in both theories every charge
$\gamma\in\Gamma_b$  must appear with multiplicity one, because
otherwise by fusing
$[V^i_k[\gamma]]\times[V^i_k[-\gamma]]=[\id^i_k]$ we find two
states $\id^i_1,\id^i_2\in {\cal H}^i_b$ with vanishing charges
under a total $u(1)^c\subset{\cal A}$ in contradiction to  uniqueness 
of the vacuum. Now for any 
$\alpha=(\alpha_l;\alpha_r),\beta=(\beta_l;\beta_r)\in\Gamma_b$ we have
$$
V^i[\alpha](z)\;V^i[\beta](w)
\sim 
c^i_{\alpha,\beta}(z-w)^{\alpha_l\beta_l}(\qu{z}-\qu{w})^{\alpha_r\beta_r}
\; V^i[\alpha+\beta](w) + \cdots,
$$
so it remains to be shown that we can arrange 
$c^1_{\alpha,\beta}=c^2_{\alpha,\beta}$ for all $\alpha,\beta\in\Gamma_b$
by normalizing the primary fields appropriately. 
In other words, we must find
constants $d_\gamma\in\R$ for any $\gamma\in\Gamma_b$ 
such that $\fa\alpha,\beta\in\Gamma_b: 
c^2_{\alpha,\beta} = d_\alpha d_\beta c^1_{\alpha,\beta}$.
This is possible, because having fixed
$d_\alpha, d_\beta, d_\gamma, d_\delta\in\R$ such that 
$$
c^2_{\alpha,\beta} = d_\alpha d_\beta c^1_{\alpha,\beta},\;
c^2_{\alpha,\gamma} = d_\alpha d_\gamma c^1_{\alpha,\gamma},\;
c^2_{\alpha,\delta} = d_\alpha d_\delta c^1_{\alpha,\delta},\;
c^2_{\beta,\gamma} = d_\beta d_\gamma c^1_{\beta,\gamma}
$$
for four nonzero twopoint functions 
$c^i_{\alpha,\beta},c^i_{\alpha,\gamma},c^i_{\alpha,\delta},
c^i_{\beta, \delta}$, by the crossing symmetries
$$
{c^1_{\alpha,\beta} c^1_{\gamma, \delta}\over 
c^1_{\alpha,\gamma} c^1_{\beta, \delta} }
= {c^2_{\alpha,\beta} c^2_{\gamma, \delta}\over 
c^2_{\alpha,\gamma} c^2_{\beta, \delta} }
\quad\mbox{ and }\quad
{c^1_{\alpha,\gamma} c^1_{\beta, \delta}\over 
c^1_{\alpha,\delta} c^1_{\beta, \gamma} }
= {c^2_{\alpha,\gamma} c^2_{\beta, \delta}\over 
c^2_{\alpha,\delta} c^2_{\beta, \gamma} }
$$
etc. we automatically have 
$c^2_{\gamma,\delta}= d_\gamma d_\delta c^1_{\gamma,\delta}$
and
$c^2_{\beta,\delta} = d_\beta d_\delta c^1_{\beta,\delta}$.
If more than two of the six twopoint functions vanish, then by
similar arguments  the normalization of one
of the primaries is independent of the three others
and a consistent choice of 
$d_\alpha, d_\beta, d_\gamma, d_\delta\in\R$ is therefore possible, too.
The proof of ${\cal H}^1_b\cong{\cal H}^2_b\cong{\cal H}_b$ is now complete.

Because $\Gamma_b$ is self dual, for
any state $|\phi\rangle \in {\cal H}^i$ carrying 
charge $\gamma$ with respect to
${\cal J}$ we have $\gamma\in \Gamma_b$ and thus find vertex operators
$V^i[\pm\gamma]\in {\cal H}_b^i$. By ii. and iii. 
$T:={1\over 2}\sum_{k=1}^{c} (j^k)^2$ acts as Virasoro field $T^i$ on each
of the theories  (check that $T-T^i$ has dimensions $h=\qu{h}=0$ 
with respect to $T^i$). Thus the
restriction of the Virasoro field $T^i$ to  ${\cal H}^i_b$
is given by $T_b^i:={1\over 2}\sum_{k=1}^{d} (j^k)^2$, and  by
picking suitable combinations $P$ of descendants $j^k_{-n}$
and $\wt{P}$ of ascendants  $j^k_{n}, n\geq 0, k\in\{1,\dots,d\}$,
we find $|\psi\rangle:=P\,V^i[-\gamma]|\phi\rangle$ such that
$$
|\phi\rangle=|\psi\rangle \otimes  V^i[\gamma]\,\wt{P}\,|0\rangle_b
\quad\mbox{ and }\quad
|\psi\rangle\in{\cal H}_f^i := 
\left\{\, |\chi\rangle\in{\cal H}^i \,\mid\, T_b^i |\chi\rangle = 0 \right\}.
$$
This shows ${\cal H}^i\cong {\cal H}_f^i\otimes {\cal H}_b$ for $i=1,2$.
${\cal H}_f^1$ and ${\cal H}_f^2$ are  
representations of ${\cal A}_f=u(1)$ (for $d=2$) or 
${\cal A}_f=su(2)_1^{\,2}$ (for $d=4$) which are completely determined
by charge and dimension of the lowest weight states. 
Because by ii. ${\cal A}_f$
contains the $U(1)$-current $J$ of the total $N=2$ superconformal algebra,
the partition functions of our theories agree by i., and we already know
${\cal H}^i\cong {\cal H}_f^i\otimes {\cal H}_b$ for $i=1,2$,
we may conclude
${\cal H}_f^1\cong {\cal H}_f^2$.
\epr
Let's see how the procedure described above works:
\btheo{2hoch2id}
Gepner's model ${\cal C}^1=(2)^2$ has a nonlinear $\sigma$ model description
${\cal C}^2$ on the two dimensional torus  $T_{SU(2)_1^{\,2}}$
with $SU(2)_1^{\,2}$ lattice $\Lambda=\Z^2$ and B-field
$B=0$.
\etheo
\bpr
If we can prove i.-iii. in the above list, by theorem \ref{idproof} 
we are done. 
\begin{enumerate}
\item
Using \req{geppartfctn} for computing the partition function of $(2)^2$ on 
one hand and \req{torpartition} for
the partition function of the $\sigma$ model on $T_{SU(2)_1^{\,2}}$  
with $B=0$ on the other, we find
$$
Z_{NS}(\tau, z)
={1\over 2}\left[ \left| {\theta_2\over \eta}\right|^4
+\left| {\theta_3\over \eta}\right|^4 + \left| {\theta_4\over \eta}\right|^4
\right] \left| {\theta_3(z)\over \eta}\right|^2
$$
for both theories.
\item
The nonlinear $\sigma$ model on $T_{SU(2)_1^{\,2}}$ has two 
rightmoving abelian currents $j_1,j_2$ which we normalize to 
$$
j_\alpha(z)\;j_\beta(w) \sim 
{{1\over 2}\delta_{\alpha\beta}\over (z-w)^2 }.
$$
Their superpartners are free Majorana fermions 
$\psi_1,\psi_2$ with coupled boundary conditions.
By $e_1,e_2$ we denote the generators of the lattice 
$\Lambda=\Lambda^\ast=\Z^2$ which defines our torus.
Then 
the $(1,0)$-fields in the nonlinear $\sigma$ model are given
by the three abelian currents $J=i\psi_2\psi_1$ (the $U(1)$ current of 
the $N=2$ superconformal algebra), $Q=j_1+j_2, R=j_1-j_2$, 
and the four vertex operators $V_{\pm e_i,\pm e_i}, i=1,2$.

In the Gepner model $(2)^2$ we have an abelian current $j,j^\prime$
from each minimal model factor along with Majorana fermions
$\psi,\psi^\prime$, where by \req{statesin2}
$\psi\psi^\prime = \Phi^0_{4,2;0,0}\otimes  \Phi^0_{4,2;0,0}$.
The $U(1)$ current of 
the total $N=2$ superconformal algebra is $J=j+j^\prime$, and
comparing $J,Q,R$-charges we can make the following identifications:
\begin{eqnarray*}
i\psi_2\psi_1 \>=\> J \;=\; j+j^\prime, \quad
j_1+j_2 \;=\; Q \;=\; j-j^\prime, \quad
j_1-j_2 \;=\; R \;=\; i\psi\psi^\prime, \e
V_{e_1,e_1} \>=\> \Phi^0_{2,0;0,0}\otimes \Phi^0_{2,2;0,0}
+\Phi^0_{-2,0;0,0}\otimes \Phi^0_{-2,2;0,0}, \e
V_{e_2,e_2} \>=\> \Phi^0_{2,0;0,0}\otimes \Phi^0_{2,2;0,0}
-\Phi^0_{-2,0;0,0}\otimes \Phi^0_{-2,2;0,0}, \e
V_{-e_1,-e_1} \>=\> \Phi^0_{2,2;0,0}\otimes \Phi^0_{2,0;0,0}
+\Phi^0_{-2,2;0,0}\otimes \Phi^0_{-2,0;0,0}, \e
V_{-e_2,-e_2} \>=\> \Phi^0_{2,2;0,0}\otimes \Phi^0_{2,0;0,0}
-\Phi^0_{-2,2;0,0}\otimes \Phi^0_{-2,0;0,0}.
\end{eqnarray*}
Thus the $(1,0)$-fields in the two theories generate the same
algebra ${\cal A}=u(1)\oplus su(2)_1^{\,2}={\cal A}_f\oplus {\cal A}_b$.
Obviously, the same structure arises on the right handed
sides.
\item
The space ${\cal H}^1_b$ for the $\sigma$ model is just the bosonic part of
the theory. The charge lattice $\Gamma_b$ with respect to the currents
${\cal J}:=(Q,R;\qu{Q},\qu{R})
=(j_1+j_2,j_1-j_2;\qu{\jmath}_1+\qu{\jmath}_2,\qu{\jmath}_1-\qu{\jmath}_2)$ 
thus contains the charges 
$M:=\left\{
{1\over 2}(\eps;\pm\eps), \eps\in\{\pm 1\}^2 \right\}$,
carried by vertex operators $V_{\pm e_i,0},V_{0,\pm e_i}, i=1,2$.
$M$ generates the self dual lattice
$\left\{ {1\over 2}(a+b;a-b)\mid\right.$ $ a,b\in\Z^2, 
\sum_{k=1}^2 a_k\equiv\sum_{k=1}^2 b_k$ 
$\left.\vphantom{{1\over 2}}\equiv 0\,(2)\right\}=\Gamma_b$.  

To complete the proof of iii. we  observe that 
in the Gepner model the fields 
$\Phi^1_{n,0;n,0}\otimes \Phi^1_{-n,0;-n,0}\pm
\Phi^1_{-3n,2;n,0}\otimes \Phi^1_{3n,2;-n,0}$ and
$\Phi^1_{n,0;-n,0}\otimes \Phi^1_{3n,2;n,0}$\linebreak
$\pm\Phi^1_{-3n,2;-n,0}\otimes \Phi^1_{-n,0;n,0}, n\in\{\pm 1\}$,
are uncharged  with respect to $J$ and carry
${\cal J}
=(j-j^\prime,i\psi\psi^\prime;
\qu{\jmath}-\qu{\jmath}^\prime,i\qu{\psi}\,\qu{\psi}^\prime)$-charges
$M=\left\{{1\over 2}(\eps;\pm\eps), \eps\in\{\pm 1\}^2 \right\}$
generating $\Gamma_b$.
\vspace{-1.5em} 
\end{enumerate}
\epr
\subsection{Gepner type description of $SU(2)_1^{\,4}/\Z_2$}
\hspace*{\fill}\\[-1em]
\btheo{su2hoch4id}
Let ${\cal C}^1=(\wh{2})^4$ denote the Gepner type model
which is obtained as orbifold of $(2)^4$ by the group
$\Z_2\cong\langle [2,2,0,0]\rangle\subset {\cal G}_{ab}^{alg}$.
Then ${\cal C}^1$ admits a nonlinear $\sigma$ model description 
${\cal C}^2={\cal K}(\Z^4,0)$
on the Kummer surface $\mathcal K(\Lambda)$
associated to the torus $T_{SU(2)_1^{\,4}}$
with $SU(2)_1^{\,4}$ lattice $\Lambda=\Z^4$ and vanishing B-field.
\etheo
\bpr
We prove conditions i.-iii. of section \ref{protoproof} and then
use theorem \ref{idproof}.
\begin{enumerate}
\item
From \req{torpartition}
one finds
\beq{su2h4part}
Z_{\Lambda=\Z^4, B_T=0}(\tau) =
\left[ {1\over 2} \left( 
\left| {\theta_2\over \eta}\right|^4 + \left| {\theta_3\over \eta}\right|^4
+ \left| {\theta_4\over \eta}\right|^4 \right) \right]^2 .
\eeq
Applying the orbifold procedure for the $\Z_2$-action of 
$[2,2,0,0]\in {\cal G}_{ab}^{alg}$ to the partition function
\req{geppartfctn} of the Gepner model $(2)^4$ \cite{fkss90}
one checks that  ${\cal C}^1$
and ${\cal C}^2$ have the same partition function 
obtained by inserting \req{su2h4part} into \req{kummerpart}.
\item
In the nonlinear $\sigma$ model ${\cal C}^2$
the current algebra \req{su2h2curalg} is enhanced to
$u(1)^4\oplus su(2)_1^{\,2}$.
The additional $U(1)$-currents are
$U_i := V_{e_i,e_i}+V_{-e_i,-e_i}, i=1,\dots,4$, where the $e_i$
are the standard generators of $\Lambda=\Lambda^\ast=\Z^4$.

In the Gepner type model ${\cal C}^1=(\wh{2})^4$, apart from the
$U(1)$-currents $J_1,\dots,J_4$ from the factor theories, where
$J=J_1+\cdots+J_4$, we find four additional fields with dimensions
$(h,\qu{h})=(1,0)$; comparing the respective operator product expansions
the following identifications can be made:
\beqn{su2h4idofcur}
J \>=\> J_1+J_2+J_3+J_4,\quad
J^\pm \;=\; \left(\Phi^0_{\mp2,2;0,0}\right)^{\otimes 4};\e
A \>=\>  J_1+J_2-J_3-J_4,\quad
A^\pm \;=\; \left(\Phi^0_{\mp 2,2;0,0}\right)^{\otimes 2}
\otimes\left(\Phi^0_{\pm 2,2;0,0}\right)^{\otimes 2};\\[4pt]
\hline\\[-4pt]\ds
\inv{2} \left( U_1+U_2 \right) \>=\>   P \;=\; J_1-J_2;\e
\inv{2} \left( U_3+U_4 \right) \>=\>   Q \;=\; J_3-J_4;\e
\inv{2} \left( U_1-U_2 \right) 
\>=\>  R \;=\; i\left(\Phi^0_{4,2;0,0}\right)^{\otimes 2}
\otimes\left(\Phi^0_{0,0;0,0}\right)^{\otimes 2};\e
\inv{2} \left( U_3-U_4 \right) 
\>=\> S \;=\; i\left(\Phi^0_{0,0;0,0}\right)^{\otimes 2}
\otimes\left(\Phi^0_{4,2;0,0}\right)^{\otimes 2}.
\eeqn
Thus the $(1,0)$-fields in the two theories generate the same
algebra ${\cal A}=su(2)_1^{\,2}\oplus u(1)^4={\cal A}_f\oplus {\cal A}_b$.
Obviously, the same structure arises on the right handed
sides.
\item
We  show that ${\cal H}^1_b$ and ${\cal H}^2_b$ both have 
self dual ${\cal J}:=(P,Q,R,S;\qu{P},\qu{Q},\qu{R},\qu{S})$-charge 
lattice\footnote{In our coordinates
$D_4=\{x\in\Z^4\mid\sum_{i=1}^4x_i\equiv 0 \,(2)\}$
and $D_4^\ast=\Z^4+(\Z+1/2)^4$.}
\beq{su2h4lattice}
\Gamma_b = \left\{ (x+y;x-y)\left| x\in \inv{2} D_4, y\in D_4^\ast
\right.\right\},
\eeq
generated by 
\begin{eqnarray*}
M_{tw} &:=& \left\{\left.  \inv{2}(x;x)\in\R^{4,4} \right|
x\in\{ (\eps_1,\eps_2,0,0), (0,0,\eps_1,\eps_2),
\right. \\
&& \hphantom{\inv{2}(x;x)\in\R^{4,4}|xx} 
\left. \vphantom{\inv{2}}
(0,\eps_1,\eps_2,0),(\eps_1,0,0,\eps_2),
\eps_i\in\{\pm 1\}\} \right\}
\\
\mbox{ and }M_{inv} &:=& \left\{ (\eps;0) \left| \eps\in\{\pm 1\}^4 \right. \right\} .
\end{eqnarray*}
In the $\sigma$ model ${\cal C}^2$ we denote by 
$\Sigma_\delta,\delta\in\F_2^4$ the
twist field corresponding to the fixed point 
$p_\delta={1\over 2}\sum_{i=1}^4\delta_i e_i$ of the $\Z_2$ orbifold.
To determine the action of $U_i$ on  twist fields notice that 
by definition, $\Sigma_\delta$ introduces a cut on 
the configuration space $Z$ to establish
the boundary condition $\phi(\sigma_0+1,\sigma_1)=-\phi(\sigma_0,\sigma_1)$ 
for  fields $\phi$ in the corresponding twisted sector, 
i.e. $\phi(0,0)=p_\delta$
(see section \ref{orbifolds}). 
Action of a vertex operator with winding mode $\lambda$
will shift the constant mode $p_\delta$ of each twisted field 
by ${\lambda\over2}$ \cite{hava87}. Hence,
\beq{uaction}
U_i(z) \; \Sigma_\delta(w) \sim {1/2\over z-w}\, \Sigma_{\delta+e_i}(w),
\eeq
where the factor ${1\over 2}$ is determined up to phases by observing
$T_f^2|\Sigma_\delta\rangle =0,
T_b^2={1\over 4}\sum_{i=1}^4\left( U_i\right)^2$, 
and $h=\qu{h}={1\over 4}$ for twist fields. 
The phases are fixed by appropriately normalizing the twist fields.
One now  checks that 
$$
\fa\eps\in\{\pm 1\}^4:\quad
s_\eps := \sum_{\delta\in\F_2^4} 
 \prod_{i=1}^4(\eps_i)^{\delta_i} \; \Sigma_\delta
$$
are uncharged under $(J;\qu{J})$ and $(A;\qu{A})$ and carry 
${\cal J}$-charges  $M_{tw}$.
For $\eps,\delta\in\{\pm 1\}$ and $k,l\in\{1,\dots,4\}$ 
we define 
$$
E_{kl}^{\eps\delta}
\;:=\;
\left( j_k - \inv[\delta]{2}\left(V_{e_k,e_k}-V_{-e_k,-e_k}\right)\right)
\left( j_l - \inv[\eps]{2}\left(V_{e_l,e_l}-V_{-e_l,-e_l}\right)\right).
$$
Then $E^{\eps\delta}_{13},E^{\eps\delta}_{14},
E^{\eps\delta}_{23},E^{\eps\delta}_{24}$ are $(J,A;\qu{J},\qu{A})$-uncharged 
and carry ${\cal J}$-charges  $M_{inv}$.

In the Gepner model, introducing 
${\cal O}(n_1):=$ 
$\left(\Phi^1_{2,1;2n_1,n_1}\right)^{\otimes 2},$
${\cal P}(n_2) :=$\linebreak
$\Phi^0_{n_2,n_2;n_2,n_2}\otimes\Phi^0_{-n_2,-n_2;-n_2,-n_2}$
($n_i\in\{\pm 1\}$) as shorthand notation
we  find $(J,A;\qu{J},\qu{A})$-uncharged fields
$
{\cal O}(n_1) \otimes{\cal O}(n_2) ,\;
{\cal O}(n_1) \otimes{\cal P}(n_2) ,\;
{\cal P}(n_1) \otimes{\cal O}(n_2) ,$
${\cal P}(n_1) \otimes{\cal P}(n_2) $
which after diagonalization with respect to the ${\cal J}$-action
carry charges $M_{tw}$.

Similarly, setting
${\cal Q}(n,s) := \Phi^0_{2n,s;0,0}\otimes\Phi^0_{2n,s+2;0,0}$,
the fields ${\cal Q}(n_1,s_1)\otimes{\cal Q}(n_2,s_2), n_i\in\{\pm 1\},
s_i\in\{0,2\}$ after diagonalization have charges $M_{inv}$.
\end{enumerate}
For later reference we note that by what was said in section
\ref{modulispace} 
there are eight more fields in the Ramond sector
with dimensions $h=\qu{h}={1\over 4}$. Each of them  is 
uncharged under ${\cal J}$
and either $(A;\qu{A})$ or $(J;\qu{J})$. We denote by
$W^J_{\eps_1,\eps_2},W^A_{\eps_1,\eps_2} ,\eps_i\in\{\pm 1\}$ 
the fields corresponding to the lowest weight
states of $su(2)_1\cong\langle J,J^\pm\rangle$ or 
$su(2)_1\cong\langle A,A^\pm\rangle$, 
with $(J;\qu{J})$ or $(A;\qu{A})$-charge
$(\eps_1;\eps_2)$ respectively and identify
\beqn{torusforms}
W^J_{\eps_1,\eps_2} \>=\>
\left(\Phi^0_{-\eps_1,-\eps_1;-\eps_2,-\eps_2}\right)^{\otimes 4}\e
W^A_{\eps_1,\eps_2} \>=\>
\left(\Phi^0_{-\eps_1,-\eps_1;-\eps_2,-\eps_2}\right)^{\otimes 2}
\otimes\left(\Phi^0_{\eps_1,\eps_1;\eps_2,\eps_2}\right)^{\otimes 2}.
\eeqn
In $\sigma$ model language and by the discussion in section \ref{modulispace}, 
by applying 
left and right  handed spectral flow to the $J$-uncharged  
$W^A_{\eps_1,\eps_2}$ we obtain $({1\over 2},{1\over 2})$-fields
in ${\cal F}_{1/2}$, the real and imaginary parts of 
whose $(1,1)$-superpartners
describe infinitesimal deformations of the torus 
$T_{SU(2)_1^{\,4}}$ our Kummer surface is 
associated to.

Summarizing, we can now obtain a list of all fields needed to generate
${\cal H}^1$ and ${\cal H}^2$ as well as a complete field by field 
identification by comparison of charges; 
for the resulting list of 
$({1\over 4},{1\over 4})$-fields see appendix \ref{fieldlist}.
\epr
Note that because $D_4\cong\sqrt2 D_4^\ast$  for the
${\cal J}$-charge lattice \req{su2h4lattice}
$$
\Gamma_b \cong \left\{\left. \inv{\sqrt2} (\mu+\lambda,\mu-\lambda) \right|
\mu\in D_4^\ast, \lambda\in D_4 \right\}.
$$
Thus $\Gamma_b$ is the charge lattice of the bosonic part of the 
$\sigma$ model ${\cal C}^3={\cal T}(D_4,0)$. Theory
${\cal C}^1$ was obtained by taking the ordinary $\Z_2$ orbifold of the torus
model on $T_{SU(2)_1^{\,4}}$, but as pointed out
in \cite{kosa88}, for the bosonic part of the theory this is equivalent
to taking the $\Z_2$ orbifold associated to a shift 
$\delta={1\over2\sqrt2}(\mu_0;\mu_0), 
\mu_0=\sum_i e_i\in\Lambda^\ast$ on the charge lattice of $T_{SU(2)_1^{\,4}}$.
Under this \textsl{shift orbifold}, the lattices 
$\Lambda=\Lambda^\ast=\Z^4$ are transformed by
$$
\Lambda^\ast\mapsto\Lambda^\ast + \left(\Lambda^\ast + \inv{2}\mu_0\right)
= D_4^\ast,\quad
\Lambda\mapsto\left\{ \lambda\in\Lambda\left| 
\langle\mu_0,\lambda\rangle \equiv 0\, (2) \right.\right\} = D_4,
$$
so the bosonic part of the resulting theory indeed is that of ${\cal C}^3$.
The entire bosonic sector of ${\cal C}^1={\cal C}^2$  agrees
with that of ${\cal C}^3$, 
because the shift acts trivially on  fermions,
and the ordinary $\Z_2$ orbifold just interchanges twisted and untwisted
boundary conditions of the fermions in time direction.
The difference between the theories merely amounts in opposite assignments
of Ramond and Neveu-Schwarz sector on the twisted states
resulting in different elliptic genera
for the $K3$-model ${\cal C}^1={\cal C}^2$ and  the 
torus model ${\cal C}^3$. 
The fact that the partition functions actually do not agree before 
projection onto even fermion numbers is not relevant here because 
locality is violated before the projection is carried out.
So, on the level of conformal field theory:
\brem{meetingpoint}
The Gepner type model ${\cal C}^1=(\wh{2})^4$ viewed as nonlinear 
$\sigma$ model
${\cal C}^2$ on the Kummer surface ${\cal K}(\Z^4,0)$ is located
at a meeting point of the moduli
spaces of theories associated to $K3$ surfaces and tori, respectively.
Namely, its bosonic sector is identical with that of the nonlinear
$\sigma$ model ${\cal C}^3={\cal T}(D_4,0)$.
\erem
This property does not translate to the stringy interpretation
of our conformal field theories, though. When we take external degrees 
of freedom into account, spin statistics theorem dictates in which
representations of $SO(4)$ the external free fields may couple
to internal Neveu-Schwarz or Ramond fields, respectively. The theories
${\cal C}^1={\cal C}^2$ and ${\cal C}^3$ therefore correspond to 
different compactifications of the type IIA string.
\subsection{Gepner's description for $SU(2)_1^{\,4}/\Z_4$}\label{gepsu2h4}
\hspace*{\fill}\\[-1em]
\btheo{su2h4mod4id}
The Gepner model ${\cal C}^I=(2)^4$ admits a nonlinear $\sigma$ model
description ${\cal C}^{II}$ on the $\Z_4$ orbifold of the torus
$T_{SU(2)_1^{\,4}}$ with $SU(2)_1^{\,4}$-lattice $\Lambda=\Z^4$
and vanishing B-field.
\etheo
\bpr
It is clear that 
${\cal C}^I=(2)^4$ can be obtained from 
${\cal C}^1=(\wh{2})^4$, for which we already have a $\sigma$ model 
description by theorem \ref{su2hoch4id}, 
by the $\Z_2$ orbifold procedure which revokes the orbifold
used to construct ${\cal C}^1$. 
The corresponding action is multiplication by $-1$ on 
$\langle[2,2,0,0]\rangle$-twisted states, i.e.
\beq{gpbackwards}
\left[ 2^\prime, 2^\prime, 0,0 \right]: \quad
\bigotimes_{i=1}^4\Phi^{l_i}_{m_i,s_i;\qu{m}_i,\qu{s}_i}
\longmapsto 
e^{{2\pi i\over8}(\qu{m}_1-m_1-\qu{m}_3+m_3)}
\bigotimes_{i=1}^4\Phi^{l_i}_{m_i,s_i;\qu{m}_i,\qu{s}_i}.
\eeq
Among the $(1,0)$-fields the following are invariant under 
$[ 2^\prime, 2^\prime, 0,0]$
(use \req{su2h2curalg} and \req{su2h4idofcur}):
\beqn[rclrclrcl]{surv10}
J \>=\>
\psi_+^{(1)}\psi_-^{(1)}+\psi_+^{(2)}\psi_-^{(2)},
\> J^+ \>=\>  \psi_+^{(1)}\psi_+^{(2)}, \>
J^- \>=\> \psi_-^{(2)}\psi_-^{(1)}; \e
A \>=\>
 \psi_+^{(1)}\psi_-^{(1)}-\psi_+^{(2)}\psi_-^{(2)};
\> P\>=\> \inv{2} \left( U_1+U_2 \right) ;
\> Q\>=\> \inv{2} \left( U_3+U_4 \right).
\eeqn
Hence we have a surviving $su(2)_1\oplus u(1)^3$ subalgebra of our 
holomorphic W-algebra.
In appendix \ref{fieldlist} we give a
list of all $({1\over 4},{1\over 4})$-fields in
${\cal C}^1=(\wh{2})^4$ together with their  description in the $\sigma$ model
${\cal C}^2$ on
the $\Z_2$ orbifold ${\cal K}(\Z^4,0)$. A similar list can
be obtained for the $(2,0)$-fields as discussed in the proof of theorem
\ref{su2hoch4id}. From these lists and \req{surv10} one  readily 
reads off that the states invariant under \req{gpbackwards} coincide with
those invariant under the automorphism $r_{12}$ on ${\cal K}(\Z^4,0)$
(see theorem \ref{su2h4autos})
which is induced by the $\Z_4$ action 
$(j_1,j_2,j_3,j_4)\mapsto (-j_2,j_1,j_4,-j_3)$, i.e.
$(\psi_\pm^{(1)},\psi_\pm^{(2)})\mapsto 
(\pm i\psi_\pm^{(1)},\mp i\psi_\pm^{(2)})$ on the underlying torus
$T_{SU(2)_1^{\,4}}$. The appertaining permutation of
exceptional divisors in the
$\Z_2$ fixed points is depicted in figure \ref{z4gen}.
The action of $r_{12}$ and that induced by
\req{gpbackwards} agree on the algebra ${\cal A}$ of $(1,0)$-fields 
and a set of states generating the entire space of states, thus they 
are the same. Because of ${\cal C}^1={\cal C}^2$ (theorem \ref{su2hoch4id})
and the fact that ${\cal C}^I=(2)^4$ is obtained from ${\cal C}^1$ by
modding out \req{gpbackwards}, it is clear that
modding out ${\cal K}(\Z^4,0)$
by the algebraic automorphism $r_{12}$ 
will lead to a $\sigma$ model description of $(2)^4$.
As shown in theorem \ref{z4production} the result is the $\Z_4$ orbifold 
${\cal C}^{II}$ of $T_{SU(2)_1^{\,4}}$. 
\epr
Theorem \ref{su2h4mod4id} has been conjectured in  \cite{eoty89}
because of  agreement of the partition functions
of ${\cal C}^I$ and ${\cal C}^{II}$. This of course is
only part of the proof as can be seen from our argumentation
in section \ref{z4orbifold}. There we showed that 
$SU(2)_1^{\,4}/\Z_4$ does not admit a $\sigma$ model description on a 
Kummer surface although its 
partition function by \cite{eoty89}
agrees with that of ${\cal K}(D_4,0)$, too.

From theorem \ref{quartic} and theorem \ref{su2h4mod4id} we  conclude:
\bcor{gepquart}
The Gepner model $(2)^4$ admits a geometric interpretation on the Fermat
quartic \req{kuqua} in $\C\Pn^3$ with volume $V_{\mathcal Q}={1\over2}$.
\ecor
Let $(\Sigma, V,B)$
denote the geometric interpretation of $(2)^4$ we gain from theorem 
\ref{su2h4mod4id}.
By the proof of theorem \ref{su2hoch4id} we know the moduli 
$V_{\delta,\eps}^\pm+V_{-\delta,-\eps}^\pm$ and  
$i(V_{\delta,\eps}^\pm-V_{-\delta,-\eps}^\pm), \delta,\eps\in\{\pm 1\}$
for volume and B-field deformation in direction of $\Sigma$
of the underlying torus 
$T_{SU(2)_1^{\,4}}$ of our $\Z_4$ orbifold:
We apply left and right handed spectral flows to 
$W^A_{1,1},W^A_{-1,-1}$ 
as given in \req{torusforms} and then compute the corresponding 
$(1,1)$-superpartners. In terms of Gepner fields this means
\beqn{volmod}
V_{\delta,\eps}^+\>=\>
\Phi^2_{2\delta,2;2\eps,2}\otimes\Phi^2_{2\delta,0;2\eps,0}\otimes
\left(\Phi^0_{0,0;0,0}\right)^{\otimes 2} \e
\>\>\quad
+\Phi^2_{2\delta,0;2\eps,0}\otimes\Phi^2_{2\delta,2;2\eps,2}\otimes
\left(\Phi^0_{0,0;0,0}\right)^{\otimes 2}\;, \e
V_{\delta,\eps}^-\>=\>
\left(\Phi^0_{0,0;0,0}\right)^{\otimes 2} \otimes
\Phi^2_{2\delta,2;2\eps,2}\otimes\Phi^2_{2\delta,0;2\eps,0}\e
\>\>\quad
+ \left(\Phi^0_{0,0;0,0}\right)^{\otimes 2}\otimes
\Phi^2_{2\delta,0;2\eps,0}\otimes\Phi^2_{2\delta,2;2\eps,2}\;.
\eeqn
Indeed,  $V_{\delta,\eps}^\pm$ are uncharged under 
$J$ and $A$ as they should, because
both $U(1)$-currents must survive  deformations within
the moduli space  of $\Z_4$ orbifold conformal field theories.
On the other hand by our discussion in section \ref{gepsym}
the $(1,1)$-superpartners of
$(\Phi^1_{\pm1,0;\pm3,2})^{\otimes 4}$,
$(\Phi^1_{\pm1,0;\pm1,0})^{\otimes 4}$,
which carry $(A;\qu{A})$-charges $\mp (1;1)$, give the
moduli of volume and corresponding B-field 
deformation if we choose the quartic hypersurface 
\req{kuqua}
as geometric interpretation of Gepner's model $(2)^4$. 
Hence along the ``quartic line'' we generically only have an 
$su(2)_1$-algebra of $(1,0)$-fields.
This agrees with the analogous picture for $c=9$ and the Gepner model 
$(3)^5$ where all additional $U(1)$-currents vanish upon deformation
along the quintic line
\cite{digr88}.
\subsubsection*{Symmetries and algebraic automorphisms revised: 
$(2)^4$ and $(\wh{2})^4$}
Among the algebraic symmetries $\Z_4^2\rtimes {\cal S}_4$
of the Gepner model $(2)^4$ all the phase symmetries
$\Z_4^2$ commute with the action of $[2,2,0,0]$ which we mod out
to obtain $(\wh{2})^4$. The residual $\Z_2\times\Z_4$ 
has a straightforward continuation to $(\wh{2})^4$ (i.e. to 
the twisted states). 
Moreover, $[2^\prime, 2^\prime,0,0]$ as given in 
\req{gpbackwards} which reverts the orbifold with respect to $[2,2,0,0]$
must belong to the algebraic symmetry group
$\wh{\cal G}^{alg}$ of $(\wh{2})^4$.
Nevertheless, one notices that
$\Z_2\times\Z_2\cong \langle [2^\prime, 2^\prime,0,0], [1,3,0,0] \rangle$
leaves $6\neq 8 = \mu(\Z_2\times\Z_2)-4$ states invariant and thus
does not act algebraically by \req{invdim}. We temporarilly leave the symmetry
$[1,3,0,0]$ out of  discussion, 
because then by the methods described
in section \ref{gepsym} we find a consistent algebraic action of
$(\Z_2\times\Z_4)\rtimes D_4$ on $(\wh{2})^4$, where
$\Z_2\times\Z_4=\langle[2^\prime, 2^\prime,0,0], [1,0,3,0]\rangle$
and  $D_4=\langle (12),(13)(24)\rangle \subset{\cal S}_4$ 
is the commutant of $[2,2,0,0]$. 

Let us compare to the $\sigma$ model description ${\cal K}(\Z^4,0)$
of $(\wh{2})^4$: 
In theorem  \ref{su2h4autos} the group of algebraic
automorphisms of ${\cal K}(\Z^4,0)$ 
which leave the orbifold singular metric invariant was determined to 
${\cal G}_{Kummer}^+=\Z_2^2\ltimes\F_2^4$. Although it is isomorphic
to the algebraic symmetry group $(\Z_2\times\Z_4)\rtimes D_4$ 
of $(\wh{2})^4$ found so far, 
${\cal G}_{Kummer}^+$
must act differently on $(\wh{2})^4$.
Namely, from the proof of theorem
\ref{su2h4mod4id} we know that the $\sigma$ model equivalent of 
$[2^\prime, 2^\prime,0,0]$ is $r_{12}\in{\cal G}_{Kummer}^+$. 
Thus only the commutant
${\cal H}\subset{\cal G}_{Kummer}^+$ of 
$r_{12}$ can comprise residual symmetries descending from 
the $\Z_4$ orbifold description on $(2)^4$. This is no contradiction, because
by the discussion in section \ref{k3mod} different subgroups of the
entire algebraic symmetry group of $(\wh{2})^4$ may leave
the respective nullvector $\upsilon$ invariant which
defines the geometric interpretation.
By what was said in section \ref{modulispace} it is actually no surprise
to find symmetries of conformal field theories which do not descend to 
classical symmetries of a given geometric interpretation. The Gepner type
model $(\wh{2})^4$ is an example where the existence of such symmetries
can be checked explicitly.

By the results of section \ref{kummerautos} we find 
${\cal H}=\Z_2\times D_4=\langle r_{12}, r_{13},t_{1100}\rangle$
(see also theorem \ref{z4autos}). 
We now use our state by state identification  obtained in the proof of
theorem \ref{su2hoch4id} (see appendix \ref{fieldlist}) to determine
the corresponding elements of $\wh{\cal G}^{alg}$ and find
\beqn[rclcl]{symmid}
r_{13} \>=\> (13)(24) \>\in\> {\cal S}_4 \e
t_{1100} \>=\> \xi\circ [1,3,0,0] \>=:\> [1^\prime,3^\prime,0,0].
\eeqn
Here $\xi$ acts by multiplication with $-1$ on those Gepner states
corresponding to the $16$ twist fields $\Sigma_\delta$ of the Kummer surface
and trivially on all the other generating fields of the space of states
we discussed in the proof of theorem \ref{su2hoch4id}. 
Note that $\xi$ is a symmetry of the theory because by the selection rules for
amplitudes of twist fields any $n$-point function containing an odd number
of twist fields will vanish. The geometric interpretation tells us
that modding out $(\wh{2})^4$ by $\xi$ will revoke the ordinary $\Z_2$ orbifold
procedure i.e. produce ${\cal T}(\Z^4,0)$.
We conclude remarking that
by the modification \req{symmid} of the $[1,3,0,0]$-action
the full group $\wh{\cal G}^{alg}=(\Z_2^2\times\Z_4)\rtimes D_4$
acts algebraically on $(\wh{2})^4$. The
subgroup ${\cal H}$ consists of all the
residual symmetries of $(2)^4$ surviving both deformations along the
quartic and the $\Z_4$ orbifold line and acting classically in 
both geometric interpretations of $(2)^4$ known so far,
the $\Z_4$ orbifold and the quartic one. 
\subsection{Gepner type description of $SO(8)_1/\Z_2$}\label{gepso81}
\hspace*{\fill}\\[-1.5em]
\btheo{so8id}
Let $\wt{\cal C}^1=(\wt{2})^4$ denote the Gepner type model
which is obtained as orbifold of $(2)^4$ by the group
$\Z_2\times\Z_2\cong\langle [2,2,0,0],[2,0,2,0]\rangle
\subset {\cal G}_{ab}^{alg}$.
This model admits a nonlinear $\sigma$ model description $\wt{\cal C}^2$
on the Kummer surface ${\cal K}({1\over\sqrt2}D_4, B^\ast)$
associated to the torus $T_{SO(8)_1}$
with $SO(8)_1$-lattice $\Lambda=\inv{\sqrt2}D_4$ and B-field 
value $B^\ast$ for which
the theory has enhanced symmetry by the Frenkel-Kac mechanism.
\etheo
\bpr
Let $e_1,\dots, e_4$ denote the standard basis of $\Z^4$.
With respect to this basis
the B-field which leads to a full $SO(8)_1$ symmetry for the 
$\sigma$ model on $T_{SO(8)_1}$ is
\beq{fkbfield}
B^\ast =
\left( \begin{array}{c|c} \begin{array}{cc} 0&1\\-1&0 \end{array} & 0 \\
\hline 0 & \begin{array}{cc} 0&1\\-1&0 \end{array} \end{array} \right)
: \Lambda\otimes\R \longrightarrow \Lambda^\ast \otimes \R\; ,
\eeq
a twotorsion point in $H^2(T_{SO(8)_1},\R)/H^2(T_{SO(8)_1},\Z)$.

We are now ready to use theorem \ref{idproof} if we can prove i.-iii.
of section \ref{protoproof}.
\begin{enumerate}
\item
From \req{torpartition}
we find
\beq{so81part}
Z_{{1\over\sqrt2}D_4, B^\ast}(\tau) =
{1\over 2} \left( 
\left| {\theta_2\over \eta}\right|^8 + \left| {\theta_3\over \eta}\right|^8
+ \left| {\theta_4\over \eta}\right|^8 \right).
\eeq
Applying the orbifold procedure for the $\Z_2\times\Z_2$ action of 
$\langle[2,2,0,0], [2,0,2,0]\rangle$ $\subset {\cal G}_{ab}^{alg}$ 
to the partition function
\req{geppartfctn} of the Gepner model $(2)^4$ \cite{fkss90}
one checks that  $\wt{\cal C}^1$
and $\wt{\cal C}^2$ have the same partition function 
obtained by inserting \req{so81part} into \req{kummerpart}.
\item
We have an enhancement of
the current algebra \req{su2h2curalg} of the nonlinear $\sigma$ model 
$\wt{\cal C}^2$ to $su(2)_1^{\,6}$.
The $12$ additional $(1,0)$-fields are 
$U_\alpha:=\inv{\sqrt2}\left( V_{\alpha,\alpha+B^\ast\alpha} \right.$
$\left. + V_{-\alpha,-\alpha-B^\ast\alpha} \right)$,
where $\alpha$ belongs to the $D_4$ rootsystem 
$\{ \pm\inv{\sqrt2}e_i\pm\inv{\sqrt2}e_j \}$. We set 
$$
W_{i,j}^\pm := \inv{2}\left( 
U_{{1\over\sqrt2}(e_i+e_j)}\pm U_{{1\over\sqrt2}(e_i-e_j)} \right)
$$
to see that upon a consistent choice of cocycle factors for the 
vertex operators
these fields indeed comprise an extra $su(2)_1^{\,4}$:
\beqn[rclrcl]{extrasu2}
P \>:=\> W_{1,4}^+ + W_{2,3}^+,
\> P^\pm \>:=\> \inv{\sqrt2}\left(W_{1,2}^+ + W_{3,4}^+\right) 
\pm\inv{\sqrt2}\left( W_{2,4}^+ +W_{1,3}^+\right),\e
Q \>:=\> W_{1,2}^+ - W_{3,4}^+,
\> Q^\pm \>:=\> \inv{\sqrt2}\left(W_{1,3}^+ - W_{2,4}^+\right) 
\pm\inv{\sqrt2}\left( W_{1,4}^+ -W_{2,3}^+\right),\e
R \>:=\> iW_{2,4}^- - iW_{1,3}^-,
\> R^\pm \>:=\> \inv{\sqrt2}\left(W_{1,4}^- - W_{2,3}^-\right) 
\pm\inv{\sqrt2}\left( W_{1,2}^- -W_{3,4}^-\right),\e
S \>:=\> W_{1,4}^- + W_{2,3}^-,
\> S^\pm \>:=\>\inv{\sqrt2}\left(W_{1,2}^- + W_{3,4}^-\right) 
\pm\inv{\sqrt2}\left( W_{2,4}^- +W_{1,3}^-\right) .
\eeqn
For the Gepner type model $\wt{\cal C}^2=(\wt{2})^4$ we use $X_{ij}$
as a shorthand notation  for the field having factors 
$\Phi^0_{4,2;0,0}$ in the $i$th and $j$th position and factors 
$\Phi^0_{0,0;0,0}$ otherwise, and $Y_{ij}$ for the field having factors
$\Phi^0_{-2,2;0,0}$ in the $i$th and $j$th position and factors 
$\Phi^0_{2,2;0,0}$ otherwise.
By comparison of operator product expansions one then checks that the following
identifications can be made:
$$
\begin{array}{rclrcl}
J \>=\> J_1+J_2+J_3+J_4,\>
J^\pm \>=\> \left(\Phi^0_{\mp2,2;0,0}\right)^{\otimes 4};\e
A \>=\>  J_1+J_2-J_3-J_4,\>
A^+ \>=\> Y_{12}, \;\; A^- \;=\; Y_{34};\\[4pt]
\hline\\[-4pt] \ds
P\>=\>\inv{\sqrt2} \left( J_1-J_2+J_3 -J_4\right),
\> P^+ \>=\> Y_{13}, \;\; P^- \;=\; Y_{24};\e
Q\>=\>\inv{\sqrt2} \left( J_1-J_2-J_3 +J_4\right),
\> Q^+ \>=\> Y_{14}, \;\; Q^- \;=\; Y_{23};\e
R\>=\> \inv[i]{\sqrt2}\left(X_{13}- X_{24}\right),
\> R^\pm \>=\>
\mp\inv{2}\left(X_{12}+X_{34}\right) 
+ \inv[i]{2}\left( X_{14}+X_{23} \right);\e
S\>=\> \inv[i]{\sqrt2}\left(X_{13}+ X_{24}\right),
\> S^\pm \>=\>
\pm\inv{2}\left(X_{12}-X_{34}\right) + \inv[i]{2}\left( X_{14}-X_{23} \right).
\end{array}
$$
Thus the $(1,0)$-fields in the two theories generate the same
algebra 
${\cal A}=su(2)_1^{\,2}\oplus su(2)_1^{\,4}={\cal A}_f\oplus {\cal A}_b$.
Obviously,  the same structure arises on the right handed
sides.
\item
We will show that the spaces of states $\wt{\cal H}^1_b$ and $\wt{\cal H}^2_b$ 
of $\wt{\cal C}^1$ and $\wt{\cal C}^2$ both have 
self dual ${\cal J}:=(P,Q,R,S;\qu{P},\qu{Q},\qu{R},\qu{S})$-charge 
lattice
\beq{so81lattice}
\wt{\Gamma}_b = \left\{ \inv{\sqrt2}(x+y;x-y)\left| x,y\in \Z^4
\right.\right\}.
\eeq
In the Gepner type model $\wt{\cal C}^1=(\wt{2})^4$ we find $16$ fields
with dimensions $h=\qu{h}={1\over 4}$ which are uncharged under 
$(J,A;\qu{J},\qu{A})$; diagonalizing them with respect to the
${\cal J}$-action for $j\in\{P,Q,R,S\}$
we obtain fields $E^\pm_j, F^\pm_j$ uncharged under 
all $U(1)$-currents apart from $j$ and with $(j,\qu{\jmath})$-charge
${1\over\sqrt2}(\pm1,\pm1)$ and ${1\over\sqrt2}(\pm1,\mp1)$, respectively.
Namely,
\begin{eqnarray*}
E_P^\pm &=& 
\Phi^0_{\mp1,\mp1;\mp1,\mp1}\otimes\Phi^0_{\pm1,\pm1;\pm1,\pm1}\otimes
\Phi^0_{\mp1,\mp1;\mp1,\mp1}\otimes\Phi^0_{\pm1,\pm1;\pm1,\pm1}, \\
F_P^\pm &=& 
\Phi^0_{\mp1,\mp1;\pm1,\pm1}\otimes\Phi^0_{\pm1,\pm1;\mp1,\mp1}\otimes
\Phi^0_{\mp1,\mp1;\pm1,\pm1}\otimes\Phi^0_{\pm1,\pm1;\mp1,\mp1}, \\
E_Q^\pm &=& 
\Phi^0_{\mp1,\mp1;\mp1,\mp1}\otimes\Phi^0_{\pm1,\pm1;\pm1,\pm1}\otimes
\Phi^0_{\pm1,\pm1;\pm1,\pm1}\otimes\Phi^0_{\mp1,\mp1;\mp1,\mp1}, \\
F_Q^\pm &=& 
\Phi^0_{\mp1,\mp1;\pm1,\pm1}\otimes\Phi^0_{\pm1,\pm1;\mp1,\mp1}\otimes
\Phi^0_{\pm1,\pm1;\mp1,\mp1}\otimes\Phi^0_{\mp1,\mp1;\pm1,\pm1}, \\
\end{eqnarray*}
and with $\eps_R:=-1, \eps_S:=1$ for $j\in\{R,S\}$
\begin{eqnarray*}
E_j^\pm &=&
\left(\Phi^1_{2,1;2,1}\right)^{\otimes 4} 
+ \eps_j\left(\Phi^1_{2,1;-2,-1}\right)^{\otimes 4}\\
&&\pm \left[
\Phi^1_{2,1;-2,-1}\otimes\Phi^1_{2,1;2,1}
\otimes\Phi^1_{2,1;-2,-1}\otimes\Phi^1_{2,1;2,1}\right.\\
&&\left.\quad\quad
+ \eps_j\, \Phi^1_{2,1;2,1}\otimes\Phi^1_{2,1;-2,-1}\otimes
\Phi^1_{2,1;2,1}\otimes\Phi^1_{2,1;-2,-1}
\right],\\
F_j^\pm &=&
\left(\Phi^1_{2,1;2,1}\right)^{\otimes 2} \otimes
\left(\Phi^1_{2,1;-2,-1}\right)^{\otimes 2} 
+ \eps_j\left(\Phi^1_{2,1;-2,-1}\right)^{\otimes 2}\otimes
\left(\Phi^1_{2,1;2,1}\right)^{\otimes 2}\\
&&\pm \left[
\Phi^1_{2,1;-2,-1}\otimes\Phi^1_{2,1;2,1}
\otimes\Phi^1_{2,1;2,1}\otimes\Phi^1_{2,1;-2,-1}\right.\\
&&\left.\quad\quad
+ \eps_j\, \Phi^1_{2,1;2,1}\otimes\Phi^1_{2,1;-2,-1}\otimes
\Phi^1_{2,1;-2,-1}\otimes\Phi^1_{2,1;2,1}
\right].
\end{eqnarray*}
Among the corresponding charges under ${\cal J}$ we find
$\smash{{1\over\sqrt2} (e_i;\pm e_i)}$ generating $\wt{\Gamma}_b$.

In the sigma model $\smash{\wt{\cal C}^1}$ we set
$$
\begin{array}{rclrcl}\ds
\alpha_1 \>:=\> \inv{\sqrt2}\left( e_1+e_2 \right), \quad 
\> \alpha_2 \>:=\> \inv{\sqrt2}\left( e_2-e_1 \right), \e
\alpha_3 \>:=\> \inv{\sqrt2}\left( e_1+e_3 \right), \quad
\>\alpha_4 \>:=\>  \inv{\sqrt2}\left( e_4-e_2 \right).
\end{array}
$$
Let $\Sigma_\delta$ with $\delta\in\F_2^4$ 
denote the twist field corresponding
to the fixed point ${1\over 2}\sum_{i=1}^4 \delta_i\alpha_i$. 
The action of $P,Q,R,S$ and their right handed partners is determined
as in \req{uaction}. Then by normalizing appropriately and matching
$({\cal J},\qu{\cal J})$-charges we find that the following identifications
can be made (sums run over $\delta\in\F_2^4$ with the indicated restrictions):
\begin{eqnarray*}
E_P^\pm &=& 
\sum_{\delta_1=\delta_2, \delta_3=\delta_4}
\Sigma_\delta 
\pm \sum_{\delta_1\neq\delta_2, \delta_3\neq\delta_4} \Sigma_\delta ,\\
F_P^\pm &=& 
\sum_{\delta_1\neq\delta_2, \delta_3=\delta_4}
\Sigma_\delta 
\pm \sum_{\delta_1=\delta_2, \delta_3\neq\delta_4} \Sigma_\delta ,\\
E_Q^\pm &=& 
\sum_{\delta_1=\delta_2, \delta_3=\delta_4}
(-1)^{\delta_4}\Sigma_\delta 
\pm \sum_{\delta_1\neq\delta_2, \delta_3=\delta_4} 
(-1)^{\delta_4}\Sigma_\delta ,\\
F_Q^\pm &=& 
\sum_{\delta_1\neq\delta_2, \delta_3\neq\delta_4}
(-1)^{\delta_3}\Sigma_\delta 
\pm \sum_{\delta_1=\delta_2, \delta_3\neq\delta_4} 
(-1)^{\delta_3}\Sigma_\delta ,\\
E_R^\pm &=& 
\sum_{\delta_1=\delta_2, \delta_3=\delta_4}
(-1)^{\delta_1}\Sigma_\delta 
\pm \sum_{\delta_1=\delta_2, \delta_3\neq\delta_4} 
(-1)^{\delta_1}\Sigma_\delta ,\\
F_R^\pm &=& 
\sum_{\delta_1\neq\delta_2, \delta_3\neq\delta_4}
(-1)^{\delta_2}\Sigma_\delta 
\pm \sum_{\delta_1\neq\delta_2, \delta_3=\delta_4} 
(-1)^{\delta_2}\Sigma_\delta ,\\
E_S^\pm &=& 
\sum_{\delta_1=\delta_2, \delta_3=\delta_4}
(-1)^{\delta_2+\delta_3}\Sigma_\delta 
\pm \sum_{\delta_1\neq\delta_2, \delta_3\neq\delta_4} 
(-1)^{\delta_2+\delta_3}\Sigma_\delta ,\\
F_S^\pm &=& 
\sum_{\delta_1\neq\delta_2, \delta_3=\delta_4}
(-1)^{\delta_2+\delta_3}\Sigma_\delta 
\pm \sum_{\delta_1=\delta_2, \delta_3\neq\delta_4} 
(-1)^{\delta_2+\delta_3}\Sigma_\delta  .
\end{eqnarray*}
In particular, the corresponding $({\cal J},\qu{\cal J})$--charges
generate $\wt{\Gamma}_b$.
\vspace{-1em} 
\end{enumerate}
\epr
Recall the Greene-Plesser construction for mirror symmetry \cite{grpl90}
to observe that the $\Z_2\times\Z_2$ orbifold $(\wt{2})^4$
of $(2)^4$ is invariant
under mirror symmetry. This can be regarded as an explanation for the 
high degree of symmetry found for $(\wt{2})^4=\wt{\cal C}^1$.

In view of \req{so81lattice} it is clear that the same phenomenon as
described in remark \ref{meetingpoint} appears for the theory discussed
above:
\brem{meetingpointtwo}
The Gepner type model $\wt{\cal C}^1=(\wt{2})^4$, or equivalently the
nonlinear $\sigma$ model
$\wt{\cal C}^2={\cal K}({1\over\sqrt2}D_4, B^\ast)$,
$B^\ast$ given by \req{fkbfield}, is located
at a meeting point of the moduli
spaces of theories associated to $K3$ surfaces and tori, respectively.
Namely, its 
bosonic sector is identical with that of the nonlinear
$\sigma$ model $\wt{\cal C}^3$ on the $SU(2)_1^{\,4}$-torus with 
vanishing B-field.
\erem
This again can be deduced from
the results in \cite{kosa88} once one observes that the lattice
denoted by $\Lambda_{O(n)\times O(n)}$ there in the case $n=4$ is
isomorphic to $\wt{\Gamma}_b$ as defined in \req{so81lattice}.
The relation between the two meeting points 
$(2)^4={\cal C}^1={\cal C}^2\cong{\cal C}^3$ and 
$(\wt{2})^4=\wt{\cal C}^1=\wt{\cal C}^2\cong\wt{\cal C}^3$
of the moduli spaces found so far is best understood by observing that 
$\wt{\cal C}^1=(\wt{2})^4$ can be constructed from 
${\cal C}^1=(\wh{2})^4$ by modding out 
$\Z_2\cong\langle[2,0,2,0]\rangle\subset{\cal G}_{alg}^{ab}$. 
If we formulate the 
orbifold procedure in terms of the charge lattice $\Gamma_b$
of ${\cal C}^1=(\wh{2})^4$  as described in \cite{grpl90}, 
this amounts to a shift orbifold by 
the vector $\delta={1\over2}(-1,1,0,0;1,-1,0,0)$ on $\Gamma_b$. 
Indeed, this shift
simply reverts the shift we used to explain remark \ref{meetingpoint}
and brings us back onto the torus $T_{SU(2)_1^{\,4}}$. 
But as for ${\cal C}^1={\cal C}^2$ and ${\cal C}^3$,
$\wt{\cal C}^1=\wt{\cal C}^2$ and $\wt{\cal C}^3$ will correspond to 
different compactifications of the type IIA string.

\begin{figure}[ht]
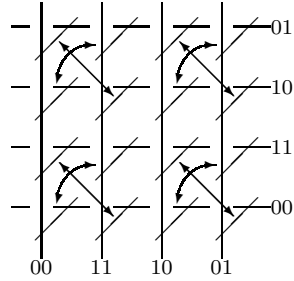

\hspace*{\fill}
\zzwei[4mm]{
\qbezier(5.5,6.4)(5.8,7.4)(6.7,7.4)
\multiput(5.6,6.5)(-4,0){2}{\vector(-1,-4){0.1}}
\multiput(6.7,7.4)(-4,0){2}{\vector(1,0){0.1}}
\multiput(6.5,6.6)(-4,0){2}{\vector(1,-1){0.9}}
\multiput(6.5,6.6)(-4,0){2}{\vector(-1,1){0.9}}
\multiput(5.6,2.5)(-4,0){2}{\vector(-1,-4){0.1}}
\multiput(6.7,3.4)(-4,0){2}{\vector(1,0){0.1}}
\multiput(6.5,2.6)(-4,0){2}{\vector(1,-1){0.9}}
\multiput(6.5,2.6)(-4,0){2}{\vector(-1,1){0.9}}
\qbezier(1.5,6.4)(1.8,7.4)(2.7,7.4)
\qbezier(5.5,2.4)(5.8,3.4)(6.7,3.4)
\qbezier(1.5,2.4)(1.8,3.4)(2.7,3.4)
}
\hspace*{\fill}
\caption{Action of the algebraic automorphism $t_{1111}$ on the Kummer lattice
$\Pi$.}\label{t1234}
\end{figure}

From \req{symmid} we are able to determine
the geometric counterpart of  $[2,0,2,0]$ on 
${\cal K}(\Z^4,0)$:
It is the unique nontrivial central element $t_{1111}$ of the algebraic
automorphism group ${\cal G}^+_{Kummer}$  depicted in figure \ref{t1234}.
Hence the commutant of $t_{1111}$ is the entire ${\cal G}_{Kummer}^+$, but it
is not clear so far how to continue the residual ${\cal G}_{Kummer}^+/\Z_2$
algebraically to the twisted sectors in $(\wt{2})^4$ with respect to the
$t_{1111}$ orbifold.

We remark that  conformal field theory also helps us to draw 
conclusions on the geometry of the Kummer surfaces under inspection:
${\cal K}({1\over\sqrt2}D_4,B^\ast)$ is obtained from ${\cal K}(\Z^4,0)$
by modding out the classical symmetry $t_{1111}$, so in terms of the
decomposition \req{decomp} we stay in the same ``chart'' of 
${\cal M}^{K3}$, i.e. choose the same nullvector $\upsilon$ for both 
theories. This means that we can explicitly relate the respective geometric
data. For both Kummer surfaces we choose the
complex structures induced by the  $N=(2,2)$ algebra
in the corresponding
Gepner models $(\wh{2})^4$ and $(\wt{2})^4$. 
Thus we identify
$J^\pm=\left(\Phi^0_{\mp2,2;\mp2,2}\right)^{\otimes4}$ 
in both theories with the twoforms 
$\pi_\ast(dz_1\wedge dz_2), \pi_\ast(d\qu{z}_1\wedge d\qu{z}_2)$ defining
the complex structure of ${\cal K}(\Lambda)$. Here 
$\pi: T_\Lambda \rightarrow {\cal K}(\Lambda)$ is the rational map of 
degree two, $\Lambda=\Z^4$ or $\Lambda={1\over\sqrt2} D_4$, respectively.
Then both ${\cal K}(\Lambda)$  are
\textsl{singular} $K3$ surfaces (see section \ref{application}).
Given the lattices of the underlying tori
one can compute the intersection form for real and
imaginary part of the  above twoforms  defining the complex structure.
One finds that they span sublattices
of the transcendental lattices with forms $\diag(4,4)$ for ${\cal K}(\Z^4)$
and $\diag(8,8)$ for ${\cal K}({1\over\sqrt2} D_4)$, respectively. 
The factor of two difference was to be expected, because $t_{1111}$ has
degree two. Nevertheless, one may check that the transcendental lattices
themselves for both surfaces have quadratic form $\diag(4,4)$. Note that
for a given algebraic automorphism in general it is hard to decide how
the transcendental lattices transform under modding out 
\cite[Cor. 1.3.3]{in76}. In our case, we could read it off thanks to the
Gepner type descriptions of our conformal field theories.
\subsection{Gepner type description of $SO(8)_1/\Z_4$}\label{gepso81z4}
\hspace*{\fill}\\[-1em]
\btheo{z2andz4}
The Gepner type model ${\cal C}^1=(\wh{2})^4$ which agrees with
${\cal C}^2={\cal K}(\Z^4,0)$ by theorem \ref{su2hoch4id} admits a nonlinear
$\sigma$ model description as $\Z_4$ orbifold of the torus model
${\cal T}({1\over\sqrt2}D_4,B^\ast)$ with $SO(8)_1$ symmetry.
\etheo
\bpr
The proof works analogously to that of theorem \ref{su2h4mod4id}.
From theorem \ref{z4production} it follows that the
$\Z_4$ orbifold of ${\cal T}({1\over\sqrt2}D_4,B^\ast)$ with 
$B^\ast$ defined by \req{fkbfield} is obtained from 
$\wt{\cal C}^2={\cal K}({1\over\sqrt2}D_4,B^\ast)$ 
by modding out the automorphism
$r_{12}$ as depicted in figure \ref{z4gen}. 
Thus we should work with the models $\wt{\cal C}^1=(\wt{2})^4$
and $\wt{\cal C}^2={\cal K}({1\over\sqrt2}D_4,B^\ast)$ which are isomorphic
by theorem \ref{so8id}.
We use the notations introduced there. Then $r_{12}$ is induced by
$e_1\mapsto e_2, e_2\mapsto -e_1, e_3\mapsto -e_4, e_4\mapsto e_3$.
Of the $su(2)_1^{\,6}$ current algebra 
of $\wt{\cal C}^2$ we find a surviving
$su(2)_1^{\,2}\oplus u(1)^4$ current algebra on the $\Z_4$ orbifold
generated by $J, J^\pm, A; P, P^\pm, Q,R,S$ (see equations \req{su2h2curalg} 
and \req{extrasu2}). The action on the generators $E_j^\pm, F_j^\pm;
j\in\{P,Q,R,S\}$ is already diagonalized. All the $E_j^\pm$ are invariant
as well as $F_P^\pm$. On the fermionic part of the space of states
of $\wt{\cal C}^2$ the identifications \req{torusforms} hold. 
The fields $W_{\eps_1,\eps_2}^J$ and $W_{\eps_1,\eps_1}^A, \eps_i\in\{\pm1\}$
are those invariant under the $\Z_4$ action. 
Our field by field identifications of theorem \ref{so8id} now allow us to
read off the induced action on the Gepner type model 
$\wt{\cal C}^1=(\wt{2})^4$. One checks that it agrees with the  symmetry
$[2^\prime,2^\prime,0,0]$ defined in \req{gpbackwards} which revokes the
orbifold by the $\Z_2$ action of $[2,2,0,0]$. Because 
$\wt{\cal C}^1=(\wt{2})^4$ was constructed from the Gepner model $(2)^4$
by modding out $\Z_2\times\Z_2\cong\langle [2,2,0,0],[2,0,2,0]\rangle
\subset {\cal G}_{ab}^{alg}$, it follows that the $\Z_4$ orbifold
of ${\cal T}({1\over\sqrt2}D_4,B^\ast)$ agrees with the Gepner type
model obtained from $(2)^4$ by modding out $\Z_2\cong\langle [2,0,2,0]\rangle$.
This clearly is isomorphic to $(\wh{2})^4$ by a permutation of the minimal
model factors.
\vspace*{-2em}
\epr
\section{Conclusions:
A panoramic picture of the moduli space}\label{conc}
\begin{figure}[ht]
\setlength{\unitlength}{1.8em}
\hbox to \hsize{\hspace*{\fill}
\begin{picture}(18,13)(-7,-3)
\multiput(-5.2,-1)(3,3){2}{\line(1,0){12.2}}
\multiput(-5.2,-1)(12.2,0){2}{\line(1,1){3}}
\qbezier[30](9.4,1.7)(9.4,2.5)(9.4,3.5)
\put(9.4,3.5){\makebox(0,0)[cb]{
\framebox(7.5,1.7){\parbox{13.5em}{
\hspace*{\fill}$\Z_2$ Orbifolds ${\cal K}(\Lambda,B_T)$, \hspace*{\fill}\\
\hspace*{\fill}$T=\R/\Lambda$, 
$B={1\over\sqrt2}B_T+{1\over2}B_\Z^{(2)}$\hspace*{\fill}
}}}}
\qbezier(-3.2,-0.8)(-2,0.5)(-1.2,-0.3)
\qbezier(-1.2,-0.3)(0,-1)(1,-0.5)
\qbezier(1,-0.5)(2.4,0.6)(3.2,-0.8)
\put(-3.6,-1.2){\line(-1,-1){1.5}}
\put(3.6,-1.2){\line(1,-1){1.5}}
\qbezier[40](-3.5,-1.8)(-5,-1.8)(-5.5,-1.8)
\put(-5.5,-1.8){\makebox(0,0)[rc]{\framebox(3.9,1.3){
\parbox{7em}{\hspace*{\fill}Tori ${\cal T}(\Lambda,B_T)$,\hspace*{\fill}\\
\hspace*{\fill} $T=\R^4/\Lambda\;$\hspace*{\fill}
}}}}
\put(-3.5,-0.5){\line(3,1){6}}
\qbezier[20](2.2,1.4)(2.7,1.9)(3.2,2.4)
\put(3.4,2.4){\makebox(0,0)[cb]{$\Lambda\sim\Z^4$}}
\put(-2,0){\circle*{0.2}}
\put(-2.2,0.2){\makebox(0,0)[rb]{$(\wh{2})^4$}}
\qbezier[80](-2,0)(-3,0)(-6,0)
\put(-6.1,0){\makebox(0,0)[rc]{\parbox{5em}{$SO(8)_1/\Z_4$\\
$\cong{\cal K}(\Z^4,0)$}}}
\qbezier[40](-2,0)(-2,-0.5)(-2,-2)
\put(-1.8,-2.2){\makebox(0,0)[ct]{\parbox{4em}{${\cal T}(D_4,0)$}}}
\multiput(-2,0.2)(0,0.8){4}{\line(0,1){0.6}}
\put(-2,0.2){\vector(0,-1){0.1}}
\put(-2,3.4){\vector(0,1){0.4}}
\put(-2.1,2.4){\makebox(0,0)[rb]{$\alpha$}}
\multiput(1.9,1.5)(0,0.4){13}{\line(0,1){0.2}}
\put(1.9,6.7){\vector(0,1){0.2}}
\multiput(5.6,1.5)(0,0.4){10}{\line(0,1){0.2}}
\put(5.6,5.4){\vector(0,1){0.2}}
\put(2.1,4.3){\makebox(0,0)[lb]{$r_{12}$}}
\put(5.5,4.3){\makebox(0,0)[rb]{$r_{12}$}}
\multiput(-1.4,0)(0.8,0){4}{\line(1,0){0.6}}
\put(-1.4,0){\vector(-1,0){0.1}}
\put(1.7,0){\vector(1,0){0.1}}
\put(0.8,0){\makebox(0,0)[rb]{$\beta$}}
\multiput(2.2,1.3)(0.4,0){9}{\line(1,0){0.2}}
\put(5.2,1.3){\vector(1,0){0.4}}
\put(3.8,1.4){\makebox(0,0)[lb]{$t_{1111}$}}
\put(0.5,-0.5){\line(3,1){6}}
\qbezier[20](6.2,1.4)(6.7,1.9)(7.2,2.4)
\put(7.4,2.4){\makebox(0,0)[cb]{$\Lambda\sim D_4$}}
\put(2,0){\circle*{0.2}}
\put(2,0.2){\makebox(0,0)[lb]{$(\wt{2})^4$}}
\qbezier[80](2,0)(4.5,-1)(7,-2)
\put(7.1,-2){\makebox(0,0)[lt]{\parbox{6em}{${\cal K}({1\over\sqrt2}D_4, 
B^\ast)$}}}
\qbezier[40](2,0)(2,-0.5)(2,-2)
\put(2.4,-2.2){\makebox(0,0)[ct]{\parbox{4em}{${\cal T}(\Z^4,0)$}}}
\put(5,1){\circle*{0.2}}
\put(5,0.8){\makebox(0,0)[lt]{\parbox{5em}{${\cal K}(D_4,0)$}}}
\qbezier[80](4.6,1)(3.5,1.5)(-1.2,0.1)
\put(4.6,1){\vector(4,-1){0.1}}
\put(3.8,1){\makebox(0,0)[rt]{$\omega$}}
\put(-4,2.5){\line(4,3){6}}
\put(-2,0){\line(4,3){9}}
\put(-2,0){\line(-1,3){1.5}}
\qbezier[80](0.4,3.675)(2.5,5.7)(4.5,7.2)
\put(5.5,7.5){\makebox(0,0)[cc]{$\Z_4$ Orbifold-plane}}
\qbezier[30](0,5.6)(0,7)(0,7.3)
\put(0,7.5){\makebox(0,0)[cc]{$\Z_4$ Orbifold-line}}
\put(-2,4){\circle*{0.2}}
\put(-2,4.2){\makebox(0,0)[cb]{$(2)^4$}}
\qbezier[80](-2,4)(-2.5,4)(-6,4)
\put(-6.1,4){\makebox(0,0)[rc]{\parbox{4em}{$\Lambda=\Z^4,$ $B_T=0$}}}
\put(-1,3.5){\line(-2,1){4}}
\qbezier[25](-4.5,6.7)(-4.5,5.9)(-4.5,5.3)
\put(-4.5,7){\makebox(0,0)[cc]{Quartic line}}
\multiput(-1.8,3.8)(1,-1){3}{\line(1,-1){0.7}}
\put(-1.8,3.8){\vector(-1,1){0.1}}
\put(1.2,0.8){\vector(1,-1){0.7}}
\put(-0.8,2.8){\makebox(0,0)[lb]{$\gamma$}}
\end{picture}
\hspace*{\fill}}
\setlength{\unitlength}{1ex}
\caption{Strata of the moduli space.}\label{modulispacepic}
\end{figure}
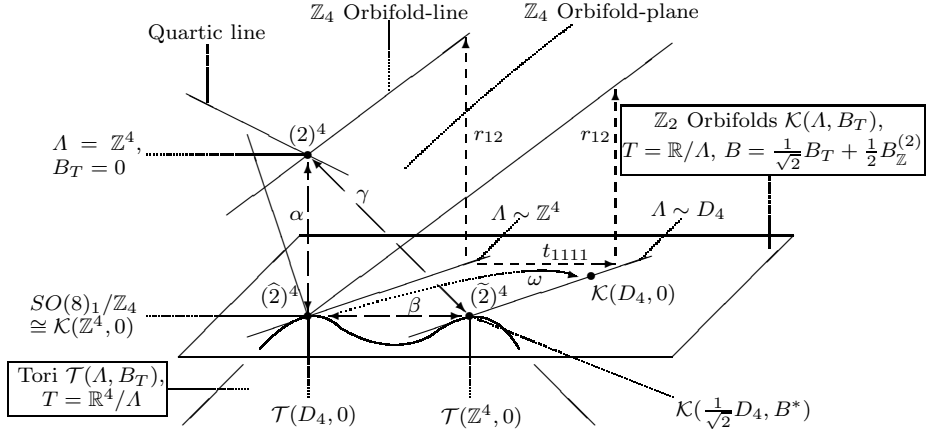

We conclude by joining  the information we gathered so far to
a panoramic picture of those strata of the moduli space we have 
fully under control now (figure \ref{modulispacepic}).

The rest of this section is devoted to a summary of what we have
learned about the various components depicted in
figure \ref{modulispacepic}. 
All the strata are defined as quaternionic submanifolds of the moduli space 
${\cal M}^{K3}$ 
consisting of theories which admit certain restricted geometric
interpretations. In other words, a suitable choice of 
$\upsilon$ as  described in section \ref{modulispace}
yields $(\Sigma, V, B)$ such
that $\Sigma, B$ have the respective  properties. 
In the following we will always tacitly assume that 
an appropriate choice of 
$\upsilon$ has been performed already.

Figure \ref{modulispacepic} contains two strata of real dimension $16$,
depicted as a 
horizontal plane and a mexican hat like object, respectively.
The horizontal plane is the 
\textsl{Kummer stratum}, the subspace
of the moduli space consisting of all theories which admit a geometric
interpretation on a Kummer surface $X$ in the orbifold limit. 
In other words, it is the $16$ dimensional
moduli space of all theories ${\cal K}(\Lambda, B_T)$ obtained
from a nonlinear $\sigma$ model on a torus $T=\R^4/\Lambda$ by applying
the ordinary $\Z_2$ orbifold procedure; the B-field takes values
$B={1\over\sqrt2}B_T+{1\over 2}B_\Z^{(2)}$, where  
$B_T\in H^2(T,\R)\hookrightarrow H^2(X,\R)$ (see the explanation after
theorem \ref{kummercriterion}), and
$B_\Z^{(2)}\in H^{even}(X,\Z)$ as described in  
theorem \ref{toremb}. We have an embedding
${\cal M}^{tori}\hookrightarrow{\cal M}^{K3}$
as quaternionic submanifold, and we know how to
locate this stratum within ${\cal M}^{K3}$.
Kummer surfaces in the orbifold limit
have a generic group $\F_2^4$ of algebraic automorphisms
which leave the metric invariant. Any conformal
field theory associated to such a Kummer surface possesses an 
$su(2)_1^{\,2}$ subalgebra \req{su2h2curalg} of the 
holomorphic W-algebra.

The mexican hat like object in figure \ref{modulispacepic} depicts the
moduli space \req{torms} of theories associated to tori.
Two meeting points with the Kummer stratum have been 
determined so far, namely $(\wh{2})^4$ and
$(\wt{2})^4$ (see remarks \ref{meetingpoint} and 
\ref{meetingpointtwo}). We found
$(\wh{2})^4={\cal K}(\Z^4, 0)={\cal T}(D_4,0)$ and
$(\wt{2})^4={\cal K}(\smash{\inv{\sqrt2}D_4},B^\ast)={\cal T}(\Z^4,0)$, where
$B^\ast$ was defined in \req{fkbfield}.

The vertical plane in figure \ref{modulispacepic} depicts
a stratum of real dimension $8$, namely the moduli space of theories
admitting a geometric interpretation as $\Z_4$ orbifold of a nonlinear
$\sigma$ model on $T=\R^4/\Lambda$. In order for the orbifold procedure
to be well defined we assume $\Lambda$ to be generated by 
$\Lambda_i\cong R_i\Z^2, R_i\in\R, i=1,2$ 
($\Lambda_1$ is not necessarily orthogonal to $\Lambda_2$) and
$B_T\in H^2(T,\R)^{\Z_4}\hookrightarrow H^2(X,\R)$ (see lemma 
\ref{bipiinh2}). The B-field then takes values 
$B={1\over2}B_T+{1\over4}B_\Z^{(4)}$ 
as described in theorem \ref{z4emb}, where 
the embedding of this stratum in ${\cal M}^{K3}$ is also explained. The 
generic group of algebraic automorphisms for $\Z_4$ orbifolds is
$\Z_2\ltimes \F_2^4$. By theorem \ref{z2andz4} there is a meeting point
with the Kummer stratum in the $\Z_4$ orbifold of
${\cal T}({1\over\sqrt2}D_4, B^\ast)$, where $B^\ast$ is given by
\req{fkbfield}, which  agrees with ${\cal K}(\Z^4,0)=(\wh{2})^4$.

The four lines in figure \ref{modulispacepic} are strata of real dimension
$4$ which are defined by restriction to theories admitting a 
geometric interpretation $(\Sigma,V,B)$ with fixed $\Sigma$ and allowed
B-field values $B\in\Sigma$.
Thus the volume is the only geometric parameter along the lines and 
we can associate a fixed hyperk\a hler structure  on $K3$ to each of them.
For all four lines it turns out that one can choose a complex structure
such that the respective $K3$ surface is  singular.
Hence $\Sigma$ can be
described by giving the quadratic form on the transcendental lattice and
the K\a hler class for this choice of complex structure. 
Specifically we have:
\begin{itemize}
\item
\textsl{$\Z^4$-line:} The subspace of the Kummer stratum given by theories
${\cal K}(\Lambda,B_T)$ with $\Lambda\sim\Z^4$ and $B_T\in\Sigma$,
which is marked by $\Lambda\sim\Z^4$ in figure \ref{modulispacepic}.
\item
\textsl{$\Z_4$ Orbifold-line:} The moduli space of all theories which
admit a geometric interpretation on a $K3$ surface obtained
from the nonlinear $\sigma$ model on a torus $T=\R^4/\Lambda$,
$\Lambda\sim\Z^4$ with B-field $B_T$ commuting with the automorphisms
listed in \req{klein}.
\item
\textsl{Quartic line:}
Though  well established in the context of Landau-Ginzburg theories,
this stratum has been
somewhat conjectural up to now. We describe it as the moduli space
of theories admitting a geometric interpretation 
$(\Sigma_{\cal Q},V_{\cal Q},B_{\cal Q})$
on the Fermat quartic
\req{kuqua} equipped with a K\a hler metric in the class of 
the Fubini-Study metric,
in order for $\Sigma_{\cal Q}$ to be invariant under 
the algebraic automorphism group
$G=\Z_4^2\rtimes {\cal S}_4$. The B-field is 
restricted to values $B_{\cal Q}\in\Sigma_{\cal Q}$,
because $\mu(G)=5$ and therefore $H^2(X,\R)^G=\Sigma_{\cal Q}$.
\item
\textsl{$D_4$-line:} The moduli space of 
theories ${\cal K}(\Lambda,B_T), \Lambda\sim D_4$
admitting as geometric interpretation a
Kummer surface  ${\cal K}(\Lambda)$ and 
$B_T\in\Sigma$. This line is labelled by $\Lambda\sim D_4$ in figure 
\ref{modulispacepic}.
\end{itemize}
The four lines are characterized by the following data\footnote{The 
quadratic form for the transcendental lattice of
quartic and the $\Z_4$ orbifold of $T=\R^4/\Z^4$
can be found in \cite{in76,shin77}.}:
\begin{center}
\begin{tabular}{l||c|c|c}
name of line & \parbox{6.4em}{associated form on the transcendental lattice
$\vphantom{X_{p}}$} 
& \parbox{9.3em}{group of algebraic automorphisms leaving the metric invariant}
& \parbox{5.9em}{generic $(1,0)$--cur\-rent algebra} \\[12pt]
\hline\hline&&&\\[-10pt]
$\Z^4$-line & $\left( \begin{array}{cc} 4&0\\0&4 \end{array} \right)$
&\vphantom{$\sum^{1\over2}$} 
$\begin{array}{l} {\cal G}_{Kummer}^+=\Z_2^2\ltimes\F_2^4\\
\;\;\cong(\Z_2\times\Z_4)\rtimes D_4\end{array}$
& $su(2)_1^{\,2}$ \\[8pt]
\hline&&&\\[-8pt]
$\Z_4$ orbifold-line & $\left( \begin{array}{cc} 2&0\\0&2 \end{array} \right)$
& $D_4$ & $su(2)_1\oplus u(1)$ \\[8pt]
\hline&&&\\[-8pt]
quartic line & $\left( \begin{array}{cc} 8&0\\0&8 \end{array} 
\right)$
& $(\Z_4\times\Z_4)\rtimes {\cal S}_4$ & $su(2)_1$ \\[8pt]
\hline&&&\\[-8pt]
$D_4$-line & $\left( \begin{array}{cc} 4&0\\0&4 \end{array} \right)$
& $\Z_2\ltimes\F_2^4$ & $su(2)_1^{\,2}$ 
\end{tabular}
\end{center}
In figure \ref{modulispacepic} we have two different
shortdashed arrows indicating
relations between lines. Consider the Kummer surface 
${\cal K}(\Z^4)$ associated to the $\Z^4$-line.
As demonstrated in theorem \ref{z4production}, the group ${\cal G}_{Kummer}^+$ 
of algebraic automorphisms of ${\cal K}(\Z^4)$ 
which leave the metric invariant
contains the automorphism $r_{12}$ of order two 
(see figure \ref{z4gen}) which upon modding out
produces the $\Z_4$ orbifold-line. 
The entire moduli space of  $\Z_4$ orbifold conformal field theories
is obtained this way from $\Z_2$ orbifold theories ${\cal K}(\Lambda,B_T)$,
where $\Lambda$ is generated by $\Lambda_i\cong R_i\Z^2, R_i\in\R, i=1,2$ and
$B_T\in H^2(T,\R)^{\Z_4}$.

Modding out
$t_{1111}\in{\cal G}_{Kummer}^+$ (see figure \ref{t1234})
on the $\Z^4$-line
produces the $D_4$-line, as argued at
the end of section \ref{gepso81}. 
Note that the
$K3$ surfaces associated to $\Z^4$- and $D_4$-lines 
have the same quadratic form
on their transcendental lattices and hence  are
identical as algebraic varieties. 
Still, the corresponding lines in moduli space are different because
different K\a hler classes are fixed. In our terminology this is expressed
by the change of lattices of the underlying tori on transition 
from one line to the other.
The $D_4$-line can also be viewed as the image of the $\Z^4$-line
upon shift orbifold on the underlying torus.

Finally, we list the zero dimensional strata shown in figure
\ref{modulispacepic}.

To construct
${\cal K}(D_4, 0)$ on the $D_4$-line,
we may as well apply the
ordinary $\Z_2$ orbifold procedure to the $D_4$-torus theory in the meeting
point $(\wh{2})^4$ (the arrow with label $\omega$ in figure 
\ref{modulispacepic}). We stress that in contrast to what was conjectured
in \cite{eoty89} this is not a meeting point with the $\Z_4$ orbifold-plane.

As demonstrated in theorem \ref{su2h4mod4id} and also conjectured in
\cite{eoty89}, Gepner's model
$(2)^4$ is the point of enhanced symmetry
$\Lambda=\Z^4, B_T=0$ on the $\Z_4$ orbifold-line. 
In section \ref{gepsym} we have studied the algebraic symmetry group
of $(2)^4$ and in corollary \ref{gepquart} proved that it
admits a geometric interpretation with Fermat quartic target space, too. 
In terms of the Gepner model, the moduli of
infinitesimal defomation along the $\Z_4$ orbifold and the quartic line
are real and imaginary parts of $V_{\delta,\eps}^\pm(\delta,\eps\in\{\pm1\})$
as in \req{volmod} and of the 
$(1,1)$-superpartners of 
$(\Phi^1_{\pm1,0;\pm3,2})^{\otimes 4}$,
$(\Phi^1_{\pm1,0;\pm1,0})^{\otimes 4}$,
respectively (see section \ref{gepsu2h4}).

The Gepner type models $(\wh{2})^4$ and $(\wt{2})^4$
which are meeting points of torus and $K3$ moduli spaces
have been mentioned above.
For all the longdash arrowed correspondences
$(2)^4 \stackrel{\alpha}{\longleftrightarrow} (\wh{2})^4
\stackrel{\beta}{\longleftrightarrow} (\wt{2})^4
\stackrel{\gamma}{\longleftrightarrow} (2)^4$ 
in figure \ref{modulispacepic} we explicitly know the
symmetries to be modded out from the Gepner (type) model as well as
the corresponding 
algebraic automorphisms on the geometric interpretations. 
For instance, 
$(\wt{2})^4\stackrel{r_{12}}{\longrightarrow}(\wh{2})^4
\stackrel{r_{12}}{\longrightarrow}(2)^4$.
Hence for these
examples we know precisely how to continue geometric symmetries to the
quantum level.

\begin{acknowledgements}
The authors would like to thank A.~Taormina for very helpful discussions on
$N=4$ superconformal field theory and
V.~Nikulin for his explanations concerning the geometry of Kummer surfaces. 
K.W. thanks F.~Rohsiepe for valuable discussions and his most efficient 
crash course in $C^{++}$. We thank M.~R\o sgen and F.~Rohsiepe for 
proof reading.

Work on this paper was supported by TMR.
\end{acknowledgements}
\begin{appendix}
\section{Minimal models and Gepner models}\label{gepner}
The $N=2$ minimal superconformal models form the discrete series
$(k), k\in\N$ of unitary representations of the $N=2$ superconformal algebra 
with central charges $c=3k/(k+2)$. 
For constructing the model $(k)$ we may start from  a $\Z_k$ parafermion 
theory and add a free bosonic field. More precisely, $(k)$ is the
coset model
\beq{minik}
{SU(2)_k \otimes U(1)_2 \over U(1)_{k+2, diag}}\; .
\eeq

The primary fields are denoted by $\Phi^l_{m,s;\qu{m},\qu{s}}(z,\qu{z})$, where
$l\in\{0,\dots,k\}$ is twice the spin of the corresponding field in the affine
$SU(2)_k$ and we have tacitly specialized to the
diagonal invariant by imposing $l=\qu{l}$. The remaining quantum numbers
$m, \qu{m}\in\Z_{2(k+2)}$ and $s, \qu{s}\in\Z_4$ 
label the representations
of $U(1)_{k+2, diag}$ and $U(1)_2$ in the decomposition \req{minik}, 
respectively, and must obey $l\equiv m+s \equiv \qu{m}+\qu{s}\; (2)$. 
Here, the fields with even (odd) $s$ create states in the 
lefthanded Neveu-Schwarz (Ramond) sector, and analogously for $\qu{s}$
and the righthanded sectors. Moreover the identification
\beq{flop}
\Phi^l_{m,s;\qu{m},\qu{s}}(z,\qu{z})
\sim
\Phi^{k-l}_{m+2+k,s+2;\qu{m}+2+k,\qu{s}+2}(z,\qu{z})
\eeq
holds. By \req{minik},
the corresponding characters $X^l_{m,s;\qu{m},\qu{s}}$ can be
obtained from the level $k$ string functions 
$c^l_j, l\in \{0,\dots,k\}, j\in\Z_{2k}$ of $SU(2)_k$ and classical
theta functions $\Theta_{a,b}, a\in\Z_{2b}$ of level $b=2k(k+2)$ by
\cite{ge88,raya87,qi87}
\beqn{minichar}
X^l_{m,s;\qu{m},\qu{s}} (\tau,z)
\> = \> \chi^l_{m,s}(\tau,z) \cdot
\chi^l_{\qu{m},\qu{s}}(\qu{\tau},\qu{z}) ,\e
\chi^l_{m,s}(\tau,z)
\> = \>
\sum_{j=1}^k c^l_{4j+s-m}(\tau) \Theta_{2m-(k+2)(4j+s), 2k(k+2)} 
\left(\smash{\tau, {z\over k+2}}\right).
\eeqn
Modular transformations act by
\beqn{minimodtrafo}
\chi^l_{m,s}(\tau +1, z)
\> = \> exp\left[ 2\pi i \left( 
{l(l+2)-m^2\over 4(k+2)} + {s^2\over 8} -{c\over 24}
\right)\right] \chi^l_{m,s}(\tau , z)
\\[2ex]\ds
\chi^l_{m,s}\left(-\inv{\tau}, \inv[z]{\tau}\right)
\> = \> \kappa(k)
\sum_{l^\prime, m^\prime, s^\prime}
\sin\left( {\pi(l+1)(l^\prime + 1)\over k+2}\right)
e^{\pi i{m m^\prime\over (k+2)}} e^{-\pi i {s s^\prime\over 2}}
\chi^{l^\prime}_{m^\prime,s^\prime}(\tau, z),
\eeqn
where $\kappa(k)$ is a constant depending only on $k$
and the summation runs over 
$l^\prime\in\{0,\dots, k\}, m^\prime\in\{-k-1,\dots,k+2\},
s^\prime\in\{-1,\dots,2\}, l^\prime + m^\prime + s^\prime \equiv 0 \; (2)$.

Let $\psi^l_{m,s}$ denote a lowest 
weight state in the irreducible representation
of the $N=2$ superconformal algebra with character $\chi^l_{m,s}$. Conformal
dimension and charge of $\psi^l_{m,s}$  then are
\beq{minidimcharge}
h^l_{m,s} = {l(l+2)-m^2\over 4(k+2)} + {s^2\over 8} \mod 1 , \quad
Q^l_{m,s} = {m\over k+2} - {s\over 2} \mod 2 . 
\eeq
The fusion--algebra is
\beq{minifusion}
\left[ \vphantom{\psi^{l^\prime}_{m^\prime,s^\prime}} \psi^l_{m,s}\right]
\times \left[ \psi^{l^\prime}_{m^\prime,s^\prime}\right]
= 
\sum_{ \stackrel{\qu{l} = |l-l^\prime| }{ 
\scriptscriptstyle\qu{l}\equiv l+l^\prime (2)} }^{
\min{(l+l^\prime, 2k-l-l^\prime) }}
\left[ \psi^{\qu{l}}_{m+m^\prime,s+s^\prime}\right].
\eeq
Note that by \req{minidimcharge} and \req{minifusion} the operators
of left and right handed spectral flow are associated to the fields
$\Phi^0_{-1,-1;0,0}=\psi^0_{-1,-1}$ and 
$\Phi^0_{0,0;-1,-1}=\qu{\psi^0_{-1,-1}}$, respectively.

The NS-part of our modular invariant partition function
is now given by
\beq{minipartfctn}
Z_{NS}(\tau,z)
= \inv{2} \hspace*{-1.5em}
\sum_{ \stackrel{ 
\stackrel{l=0,\dots, k}{m=-k-1,\dots,k+2} 
}{\scriptscriptstyle l+m\equiv 0(2)}} 
\left( \chi_m^{l,0}(\tau,z) + \chi_m^{l,2}(\tau,z) \right)
\left( \chi_{m}^{l,0}(\qu{\tau},\qu{z}) 
+ \chi_{m}^{l,2}(\qu{\tau},\qu{z}) \right),
\eeq
and expressions for the other three parts $Z_{\wt{NS}}, Z_{R}, Z_{\wt{R}}$
are obtained by flows as 
described in \req{sectors}.

In the case $k=2$ which we employ in this paper, the parafermion
algebra is nothing but the algebra satisfied by the Majorana fermion $\psi$ of
the Ising model. By inspection of the charge lattice one may confirm
that the minimal model $(2)$ can readily be constructed by tensoring the
Ising model with the one dimensional free theory which describes 
a bosonic field
$\phi$ compactified on a circle of radius $R=2$. 
The primary fields decompose as
\beqn{statesin2}
\Phi^l_{m,s;\qu{m},\qu{s}}(z,\qu{z})
\> = \>
\Xi^l_{m-s;\qu{m}-\qu{s}}(z,\qu{z})\;
e^{ {i\over 2\sqrt{2}} (-m+2s) \phi}(z)\;
e^{ {i\over 2\sqrt{2}} (-\qu{m}+2\qu{s}) \qu{\phi}}(\qu{z})
\e
\Xi^0_{j;\qu{\jmath}}(z,\qu{z}) 
\> = \> \Xi^2_{j\pm 2;\qu{\jmath}\pm 2}(z,\qu{z}) 
\; = \; \xi^0_{j}(z)\xi^0_{\qu{\jmath}}(\qu{z}), \quad
\xi^0_0 = \id, \; \xi^0_2 = \psi ,
\eeqn
and $\Xi^1_{1,1} = \Xi^1_{-1,-1}$, $\Xi^1_{1,-1} = \Xi^1_{-1,1}$
denote the ground states of the
two $h=\qu{h}={1\over 16}$ representations of the Ising model.
Indeed,  the
level 2 string functions are obtained from the characters 
of lowest
weight representations in the Ising model  by dividing by the Dedekind 
eta function.
To construct a Gepner model with central charge $c=3d/2, d\in\{2,4,6\}$, 
one first takes the (fermionic) tensor
product of $r$ minimal models $\otimes_{i=1}^r (k_i)$ such that 
the central charges add up to
$\sum_{i=1}^r 3k_i/(k_i+2) = 3d/2$.
The bosonic modes acting on different theories commute
and the fermionic modes anticommute. 
More concretely \cite[(4.5)]{fks92},
\beq{gepferm}
\Phi^{l_1}_{m_1,s_1;\qu{m}_1,\qu{s}_1}
\otimes \Phi^{l_2}_{m_2,s_2;\qu{m}_2,\qu{s}_2}
= (-1)^{{1\over 4}(s_1-\qu{s}_1)(s_2-\qu{s}_2)}
\Phi^{l_2}_{m_2,s_2;\qu{m}_2,\qu{s}_2}
\otimes\Phi^{l_1}_{m_1,s_1;\qu{m}_1,\qu{s}_1}.
\eeq
The diagonal sums $T,J,G^{\pm}$ of the
fields which generate the $N=2$ algebras of the factor theories $(k_i)$
then comprise a total $N=2$ superconformal 
algebra of central charge $c=3d/2$. 
Denote by ${\cal Z}$ the cyclic group generated by
$e^{2\pi i J_0}$, then ${\cal Z}\cong \Z_n$ with 
$n=\lcm\{2; k_i+2, i=1,\dots, r\}$. 
Now the Gepner model $\prod_{i=1}^r (k_i)$ is the orbifold of 
$\otimes_{i=1}^r (k_i)$ with respect to ${\cal Z}$. Effectively 
this means that  
$\prod_{i=1}^r (k_i)$ is obtained from $\otimes_{i=1}^r (k_i)$ by
projecting onto integer left and right charges in the $(NS+\wt{NS})$-sector,
onto integer or half integer left and right charges in the 
$(R+\wt{R})$-sector according to $c$ being even or odd,
and adding twisted 
sectors for the sake of modular invariance. In particular, the so constructed
model describes an $N=(2,2)$ superconformal field theory with central charge
$c=3d/2$ and (half) integer charges. For $d=4$ the Gepner model is thus 
associated to 
a $K3$ surface or a torus, as discussed in the introduction. 
We again decompose the partition function as in \req{sectors} and find
\beqn{geppartfctn}
Z_{NS}(\tau,z)
\>=\> \sum_{b=0}^n \sum_{(\vec{l},\vec{m})} 
\!\!\vphantom{\sum^m}^\prime \prod_{j=1}^r \left(
\inv{2}  
\left( \vphantom{\chi_{m_j+2b}^{l_j,0}(\qu{\tau},\qu{z}}
\chi_{m_j}^{l_j,0}(\tau,z) + \chi_{m_j}^{l_j,2}(\tau,z) \right)\right.\cdot\e
\>\>\quad\left.\cdot
\left( \chi_{m_j+2b}^{l_j,0}(\qu{\tau},\qu{z}) 
+ \chi_{m_j+2b}^{l_j,2}(\qu{\tau},\qu{z}) \right)
\right),
\eeqn
where $\sum_{(\vec{l},\vec{m})} ^\prime$ denotes the sum over all values 
$(\vec{l},\vec{m})\in\Z^{2r}$ with
$l_j\in \{0,\dots,k_j \}$, $m_j\in \{-k_j-1,\dots,k_j+2 \}, 
l_j+m_j\equiv 0\; (2)$ and 
$\sum_{j=1}^r {m_j\over k_j+2}$, $\sum_{j=1}^r {\qu{m}_j\over k_j+2}\in\Z$ .
We note that the field 
$\prod_{j=1}^r \Phi^{l_i}_{m_j,s_j; \qu{m}_j, \qu{s}_j}$
of the resulting Gepner model
belongs to the $b$th twisted sector with respect to the orbifold by
${\cal Z}$ iff $2b\equiv (\qu{m}_j-m_j)\mod n$ for $j=1,\dots, r$.
This means that the $(b+1)$st twisted sector is obtained from the $b$th
twisted sector by applying the twofold right handed spectral flow which
itself is associated to the primary field 
$\left(\Phi^0_{0,0;-2,2}\right)^{\otimes r}$ of our theory.
We explicitly see that for $c=6$ the fields
$\left(\Phi^0_{\mp 2,2;0,0}\right)^{\otimes r}$ belonging to the operators
of twofold lefthanded spectral flow are nothing but the $SU(2)$-currents
$J^\pm$  which extend the $N=2$ superconformal algebra to an 
$N=4$ superconformal algebra, and analogously for the 
righthanded algebra.
Moreover, to calculate $Z_{NS}(\tau,z;\qu{\tau},\qu{z})$
instead of using the closed formula \req{geppartfctn} one may proceed
as follows:
Start by multiplying the NS-parts of the partition functions of the
minimal models $(k_i), i=1,\dots, r$.  
Keep only the ${\cal Z}$-invariant i.e. integrally charged part 
of this function; let us denote the result by $F(\tau,z;\qu{\tau},\qu{z})$. 
Add the $b$th twisted sectors, $b=1,\dots, n$, by performing a
$2b$-fold righthanded spectral flow, i.e. by adding
$\qu{q}^{db^2/4} \qu{y}^{db/2} F(\tau,z;\qu{\tau},\qu{z}+b\qu{\tau})$.
This way calculations get extremely simple as soon as the characters
of the minimal models are written out in terms of classical theta functions.

We further note that to accomplish Gepner's actual construction of 
a consistent theory of superstrings in $10-d$ dimensions 
we would firstly have to take into account $8-d$ additional free superfields
representing flat (10-d)-dimensional Minkowski space in light-cone gauge,
secondly perform the GSO projection onto odd integer left and right 
charges and thirdly convert the resulting theory into a heterotic one. 
However, at the stage described above we have constructed
a consistent conformal field theory with central charge
$c=3d/2$ which for $d=4$
is associated to a $K3$ surface or a torus, so we may and 
will omit these last three steps of Gepner's construction.
\section{Explicit field identifications:
$(\wh{2})^4={\cal K}(\Z^4,0)$}\label{fieldlist}
In this appendix, we give a complete list of  
$({1\over 4},{1\over 4})$-fields in 
$(\wh{2})^4$ (see theorem \ref{su2hoch4id}) together with
their equivalents in the nonlinear $\sigma$ model on ${\cal K}(\Z^4,0)$. 
As usual, $\eps,\eps_i\in\{\pm 1\}$
and we use notations as in \req{uaction} and \req{torusforms}.
\\[3mm]
\textbf{Untwisted $({1\over 4},{1\over 4})$-fields with respect to
the $\langle[2,2,0,0]\rangle$-orbifold}:
\begin{eqnarray*}\label{quarterlistuntwa}
\left(\Phi^0_{-\eps_1,-\eps_1;-\eps_2,-\eps_2}\right)^{\otimes 4}
&=& W^J_{\eps_1,\eps_2}   \\
\left(\Phi^0_{-\eps,-\eps;-\eps,-\eps}\right)^{\otimes 2}
\otimes\left(\Phi^0_{\eps,\eps;\eps,\eps}\right)^{\otimes 2}
&=& W^A_{\eps,\eps}\\
\left(\Phi^{1}_{2,1;2,1}\right)^{\otimes 4}
&=& \Sigma_{0000}-\Sigma_{1100}+\Sigma_{1111}-\Sigma_{0011} \\
\left(\Phi^{1}_{2,1;-2,-1}\right)^{\otimes 4}
&=& \Sigma_{1010}+\Sigma_{0101}-\Sigma_{0110}-\Sigma_{1001} 
\end{eqnarray*}
$$
\begin{array}{l}\label{quarterlistuntwb}
\ds
\left(\Phi^{1}_{2,1;2,1}\right)^{\otimes 2}\otimes
\Phi^{0}_{-1,-1;-1,-1}\otimes\Phi^{0}_{1,1;1,1} \e
\quad\quad\quad= \Sigma_{0000}-\Sigma_{1100}-\Sigma_{1111}+\Sigma_{0011}
+\Sigma_{0010}+\Sigma_{0001}-\Sigma_{1101}-\Sigma_{1110} \e
\left(\Phi^{1}_{2,1;2,1}\right)^{\otimes 2}\otimes
\Phi^{0}_{1,1;1,1}\otimes\Phi^{0}_{-1,-1;-1,-1} \e
\quad\quad\quad= \Sigma_{0000}-\Sigma_{1100}-\Sigma_{1111}+\Sigma_{0011}
-\Sigma_{0010}-\Sigma_{0001}+\Sigma_{1101}+\Sigma_{1110} \e
\Phi^{0}_{-1,-1;-1,-1}\otimes\Phi^{0}_{1,1;1,1}\otimes
\left(\Phi^{1}_{2,1;2,1}\right)^{\otimes 2} \e
\quad\quad\quad= \Sigma_{0000}+\Sigma_{1100}-\Sigma_{1111}-\Sigma_{0011}
+\Sigma_{1000}+\Sigma_{0100}-\Sigma_{1011}-\Sigma_{0111} \e
\Phi^{0}_{1,1;1,1}\otimes\Phi^{0}_{-1,-1;-1,-1}\otimes
\left(\Phi^{1}_{2,1;2,1}\right)^{\otimes 2} \e
\quad\quad\quad= \Sigma_{0000}+\Sigma_{1100}-\Sigma_{1111}-\Sigma_{0011}
-\Sigma_{1000}-\Sigma_{0100}+\Sigma_{1011}+\Sigma_{0111} \e
\Phi^{0}_{-1,-1;-1,-1}\otimes\Phi^{0}_{1,1;1,1}\otimes
\Phi^{0}_{-1,-1;-1,-1}\otimes\Phi^{0}_{1,1;1,1} \e
\quad\quad\quad= \left(\Sigma_{0000}+\Sigma_{1100}+\Sigma_{1111}+\Sigma_{0011}\right)
+\left(\Sigma_{1000}+\Sigma_{0100}+\Sigma_{0111}+\Sigma_{1011}\right)\e
\quad\quad\quad +\left(\Sigma_{0010}+\Sigma_{0001}+\Sigma_{1101}+\Sigma_{1110}\right)
+\left(\Sigma_{1010}+\Sigma_{0101}+\Sigma_{0110}+\Sigma_{1001}\right) \e
\Phi^{0}_{-1,-1;-1,-1}\otimes\Phi^{0}_{1,1;1,1}\otimes
\Phi^{0}_{1,1;1,1}\otimes\Phi^{0}_{-1,-1;-1,-1} \e
\quad\quad\quad= \left(\Sigma_{0000}+\Sigma_{1100}+\Sigma_{1111}+\Sigma_{0011}\right)
+\left(\Sigma_{1000}+\Sigma_{0100}+\Sigma_{0111}+\Sigma_{1011}\right)\e
\quad\quad\quad -\left(\Sigma_{0010}+\Sigma_{0001}+\Sigma_{1101}+\Sigma_{1110}\right)
-\left(\Sigma_{1010}+\Sigma_{0101}+\Sigma_{0110}+\Sigma_{1001}\right) \e
\Phi^{0}_{1,1;1,1}\otimes\Phi^{0}_{-1,-1;-1,-1}\otimes
\Phi^{0}_{1,1;1,1}\otimes\Phi^{0}_{-1,-1;-1,-1} \e
\quad\quad\quad= \left(\Sigma_{0000}+\Sigma_{1100}+\Sigma_{1111}+\Sigma_{0011}\right)
-\left(\Sigma_{1000}+\Sigma_{0100}+\Sigma_{0111}+\Sigma_{1011}\right)\e
\quad\quad\quad -\left(\Sigma_{0010}+\Sigma_{0001}+\Sigma_{1101}+\Sigma_{1110}\right)
+\left(\Sigma_{1010}+\Sigma_{0101}+\Sigma_{0110}+\Sigma_{1001}\right) \e
\Phi^{0}_{1,1;1,1}\otimes\Phi^{0}_{-1,-1;-1,-1}\otimes
\Phi^{0}_{-1,-1;-1,-1}\otimes\Phi^{0}_{1,1;1,1} \e
\quad\quad\quad= \left(\Sigma_{0000}+\Sigma_{1100}+\Sigma_{1111}+\Sigma_{0011}\right)
-\left(\Sigma_{1000}+\Sigma_{0100}+\Sigma_{0111}+\Sigma_{1011}\right)\e
\quad\quad\quad +\left(\Sigma_{0010}+\Sigma_{0001}+\Sigma_{1101}+\Sigma_{1110}\right)
-\left(\Sigma_{1010}+\Sigma_{0101}+\Sigma_{0110}+\Sigma_{1001}\right)
\end{array}
$$
\textbf{Twisted $({1\over 4},{1\over 4})$-fields with respect to
the $\langle[2,2,0,0]\rangle$-orbifold}:
\begin{eqnarray*}\label{quarterlisttwia}
\left(\Phi^0_{-\eps,-\eps;\eps,\eps}\right)^{\otimes 2}
\otimes\left(\Phi^0_{\eps,\eps;-\eps,-\eps}\right)^{\otimes 2}
&=& W^A_{\eps,-\eps} \\
\left(\Phi^{1}_{2,1;-2,-1}\right)^{\otimes 2}\otimes
\left(\Phi^{1}_{2,1;2,1}\right)^{\otimes 2}
&=& \Sigma_{1000}-\Sigma_{0100}+\Sigma_{0111}-\Sigma_{1011} \\
\left(\Phi^{1}_{2,1;2,1}\right)^{\otimes 2}\otimes
\left(\Phi^{1}_{2,1;-2,-1}\right)^{\otimes 2}
&=& \Sigma_{0010}-\Sigma_{0001}+\Sigma_{1101}-\Sigma_{1110} 
\end{eqnarray*}
$$
\begin{array}{l}\label{quarterlisttwib}
\left(\Phi^{1}_{2,1;-2,-1}\right)^{\otimes 2}\otimes
\Phi^{0}_{-1,-1;-1,-1}\otimes\Phi^{0}_{1,1;1,1} \e
\quad\quad\quad\quad= \Sigma_{1000}-\Sigma_{0100}+\Sigma_{1011}-\Sigma_{0111}
+\Sigma_{1010}-\Sigma_{0101}+\Sigma_{1001}-\Sigma_{0110} \e
\left(\Phi^{1}_{2,1;-2,-1}\right)^{\otimes 2}\otimes
\Phi^{0}_{1,1;1,1}\otimes\Phi^{0}_{-1,-1;-1,-1} \e
\quad\quad\quad\quad= \Sigma_{1000}-\Sigma_{0100}+\Sigma_{1011}+\Sigma_{0111}
-\Sigma_{1010}+\Sigma_{0101}-\Sigma_{1001}+\Sigma_{0110} \e
\Phi^{0}_{-1,-1;-1,-1}\otimes\Phi^{0}_{1,1;1,1}\otimes 
\left(\Phi^{1}_{2,1;-2,-1}\right)^{\otimes 2} \e
\quad\quad\quad\quad= \Sigma_{0010}-\Sigma_{0001}-\Sigma_{1101}+\Sigma_{1110}
+\Sigma_{1010}-\Sigma_{0101}-\Sigma_{1001}+\Sigma_{0110} \e
\Phi^{0}_{1,1;1,1}\otimes\Phi^{0}_{-1,-1;-1,-1}\otimes
\left(\Phi^{1}_{2,1;-2,-1}\right)^{\otimes 2} \e
\quad\quad\quad\quad= \Sigma_{0010}-\Sigma_{0001}-\Sigma_{1101}+\Sigma_{1110}
-\Sigma_{1010}+\Sigma_{0101}+\Sigma_{1001}-\Sigma_{0110} \e
\end{array}
$$
\end{appendix}

\end{document}